\documentclass[%
reprint,
superscriptaddress,
nofootinbib,
 amsmath,amssymb,
 aps,
 showpacs,
 physrev,
]{revtex4-1}

\usepackage[unicode=true,pdfusetitle,
 bookmarks=true,bookmarksnumbered=false,bookmarksopen=false,
 breaklinks=false,pdfborder={0 0 1},backref=false,colorlinks=true]{hyperref}
\hypersetup{
 urlcolor=blue, citecolor=blue}
\usepackage{amsmath}
\usepackage{tabularx}
\usepackage{footnote}

\usepackage[pdftex]{color}
\usepackage[pdftex]{graphicx}
\usepackage{dcolumn}
\usepackage{bm}

\usepackage{lineno}

\usepackage[caption=false]{subfig}

\usepackage[normalem]{ulem} 
\begin{document}

\preprint{APS/123-QED}

\title{Precise Mass Measurement of the Longest Odd-Odd Chain of \boldmath $1^+$ Ground States}%


\author{B.~Liu}
\affiliation{Department of Physics and Astronomy, University of Notre Dame, Notre Dame, IN 46556, USA}
\affiliation{ Physics Division, Argonne National Laboratory, Lemont, IL 60439, USA}

\author{M.~Brodeur}
\affiliation{Department of Physics and Astronomy, University of Notre Dame, Notre Dame, IN 46556, USA}

\author{J.A.~Clark}
\affiliation{ Physics Division, Argonne National Laboratory, Lemont, IL 60439, USA}

\author{I.~Dedes}
\affiliation{Institute of Nuclear Physics Polish Academy of Sciences, PL-31 342 Krak\'ow, Poland}

\author{J.~Dudek}
\affiliation{Université de Strasbourg, CNRS, IPHC UMR 7178, F-67 000 Strasbourg, France}
\affiliation{Institute of Nuclear Physics Polish Academy of Sciences, PL-31 342 Krak\'ow, Poland}
\affiliation{Institute of Physics, Marie Curie-Sk\l odowska University, PL-20 031 Lublin, Poland}

\author{F.G.~Kondev}
\affiliation{ Physics Division, Argonne National Laboratory, Lemont, IL 60439, USA}

\author{D.~Ray}
\affiliation{Department of Physics and Astronomy, University of Manitoba, Winnipeg, MB R3T 2N2, Canada }
\affiliation{ Physics Division, Argonne National Laboratory, Lemont, IL 60439, USA}

\author{G.~Savard}
\affiliation{ Physics Division, Argonne National Laboratory, Lemont, IL 60439, USA}
\affiliation{Department of Physics, University of Chicago, Chicago, IL 60637, USA}

\author{A.A.~Valverde}
\affiliation{Department of Physics and Astronomy, University of Manitoba, Winnipeg, MB R3T 2N2, Canada }
\affiliation{ Physics Division, Argonne National Laboratory, Lemont, IL 60439, USA}

\author{A.~Baran}
\affiliation{Institute~of~Physics,~Marie Curie-Sk\l odowska University,
             PL-20\,031 Lublin, Poland}

\author{D.P.~Burdette}
\affiliation{ Physics Division, Argonne National Laboratory, Lemont, IL 60439, USA}

\author{A.M.~Houff}
\affiliation{Department of Physics and Astronomy, University of Notre Dame, Notre Dame, IN 46556, USA}

\author{R.~Orford}
\affiliation{Nuclear Science Division, Lawrence Berkeley National Laboratory, Berkeley, California 94720, USA}

\author{W.S.~Porter}
\affiliation{Department of Physics and Astronomy, University of Notre Dame, Notre Dame, IN 46556, USA}

\author{F.~Rivero}
\affiliation{Department of Physics and Astronomy, University of Notre Dame, Notre Dame, IN 46556, USA}

\author{K.S.~Sharma}
\affiliation{Department of Physics and Astronomy, University of Manitoba, Winnipeg, MB R3T 2N2, Canada }

\author{L.~Varriano}
\thanks{Present address: Center for Experimental Nuclear Physics and Astrophysics, University of Washington, Seattle, WA 98195}
\affiliation{Department of Physics, University of Chicago, Chicago, IL 60637, USA}
\affiliation{ Physics Division, Argonne National Laboratory, Lemont, IL 60439, USA}

\date{\today}

\begin{abstract}

Precise mass measurements of the ground and isomeric states of the odd-odd $^{108, 110, 112, 114, 116}$Rh were performed using the Canadian Penning Trap at Argonne National Laboratory, showing good agreement with recent JYFLTRAP measurements. A new possible isomeric state of $^{114}$Rh was also observed. These isotopes are part of the longest odd-odd chain of identical ground-state spin-parity assignment of 1$^+$, spanning $^{104-118}$Rh, despite being in a region of deformation. Realistic phenomenological mean-field calculations using ``universal'' Wood-Saxon Hamiltonian were performed, which explained this phenomenon for the first time. In addition, multi-quasiparticle blocking calculations were performed to study the configuration of low-lying states in the odd-odd Rh nuclei, elucidating anomalous isomeric yield ratio observed for $^{114}$Rh.

\end{abstract}


\maketitle



\section{\label{sec:intro}Introduction}

The nuclear shape is a fundamental property that can provide information on the underlying structure of the nucleus. Although doubly magic nuclei are spherical, the shape of nuclei with partially filled shells varies widely across the nuclear chart. The nuclear shape can change not only rapidly from one nucleus to its neighbor but also between the ground state and excited states of the same nucleus where shape co-existence could also happen at similar energies \cite{GARRETT2022}. 

Certain excited states, called spin trap isomers, often have a large spin difference compared to their ground state, resulting in suppressed $\gamma$ transitions to the ground state \cite{OrfordSpinTrap}. These isomers, usually strengthened by the complexity of the particle-hole excited structures, are useful in understanding single-particle excitation energies and spin-dependent residual interactions between unpaired nucleons, which are essential inputs to study the structure of odd-odd nuclei with deformation. However, the long lifetime of these isomers makes their $\gamma$ spectroscopic study challenging. On the other hand, such long lifetimes are ideal for performing an identification and mass measurement of these states using the phase-imaging ion-cyclotron-resonance (PI-ICR) Penning trap mass spectrometry technique \cite{OrfordSpinTrap, PhysRevLett.110.082501}.

The odd-odd Rh isotopes from $^{104}$Rh to $^{118}$Rh are particularly interesting not only because they are mid-shell, in a region of deformation, but also for a few other unique features. Firstly, the spin and parity assignment of the ground state of all these isotopes is $I^{\pi}_{gs}=1^+$ \cite{NUBASE}. Assuming that all assignments are correct, this is the longest chain of odd-odd isotopes with the same spin and parity in the entirety of the chart of the nuclides. To make it even more unusual, it is in a region of rapid change in nuclear shape (from prolate to oblate), which affects single particle orbitals and deformed-shell gaps and usually results in a change in the nuclear spin. Secondly, all of these isotopes have low-lying excited states with very similar excitation energies, where it is slightly lower for $^{110,112}$Rh. 

In this article, these peculiar aspects of neutron-rich odd-odd Rh isotopes are investigated using a multi-prong approach. Firstly, precision mass measurements of the $^{108, 110, 112, 114, 116}$Rh ground states and first isomeric states were performed, confirming recent Penning trap measurements \cite{JYFLarxiv} while discovering one more isomeric state. Secondly, detailed 
deformed Woods-Saxon Hamiltonian calculations \cite{WoodsSaxon} in its 
universal parametrization (similar to \cite{Porter2022}) of the single particle energy levels of these isotopes were performed resulting in an explanation of the peculiar spin and parity assignment of their ground states. Finally, multi-quasiparticle blocking calculations (similar to \cite{OrfordSpinTrap}) were performed to predict the configuration of low-lying states in the odd-odd Rh nuclei and investigate the anomalous yield of $^{114}$Rh. 

\section{\label{sec:exp}Experiment}

The precise mass measurement of neutron-rich Rh isotopes was conducted using the Canadian Penning Trap mass spectrometer (CPT) at the Californium Rare Isotope Breeder Upgrade (CARIBU) facility \cite{CARIBU} at Argonne National Laboratory (ANL). The CPT has performed several high-precision mass measurements over the past including ones where new isomers were discovered \cite{Hartley2018}.

\subsection{\label{subsec:setup}Experiment Setup}

The CARIBU facility uses a $^{252}$Cf fission source located at one end of a gas catcher to provides a wide range of fission products. The helium filled gas catcher collects these fission products and guides them out with the DC and RF fields applied. After the gas catcher, an isobar separator with a resolving power of $R=m / \Delta m\sim$10,000 selects the beam based on $A/q$ to remove non-isobaric contaminants \cite{Davids08}. This continuous beam is then cooled and bunched by a radio-frequency quadrupole (RFQ) cooler-buncher. After that, the bunched beam is sent to a multi-reflection time-of-flight mass spectrometer (MR-TOF) to gain further time-of-flight separation between the ion of interest and other isobaric isotope and molecules, with a resolving power of $R \sim 100,000$ \cite{CARIBU-MRTOF}. By controlling the gate time of a Bradbury-Nielsen gate (BNG), the ions of interest are selected and sent to the CPT tower. The CPT tower has a linear trap serving as the preparation trap to cool the ions to minimize energy dispersion before the CPT. A position-sensitive microchannel plate detector (PS-MCP) is located after the CPT to record the ions ejected from the CPT and measure their phase. 

\subsection{\label{subsec:PIICR}Measurement method}

In Penning trap mass spectrometry, the mass of an ion is determined from a measurement of its cyclotron frequency

\begin{equation}
\nu_c=\frac{qB}{2\pi m}
                                       \label{eq:vc1}
\end{equation}
where $B$ is the magnetic field and $q$ and $m$ are the charge and mass of the ion. This frequency can be measured using various techniques including the time-of-flight
ion-cyclotron-resonance (TOF-ICR) method~\cite{KONIG199595, Bollen:204719} and the PI-ICR  \cite{PhysRevLett.110.082501} techniques. The latter technique yields higher resolving power and better precision on short-lived isotopes produced at low rate by measuring the phase of the ion spots to determine the cyclotron frequency. The PI-ICR technique was used for the presented measurements.

The ion motion perpendicular to the axis of a Penning trap is a composition of a reduced cyclotron and a magnetron motion of different frequencies. The sum of the frequencies of these motions equals the cyclotron frequency and can be measured directly using the scheme presented in \cite{ORFORD2020491}. The ions are first excited by a dipole excitation to a certain radius and rotate with the reduced cyclotron frequency for a given amount of time, called accumulation time $t_{acc}$. Ions with different mass-to-charge ratios have different reduced cyclotron frequencies and thus acquire different phases during this time period. Then the ions are converted into magnetron motion by the quadrupole excitation before the ions are ejected. For a measurement, a reference spot is first recorded with zero accumulation time and a final spot is measured at a given non-zero $t_{acc}$. The total phase difference $\phi_{tot}$ is determined by the phase difference $\phi_\textit{diff}$ between the two spots and the additional number of turns $N$ which the final spot acquired. Finally, the cyclotron frequency $\nu_c$ can be calculated from the total phase accumulated $\phi_{tot}$ during $t_{acc}$ using

\begin{equation}
\nu_c=\frac{\phi_{tot}}{2\pi t_{acc}}=\frac{\phi_\textit{diff}+2\pi N}{2\pi t_{acc}}.
                                                                                                                              \label{eq:vc2}
\end{equation}

With longer $t_{acc}$, better precision can be achieved. By measuring the cyclotron frequency of two isotopes, the mass ratio of the two isotopes is obtained. Thus the mass of interest can be determined by this ratio together with the mass of the calibrant. 

\subsection{\label{subsec:exp}Experimental details}

To guarantee the accuracy of the measurement and avoid systematic effects from having overlapping spots of different species, prior to any measurement, great care was taken to identify the isotope of interest from all the neighboring isotopes or molecular contaminants. To accomplish this, the gate time of the BNG was scanned to inject the various neighboring isotopes in the trap. Then the cyclotron frequencies were determined over a range of accumulation times ranging from a few ms to hundreds of ms. These isotopes were identified by their measured cyclotron frequencies and comparing them to the expected values based on the AME2020 \cite{AME2020} mass excess. The abundance of each isotope was also checked against the predicted $^{252}$Cf fission yield. Observed contaminants not only included fission products but also molecular ions produced in the gas catcher.

\begin{figure}
\centering
\includegraphics[trim = 0mm 0mm 0mm 0mm, clip,width=\linewidth]{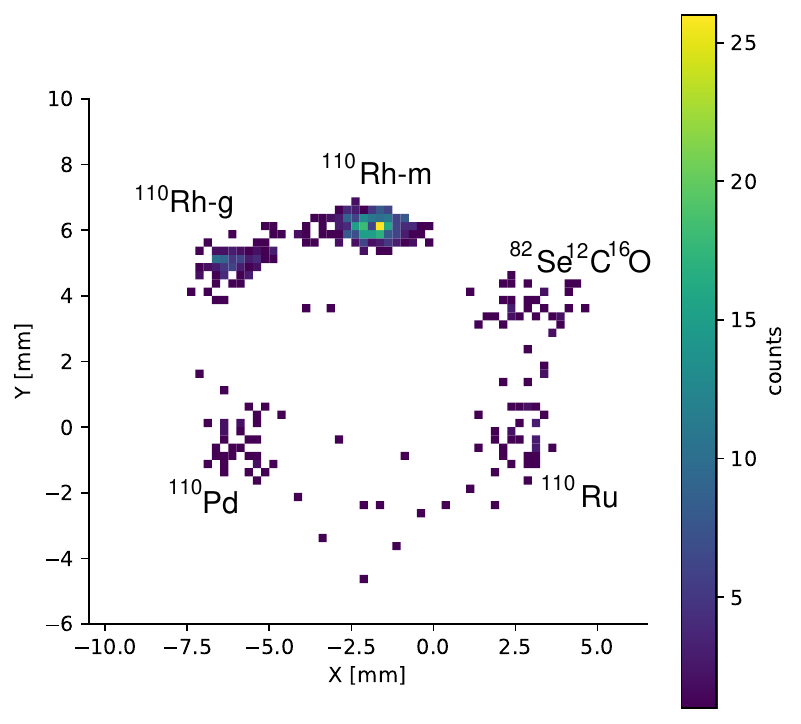}
\caption{A typical count histogram of the PS-MCP detector plane during the measurement of $^{110}$Rh with 450.614 ms accumulation time.}
\label{fig:spot} 
\end{figure}

To measure isomeric states, the $t_{acc}$ was increased slowly to unambiguously observe the separation of these different states for positive identification. Most of the measurements were conducted with a $t_{acc} \sim$ 430-450~ms, except for $^{114}$Rh, which was measured up to 700~ms to separate all three states present. Figure~\ref{fig:spot} is a typical measurement histogram plot with all the isotopes and isomeric states for $A=110$ identified.

The software SCM \cite{SCM} developed by the LEBIT group was used to search for the potential candidate molecules responsible for the observed contaminants. The software explores all possible combinations of elements, aiming to match the observed contaminant frequencies within a defined uncertainty range. The number of total element species and the number of atoms included in the search can be specified. Then the list of potential candidate molecules are examined manually to check the chemical feasibility of formation to identify the most likely molecule responsible for the observed contaminants.

Following the discussion in \cite{ORFORD2020491}, additional care was taken during measurements to minimize potential systematic effects such as ion-ion interaction, and imperfections in the electric potential. One systematic effect associated with the PI-ICR technique is the non-circular beam spot path as seen on the PS-MCP. This effect is either due to a misalignment of the extraction drift tube following CPT with the magnetic field or due to the asymmetry of the magnetic field itself. Hence, for all the measurements, the reference spot and the final spot were within 10 degrees to minimize the effect of this imperfection and to minimize our dependence on the accuracy of a precise trap center measurement. Furthermore, cyclotron frequency measurements were repeated with varying but closely spaced accumulation times to account for dependence on accumulation time. 


In addition, the cyclotron frequency has been observed to vary with the angular position of the spots. The measured cyclotron frequency $\nu$ of $^{133}$Cs$^{+}$ and $^{85}$Rb$^{+}$ from the stable ion source both present a dependency with the angle $\theta$ of the spot and can both be expressed using the empirical equation of the angle $\theta$: $\nu = \nu_0 + \nu_0 \cdot b \cdot cos(\theta + \phi)$ with the constants $b=3.89(32)\times 10^{-8}$, $ \phi=-45.7(36)^{\circ}$, where $\nu_0$ is the true cyclotron frequency. To account for this angular dependence, the measured cyclotron frequencies in this work are corrected using the empirical equation $\nu_0 = \nu / (1+ b \cdot cos(\theta + \phi))$. The changes in the frequency of rhodiums due to this angle correction are less than 0.02 Hz.

As seen in Table~\ref{tab:ME}, for all measurements except $^{114}$Rh, we used calibrant species with the same atomic mass numbers, which effectively quench most systematic uncertainties including the ones due to trap misalignment, non-homogeneity of the magnetic field and electric potential imperfections. For $^{114}$Rh, the calibrant $^{112}$Sn differs by only two atomic mass units, thereby minimizing mass-dependent systematic effects.



\section{\label{sec:result}Measurement results and discussion}

\begin{figure*}
\begin{center}
\includegraphics[trim = 0mm 0mm 0mm 0mm, clip,width=\linewidth]{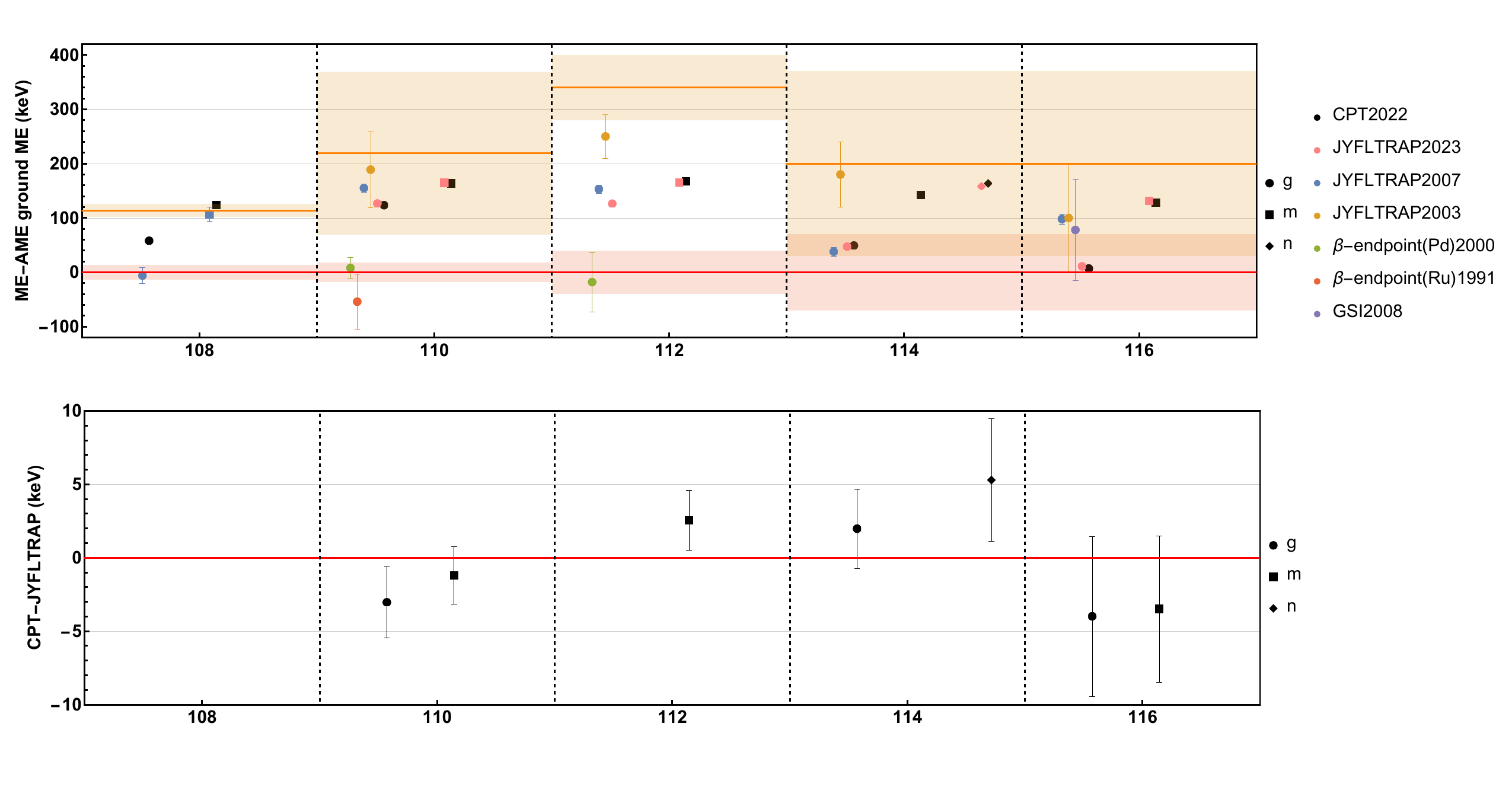}
\end{center}
\caption{\label{fig:ME} Top panel: Mass excess difference between this work and the AME ground state. The red solid line and red shaded region show the AME value and uncertainty range for the ground state. The orange solid line and orange shaded region show the AME value and uncertainty range for the isomeric state. All the experiment values adopted by the AME2020 \cite{AME2020} listed in Table~\ref{tab:ref} are shown by different colors: JYFLTRAP2007~\cite{07Ha20}, JYFLTRAP2003~\cite{03KoA}, $\beta$-endpoint(Pd)~\cite{00KrA}, $\beta$-endpoint(Ru)~\cite{91Jo11},  GSI2008~\cite{08KnA}. In addition, two JYFLTRAP measurements~\cite{07Ha20,03KoA} of $^{110}$Rh are also plotted for comparison. Different states are represented by different symbols. The recent publication from JYFLTRAP~\cite{JYFLarxiv} is plotted in pink. Bottom panel: The mass difference between this work and values from the recent JYFLTRAP measurement~\cite{JYFLarxiv} when a comparable state is present in both measurements.}
\end{figure*}

The masses of the ground state and isomeric state(s) of $^{108, 110, 112, 114, 116}$Rh were measured in the 1+ charge state using beams from CARIBU with the CPT during two separate campaigns in 2022. The results of these measurements are presented in Table~\ref{tab:ME} and shown by the black markers in Fig.~\ref{fig:ME}.

\begin{table*}
\caption{\label{tab:ME}%
Cyclotron frequency ratio and mass excess (ME) of the $^{108, 110, 112, 114, 116}$Rh states measured by the CPT and the excitation energy (Ex) calculated based on the measured mass value. The mass excess (ME$_{\textrm{AME}}$) and excitation energy (Ex$_{\textrm{AME}}$) from the AME2020 \cite{AME2020} are also listed. All the isotopes of interest and the calibrants were in the 1+ charge state. }
\begin{ruledtabular}
\begin{tabular}{ccccccc}
\textrm{Nuclide}&
\textrm{Calibration}&
\textrm{$\nu_c^{cal} / \nu_c$}&
\textrm{ME(keV)}&
\textrm{ME$_{\textrm{AME}}$(keV)}&
\textrm{Ex(keV)}&
\textrm{Ex$_{\textrm{AME}}$(keV)}\\
\colrule
\\
 $^{108}$Rh$^g$ & $^{108}$Pd &  1.000045281(11) &-84972.9(16) & -85031(14) &  & \\
$^{108}$Rh$^m$ & $^{108}$Pd & 1.000045935(13) & -84907.2(17) & -84917(12)& 65.7(23) & 115(18) \\
 $^{110}$Rh$^g$ & $^{110}$Pd & 1.0000549494(49)&-82705.43(79) & -82829(18)&   & \\
 $^{110}$Rh$^m$ & $^{110}$Pd & 1.0000553395(39)&-82665.49(73) & -82610(150)$\#$ & 39.9(11) & 220(150)$\#$\\
 $^{112}$Rh$^m$ & $^{112}$Sn & 1.0000872272(64)& -79562.64(73) & -79390(60) &  & 340(70) \\
 $^{114}$Rh$^g$ & $^{112}$Sn & 1.0179970780(66) & -75660.72(74) & -75710(70) &  & \\
 $^{114}$Rh$^m$ & $^{112}$Sn & 1.0179979702(71)& -75567.72(80) & -75510(70)$\#$ & 93.0(11) & 200(150)$\#$ \\
 $^{114}$Rh$^n$ & $^{112}$Sn & 1.0179981737(71)& -75546.51(80) & &114.2(11)\\
 $^{116}$Rh$^g$ & $^{116}$Cd &1.000166532(47)& -70733.0(51) & -70740(70) & & \\
 $^{116}$Rh$^m$ & $^{116}$Cd & 1.000167655(38)& -70611.8(41) & -70540(170)$\#$ & 121.2(65) & 200(150)$\#$ \\
 \\
\end{tabular}
\end{ruledtabular}
\end{table*}

The experimental mass values for $^{108, 110, 112, 114, 116}$Rh adopted by the AME2020 mainly originates from two JYFLTRAP measurements \cite{07Ha20,03KoA} and two $\beta$ decay endpoint measurements of Ru $\beta$ decay to Rh~\cite{91Jo11} or Rh $\beta$ decay to Pd~\cite{00KrA} together with a storage-ring measurement~\cite{08KnA}. All experimental values adopted by the AME2020 are listed in Table~\ref{tab:ref} and represented by color-coded markers in Fig.~\ref{fig:ME}. 

AME2020 includes only the ground state and one isomeric state for $^{108, 110, 112, 114, 116}$Rh and there are no experimental mass values of the isomeric states of $^{110, 112, 114, 116}$Rh. In this experiment, two states were found in $^{108}$Rh, $^{110}$Rh and $^{116}$Rh as the AME2020 predicted. However, only one state of $^{112}$Rh was observed, while three states of $^{114}$Rh were present.

JYFLTRAP also conducted mass measurements of $^{110, 112, 114, 116}$Rh recently \cite{JYFLarxiv} and those results are plotted for comparison in Fig.~\ref{fig:ME} by the coral-colored markers. 

\subsection{\texorpdfstring{$^{108}$Rh}{108Rh}}

$^{108}$Rh is the only rhodium isotope we measured that has experimental values for both the ground state and the isomeric state in AME2020. Both states are based on a 2007 JYFLTRAP measurement \cite{07Ha20} shown in blue in Fig.~\ref{fig:ME}. Our ground state mass is 58(14) keV higher than the AME value while our isomeric state, with an excitation of 65.7(23) keV, agrees with the AME2020 isomeric state mass excess. Our measurement improves the precision by an order of magnitude. The discrepancy in the ground state mass may warrant another independent measurement.

\subsection{\texorpdfstring{$^{110}$Rh}{110Rh}} 

Two states of $^{110}$Rh were observed and measured in this experiment. The measured ground state is 124(18)~keV higher than the current AME2020 value. The experimental values adopted for the AME2020 ground state are from two beta decay endpoint measurements: one from \cite{91Jo11} shown in red and a private communication \cite{00KrA} shown in green in Fig.~\ref{fig:ME}. Besides the beta-endpoint measurements, there are two Penning trap measurements by JYFLTRAP \cite{07Ha20,03KoA} which are not adopted by the AME2020 and both could be a result of a mixture of the ground state and isomeric state(s) as indicated in the publication. These are also plotted in Fig.~\ref{fig:ME} for comparison. The JYFLTRAP2007 measurement \cite{07Ha20} shown in blue lands between our ground state and isomeric state and leans closer to the latter. The JYFLTRAP2007 value would be more influenced by the isomeric state if the isomeric state was produced in greater abundance than the ground state, as was observed in this experiment. The large uncertainty in the JYFLTRAP2003 measurement \cite{03KoA}, shown in yellow, encompasses both our ground state and isomeric state. Our ground and isomeric values agree well with the most recent JYFLTRAP measurements \cite{JYFLarxiv}. 

During the measurement, we also observed another candidate isomeric state with an excitation energy of 377.4(13)~keV, which is not reported in the AME2020 nor in \cite{JYFLarxiv}. However, after a complete molecule search, a stable molecule $^{82}$Se$^{12}$C$^{16}$O$^+$, which is only 0.02Hz away from this candidate new state in frequency was found. Although this difference is greater than the measured uncertainty of the candidate new state (0.007Hz), it corresponds to just 3 degrees on the PS-MCP detector, much smaller than the typical spot spread and below our resolution limit. Furthermore, $^{80}$Se$^{12}$C$^{16}$O$^+$ was also present in the measurement of $^{108}$Rh, validating the formation of this type molecule in the system. Therefore, this ``new candidate state'' is highly likely the stable $^{82}$Se$^{12}$C$^{16}$O$^+$ molecule instead.


\subsection{\texorpdfstring{$^{112}$Rh}{112Rh}} 
Only one state of $^{112}$Rh was observed in this measurement. This observed state is 167(40)~keV higher than the current AME2020 ground state value but 173(60)~keV lower than the theoretical isomeric state value in AME2020. The current AME value for the $^{112}$Rh ground state is dominated at 66$\%$ by the beta endpoint measurements from a private communication \cite{00KrA} and the remaining contribution comes from the two JYFLTRAP measurements \cite{07Ha20,03KoA} shown by blue and yellow in Fig.~\ref{fig:ME}. The most recent JYFLTRAP measurement \cite{JYFLarxiv} reported both the ground state and the isomeric state. Both states were observed in their fission data. However the ground state production rate was too small to perform a precise measurement. As such, the ground state has also been produced by the in-trap decay of the even-even $^{112}$Ru for the mass measurement of that state ~\cite{JYFLarxiv}. The single state observed in this work has a consistent mass excess value with the isomeric state observed by JYFLTRAP. Following a further examination, the clustering algorithm identifies a hint of a possible spot in the ``tail'' of the prevailing isomeric state at a similar ground state mass to \cite{JYFLarxiv}. Because the existence of that spot was found in the analysis, the standard cyclotron frequency measurement procedure was not followed for it and as such no accurate frequency ratio corresponding to that state can be reported.


\subsection{\texorpdfstring{$^{114}$Rh}{114Rh}} 
The measured ground state agrees with the current AME2020 value and the two isomeric states are also within the range of the theoretical value as reported in NUBASE2020. The literature value of the $^{114}$Rh ground state in AME2020 come from the two prior JYFLTRAP measurements \cite{07Ha20,03KoA}. Our ground state mass departs from JYFLTRAP2003 but is only 11(8)~keV higher than the JYFLTRAP2007. Both our ground state and the isomeric state with an excitation energy of 114.2(11)~keV are consistent with the most recent JYFLTRAP measurement  \cite{JYFLarxiv}. We also observed a possible additional isomeric state with a slightly lower excitation energy of 93.0(11)~keV, which was not predicted in the AME2020 or reported in \cite{JYFLarxiv}. Note that this new state has also been observed recently by  JYFLTRAP \cite{JYFLarxiv2}.


It is worth noting that unlike all the other Rh isotopes in this measurement, where the production yield is predominantly from the isomeric state or the two states exhibit similar abundances, the production yield of $^{114}$Rh was observed to be clearly dominated by the 1$^+$ ground state instead of the (7$^-$) isomeric state. This contradicts the expectation that the isomeric state, with its higher spin state, should be more favorably produced by fission. This inversion in production yield ratio is also reported in \cite{JYFLarxiv}. The inversion in fission yield could be explained by decay loss if the half-life of the ground state is greater than the isomeric state. However only one half-life of 1.85(5)~s has been observed \cite{1988Ay02}. In \cite{2003Lh01}, the half-life of the high spin state has been measured to be 1.86(6)~s, matching the previous value  \cite{1988Ay02}. Furthermore \cite{2003Lh01} pointed out that the half-life of lower spin state 1$^+$ could be shorter than 1.86~s. Thus the current information of the half-life is not enough to explain the anomalous inversion of the yield. The other possibility is the inversion of the spin assignment of the ground state and isomeric state. This is discussed in more detail in Section \ref{subsec:Multiquasiparticle}.

\subsection{\texorpdfstring{$^{116}$Rh}{116Rh}} 
Both our ground state and isomeric state agree with the AME2020 and NUBASE2020. The $^{116}$Rh ground state AME2020 value is derived by the two JYFLTRAP measurements \cite{07Ha20,03KoA} and a GSI storage-ring experiment \cite{08KnA}. All three measurements fall between our ground state and isomeric state, which suggests that they may be the result of a mixture of two states as the publications suggested. Both our ground state and isomeric state are in close agreement with the most recent JYFLTRAP measurement in \cite{JYFLarxiv} as the bottom panel in Fig.~\ref{fig:ME} indicates.
 
\subsection{Summary} 

This section reported new Penning trap mass measurements of the ground and isomeric states of $^{108, 110, 112, 114, 116}$Rh that yielded consistent values with recent JYFLTRAP~\cite{JYFLarxiv} measurements, improving the confidence on the accuracy of the mass excess of different states in these isotopes. Both groups had the same difficulty observing two states of $^{112}$Rh directly from the fission beam. Additionally, both groups observed the inversion in yield of the ground state and isomeric state of $^{114}$Rh, in addition to the successful measurement of the ground state and one isomeric state as the AME2020 predicts. Similarly as in \cite{JYFLarxiv2}, we also report a possible additional isomeric state for $^{114}$Rh.

\begin{table}
\caption{\label{tab:ref}%
References of the experimental values adopted by the AME2020.}
\begin{ruledtabular}
\begin{tabular}{ccc}
\textrm{Nuclide}&
\textrm{Method}&
\textrm{Reference}\\
\colrule
\\
 $^{108}$Rh$^g$ & Penning Trap, TOF-ICR & \cite{07Ha20}  \\
$^{108}$Rh$^m$ & Penning Trap, TOF-ICR & \cite{07Ha20} \\
 $^{110}$Rh$^g$ & $\beta$-endpoint (Pd) & \cite{00KrA} \\
                       & $\beta$-endpoint (Ru) & \cite{91Jo11}\\
 $^{110}$Rh$^m$ & - & -\\
 $^{112}$Rh$^g$ & $\beta$-endpoint (Pd) & \cite{00KrA}  \\
                       & Penning Trap, TOF-ICR & \cite{07Ha20} \\
                       & Penning Trap, TOF-ICR & \cite{03KoA} \\
 $^{114}$Rh$^g$ & Penning Trap, TOF-ICR & \cite{07Ha20} \\
                       & Penning Trap, TOF-ICR & \cite{03KoA} \\
 $^{114}$Rh$^m$ & - & -\\
 $^{114}$Rh$^n$ & - & -\\
 $^{116}$Rh$^g$ & Penning Trap, TOF-ICR & \cite{07Ha20} \\
                       & Penning Trap, TOF-ICR & \cite{03KoA} \\
                       & Storage Ring & \cite{08KnA} \\
 $^{116}$Rh$^m$ & - & -\\
 \\
\end{tabular}
\end{ruledtabular}
\end{table}

\section{\boldmath Nuclear Mean-Field Interpretation of the longest odd-odd chain of $1^+$ ground states}
\label{Sect-Th-I}

Odd-odd nuclei $^{104-118}$Rh happen to form the longest consecutive chain of odd-odd isotopes with identical spin-parity assignments for all the ground states: $I^\pi_{\rm gs}=1^+$ for $N \in [63,71]$~\cite{NUBASE}. From the nuclear mean-field description view point, this implies that the single particle orbitals occupation of 6 consecutive odd neutrons leads to the same spin-parity values for the ground states. To discover the possible underling factors for this striking feature, we employed standard mean field techniques of analyzing multi-particle and particle-hole configurations for axially-symmetric nuclei known in the literature under the nickname `tilted Fermi surface algorithm".


The mean-field modeling used is based on a phenomenological approach with deformed ``universal'' Woods-Saxon potential\footnote{The universal Woods-Saxon Hamiltonian and associated universal parametrization have been developed in a series of articles~\cite{WS01,*WS02,*WS03,*WS04,*WS05} and continue to be used widely~\cite{Y2013-01,*Y2013-02,*Y2013-03,*Y2013-04,*Y2013-05,*Y2013-06,*Y2013-07,*Y2013-08}.}. The corresponding calculations usually involve a few typical steps, which will be shortly discussed and illustrated with the help of specifically adapted diagrams usually employed in this context. \\ 
$\bullet$ First, we construct single-nucleon diagrams as functions of quadrupole deformation to find the leading shell-gap sizes and their impacts especially on the prolate/oblate shape coexistence in axially symmetric nuclei which are abundant in the discussed zone (see Sec.~\ref{Sect-I-A});\\
$\bullet$ Second, we present the typical potential energy surfaces as functions of quadrupole deformations $(\alpha_{20},\alpha_{22})$, which allows to address the expected shape coexistence and evolution with varying nucleon numbers and angular momentum. Here we pay particular attention for possible domination of the axially symmetric configurations (see Sec.~\ref{Sect-I-B}); \\
$\bullet$ Third, we proceed to a detailed particle-hole analysis employing the Lagrange multiplier method, usually called ``tilted Fermi Surface algorithm'' (see Sec.~\ref{Sect-I-E} and Appendices).\\
$\bullet$ Since, as it will be shown, the universal Woods-Saxon mean field approach reproduces majority of experimental observations at hand, we formulate some rather general comments about possible Hamiltonian and/or parametrization selections, (see Sec. \ref{Sect-IV-D}).


\subsection{Single-Nucleon Spectra Underlying Microscopic Structure of the Nuclei of Interest}
\label{Sect-I-A}

As a starting point, we look at the deformation dependence of the orbitals in the region of $^{104-118}$Rh, between spherical magic numbers $N=50$ and 82, i.e., essentially the main-shell $N_{\rm main}=4$ and the intruder orbital $1h_{11/2}$. Concerning the mean field Hamiltonian parameters, they were optimized by applying a state-of-the-art variant of the Inverse Problem Theory paying particular attention to the removal of the parametric correlations, known to destroy the parametrization stability and what is referred to as {\em predictive power} (for details see~\cite{Ded19,Gaa2021}). The parameter optimization method was used in conjunction with Monte Carlo simulations which allows to obtain probabilities of uncertainties of parameters and consequently to estimate uncertainties of the final predictions (this will be the subject of a forthcoming publication). 

The same optimized Hamiltonian has been used for the total energy calculations and potential energy maps presented below.
Because the 9 independent parameters on which the Hamiltonian depends were optimized to the newest experimental data, we consider the presented approach ``state of the art", and since they are used without further modifications for all $\sim$3\,000 nuclei of the Atomic Mass Evaluation, we consider the parametrization in question to be ``universal'' \cite{PhDedes17,Ded19,Gaa2021}. Specific comments about experimental verifications of predictions obtained with this approach can be found in Sect. \ref{Sect-IV-D}.


\begin{figure}[h!]
\begin{center}
\includegraphics[width=0.48\textwidth,angle=-00]{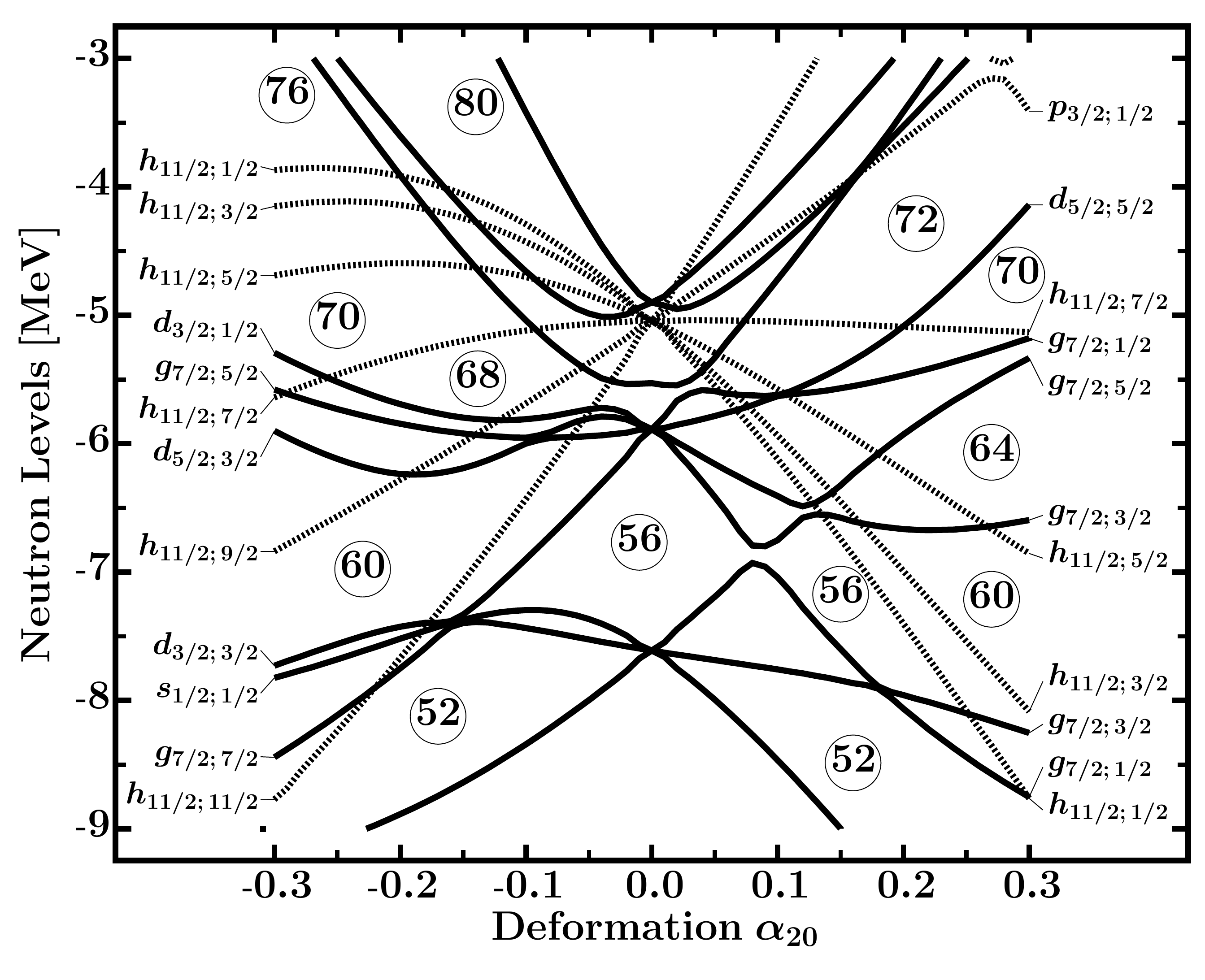}
\end{center}
\caption{Single particle neutron levels calculated employing realistic 
         phenomenological mean field Hamiltonian with the ``universal'' parameter set
         {~\cite{Gaa2021}}. Observe that for the moderate both prolate ($\alpha_{20}>0$)
         and oblate ($\alpha_{20}<0$) deformations the odd neutrons with $N \in [63,71]$ occupy levels
         with strongly varying quantum characteristics as can be seen from comparing 
         the spectroscopic labels.
                                                                     \label{Fig_01}
}
\end{figure}

One can read from the single particle diagram in~Fig.~\ref{Fig_01} that the odd-neutron in the compared isotopes can be placed in one of the orbitals with the following spectroscopic characteristics: $1h_{11/2,5/2}$, $1g_{7/2,3/2}$, $1g_{7/2,5/2}$, $1g_{7/2,1/2}$,  $1h_{11/2,7/2}$ and $2d_{5/2,5/2}$. This sequence introduces considerable variation in the microscopic structure of the wave functions occupied by the odd neutron which challenges our understanding of why the ground-state spin-parity remains constant in $^{104-116}$Rh. Furthermore, one notices the comparably large shell gaps on the prolate and oblate sides of the quadrupole axis especially close to the neutron numbers $N\sim70$ suggesting that oblate equilibrium deformations may be winning in the odd-$N$ isotopes $^{116,118}$Rh.

\begin{figure}[h!]
\begin{center}
\includegraphics[width=0.48\textwidth,angle=-00]{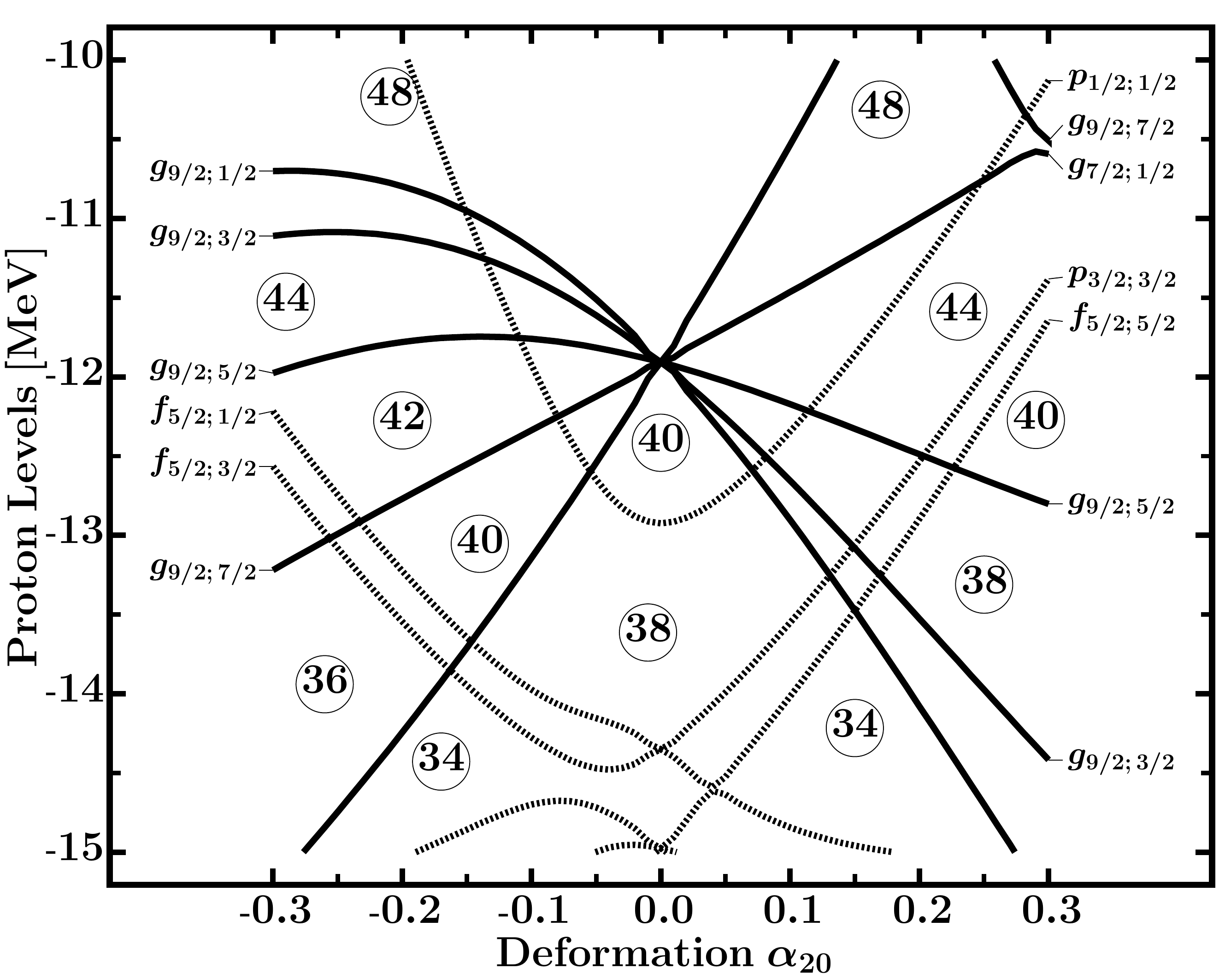}
\end{center}
\caption{Illustration similar to Fig.~\ref{Fig_01}, but for the protons. Please 
         note that the last proton in  discussed isotopes occupies the level $(Z=45)$ which passes through 
         the low level-density zones on both prolate and oblate sides, expected to 
         stimulate the prolate/oblate shape competition discussed in the text.  
         Both this diagram and Fig.~\ref{Fig_01} correspond to a selected central nucleus representative for the 
         whole series of nuclei discussed in this article.
                                                                     \label{Fig_02}
}
\end{figure}

The proton single particle energies are illustrated in Fig.~\ref{Fig_02} showing, among others, that the last occupied proton level at $Z=45$ is close, on the oblate shape side,  to the low level-density zone around $Z=40-42-44$ that has about 3 MeV in size. Such low level-density zones are known to generate strong negative shell-corrections of Strutinsky resulting in a local minima of the potential energy surfaces whose properties are discussed next.

\subsection{Prolate-Oblate-Triaxial Shape Competition Within Mean-Field Approach}
\label{Sect-I-B}

\begin{figure}[h!]
     \centering
         \centering
         \includegraphics[width=0.45\textwidth,angle=-00]{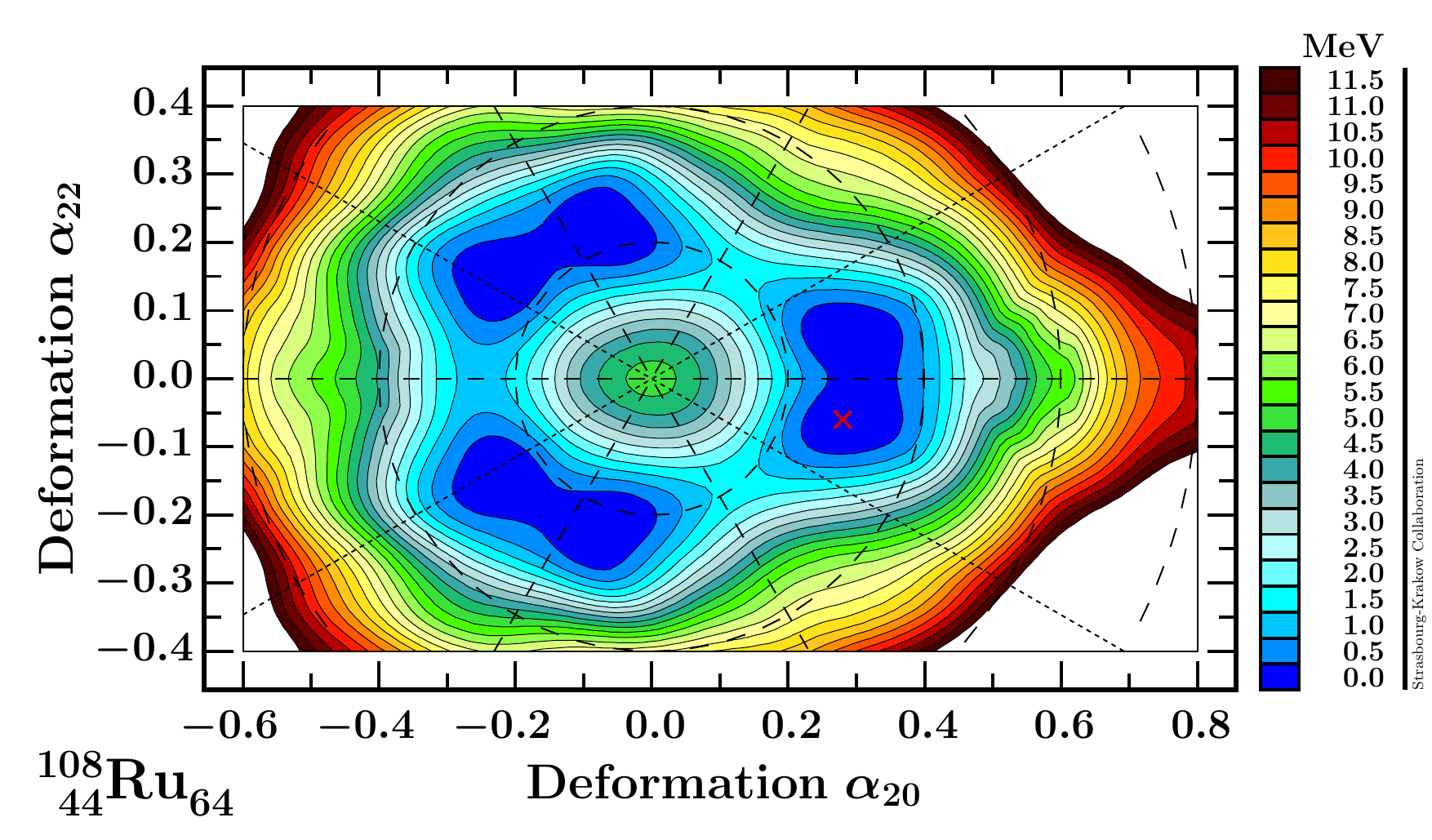}
     \hfill
         \centering
         \includegraphics[width=0.45\textwidth,angle=-00]{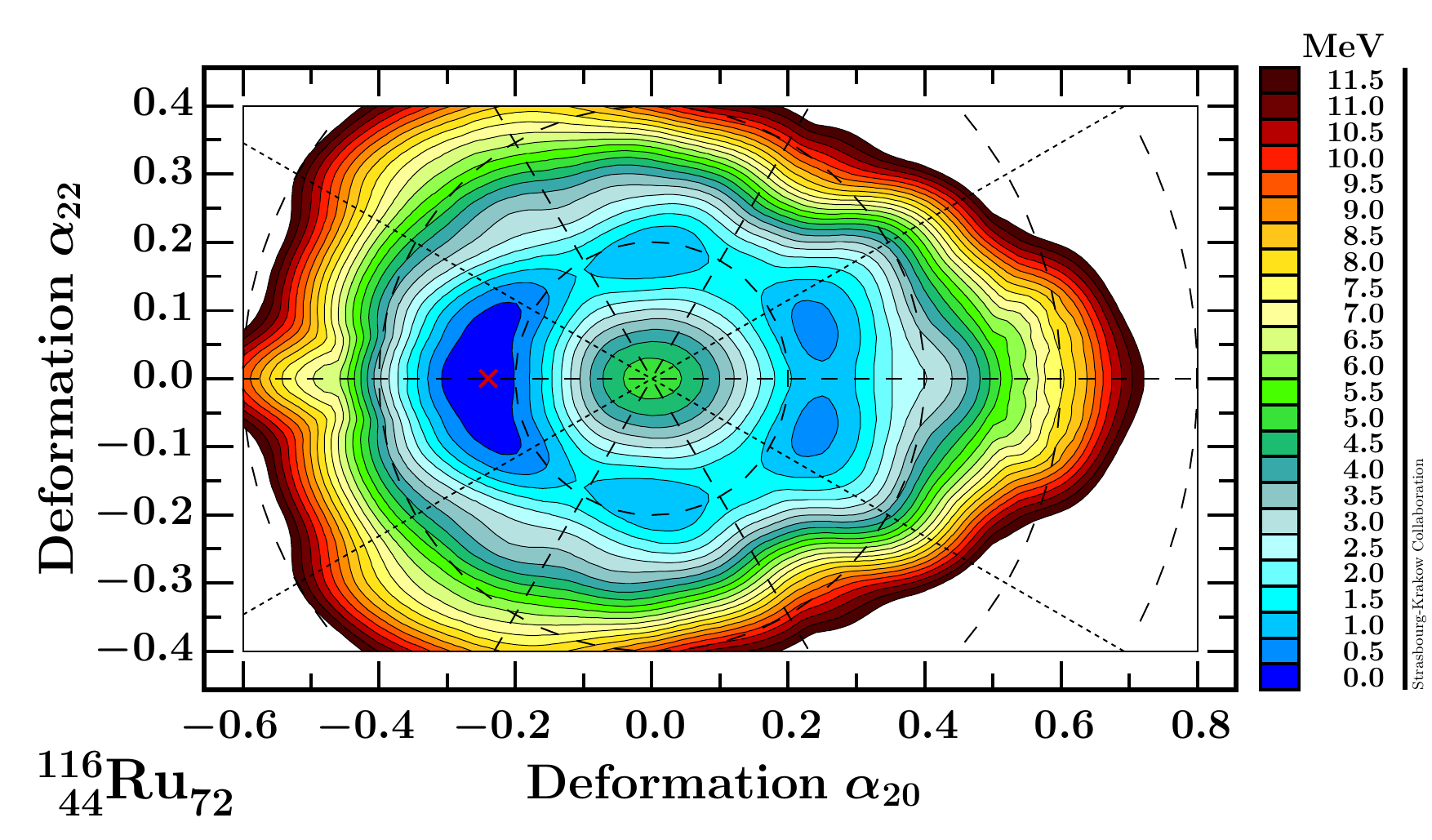}
        \caption{Potential energy surfaces projected onto the quadrupole deformation
                 plane ($\alpha_{20},\alpha_{22}$) after minimization over 
                 hexadecapole deformation $\alpha_{40}$, for $^{108}${Ru} and 
                 $^{116}${Ru} with the absolute energy minima indicated by the 
                 red crosses. The intermediate nuclei 
                 $^{102}${Ru}, $^{104}${Ru}, $^{106}${Ru}, $^{110}$Ru and $^{112}$Ru, 
                 exhibit potential energy surface structures akin to $^{108}${Ru}. 
                 Moreover, $^{114}${Ru} is similar to $^{116}${Ru}. 
                 Thus these two drawings can be seen as representative for groups 
                 of nuclei just listed. The long-dashed straight lines are placed at $\gamma=\pm 60^o$ and $\gamma=\pm 120^o$.
                 }
                                                                      \label{Fig_03}
\end{figure}

We begin by presenting the potential energy maps for selected even-even nuclei, $^{108}${Ru} and $^{116}${Ru}, neighbors of the ones of direct interest; they can be considered representative for those other ones.  The corresponding potential energy surfaces  are presented in Fig.~\ref{Fig_03}. In this particular case we limited the minimization of the energy at each $(\alpha_{20},\alpha_{22})$-point to hexadecapole deformation $\alpha_{40}$, known to influence rather strongly the final results for those nuclei.

Regarding the $(\beta,\gamma)$-plots, when only the quadrupole deformations $\{\alpha_{20}, \alpha_{22}\}\leftrightarrow \{\beta,\gamma\}$ are used and all other deformations set to zero, $\alpha_{\lambda \neq 2, \mu}=0$, one can follow the symmetry of the shapes. This implies that for $\gamma \to \gamma'= \gamma \pm k \times 60 ^{\circ} ($for $k=1,2,3$) the shapes are the same and the only difference is the relative orientation of the body with respect to the references frame. It follows that the corresponding energies are equal, and one can limit the deformation space to the ``triangle''  $0^{\circ} \leq \gamma \leq 60^{\circ}$, as often seen in the literature. We are performing the large scale calculations employing generally multidimensional deformation spaces minimizing over the selected deformations with $\lambda>2$. 
Under these circumstances we should take into account various orientations of the quadrupole shapes and other multipole shapes and consider $0^{\circ} \leq \gamma \leq 360^{\circ}$ as seen in Fig.~\ref{Fig_03}. 

The shapes corresponding to  $\gamma=\pm 60^{\circ}$ and $\gamma=180^{\circ}$ are oblate differing only in terms of the orientation of the symmetry axis, whereas the shapes corresponding to $\gamma=0^{\circ}$ and to $\gamma= \pm120^{\circ}$ are prolate. 
Various local minima visible in Fig.~\ref{Fig_03} correspond to various combinations and/or variations in terms of stronger or weaker shape deformations combined with evolving shape orientations. 





The positions of absolute minima which are marked by the red crosses indicate the most likely nuclear deformation for each given isotope. Based on our calculations with the sample results presented in Fig.~\ref{Fig_03}, we may conclude that the lightest nuclei from $^{104}${Rh} to $^{114}${Rh}, are expected to be prolate deformed, whereas $^{116}${Rh} and $^{118}${Rh} -- oblate. Taking into account clearly distinct shapes, it is even more counter-intuitive trying to understand identical ground-state spin-parity identification $I^\pi=1^+$ within the whole series of considered isotopes.





\subsection{Ground-State Configurations\, in\, \boldmath $^{104-116}$Ru Interpreted With the Tilted Fermi Surface Method}
\label{Sect-I-E}

Next we employed the multi-particle multi-hole analysis based on the Lagrange multiplier method applicable to axially symmetric nuclei. The implied minimization technique is also referred to as ``tilted Fermi surface algorithm''; it allows to minimize the energy of the nucleus at a given angular momentum by varying the nucleonic occupation scheme. This approach has been successful in the past in describing the yrast isomers and the so-called yrast traps in axially symmetric nuclei, but also the low-lying excited states especially at increasing angular momenta. We implement it employing realistic phenomenological mean-field approach with the ``universal'' Woods-Saxon Hamiltonian (for details see Ref.~\cite{Gaa2021} and references therein). A detailed description of the method can be found in the Appendix 1.

The calculations of the nuclei of interest were performed numerically for each odd-odd isotope along the $^{104-118}$Rh chain, which all have ground-states with spin-parity values found experimentally as  $I_{\rm gs}^\pi = 1^+$. As done in Sec.~\ref{Sect-I-B}, we will discuss in detail in this Section only the cases of two isotopes, $^{110}$Rh and $^{118}$Rh, which are representatives of prolate and oblate deformed nuclei, respectively.

\begin{figure}[h!]
\begin{center}
         \includegraphics[width=0.50\textwidth,angle=-00]{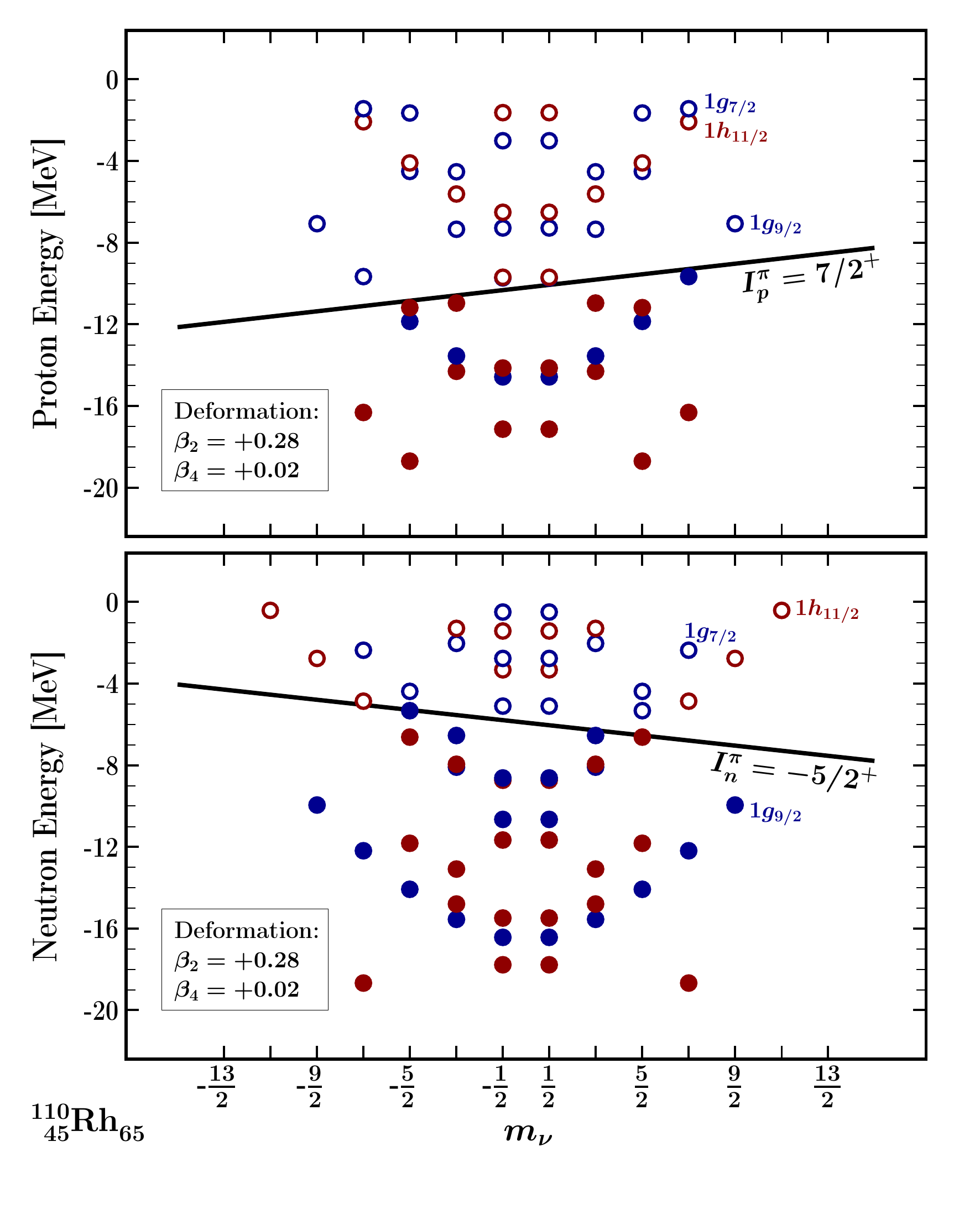}
         \end{center}
\caption{Single particle energies, $e_\nu$ vs.~angular momentum projection $m_\nu$ for protons, top, and neutrons, bottom. Notice that the resulting proton projection is $M=7/2$ originating from the unpaired $g_{7/2,m=+7/2}$ orbital whereas the 65th neutron occupying $g_{7/2,m=-5/2}$ orbital resulting in $M=-5/2$. Both configurations being of positive parity result in the positive parity of the final product wave function $\varphi=\varphi_{prot.} \cdot \varphi_{neut.}$. 
}

                                                              \label{Umb_110Rh}
\end{figure}

Calculations for $^{110}$Rh, according to the tilted Fermi surface diagrams in Figs.~\ref{Umb_110Rh}, lead to the expected characteristics of the ground-state with the spin-parity configuration obtained through permitted minimizations with no parameter  adjustments. According to our mean-field calculations, the 65th neutron occupies the $g_{7/2,m=-5/2}$ orbital, whereas in the present case the couple of the 61st and 62nd neutrons occupy the pair of $h_{{11/2},m=\pm5/2}$ orbitals, and the next one, the 63rd and 64th neutrons occupy the pair of $g_{7/2,m=\pm 3/2}$ thus, all 4 mentioned valence neutrons bring null contribution to the final projection value.
In agreement with the direct tilted Fermi surface picture the 45th proton occupies  the $g_{7/2,m=+7/2}$ orbital whereas its negative projection partner remains empty. Finally, as the result of the occupation of the single particle orbitals with opposite sign projections we find: $M = | 7/2 - 5/2 |=1 \leftrightarrow I = 1 $ and $I^\pi_{\rm g.s.} = 1^+$.

Fig.~\ref{Umb_110Rh} shows the upward concave umbrella plot characteristic of a prolate-deformed nuclei. Our calculations indicate that this type of deformation is present from $^{104-114}$Rh.

The calculation results of $^{110}$Rh represent an ideal case. 
When examining theoretical ground-state properties of an odd-even, even-odd or odd-odd nucleus we are particularly interested in obtaining the expected spin-parity combination at the lowest energy in the calculated energy vs.~spin sequence so that we can directly interpret such a state as the ground-state. Consequently, we wish to find out whether according to our calculations a deformation exists for which the experimental spin-parity combination is reproduced at the lowest energy. Note that this method contains no adjustable parameters -- so that predictions of this kind must follow from the minimization algorithms (i.e. minimization over the configuration and the deformation) as the only liberty. A successful reproduction could be considered as a strong argument in favor of the quality of the mean-field employed. It should be noted that this ``best expected'' result appeared not only in $^{110}$Rh, but also in $^{104,108,112,118}$Rh.



\begin{figure}[h!]
\begin{center}
         \includegraphics[width=0.50\textwidth,angle=-00]{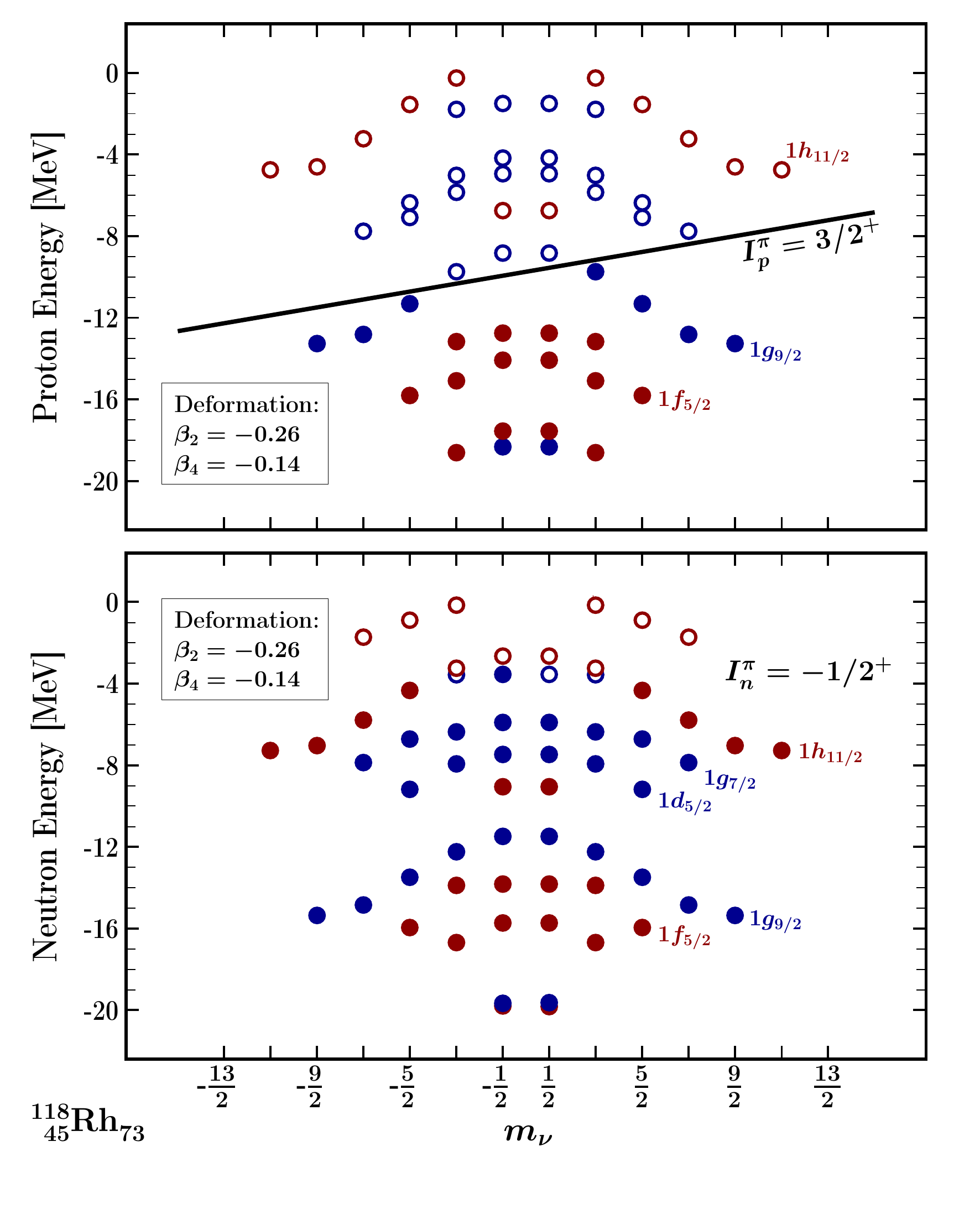}
         \end{center}
\caption{This is an oblate shape case in the present series with $\beta_2=-0.26$ compared with the one to come,  $\beta_2=-0.22$ in Fig.~\ref{Umb_116Rh} for $^{116}$Rh. In the case of protons, with the slightly less dense collection of single particle energies we obtain the exact title Fermi surface solution as illustrated in the top part. In contrast, the solution for the neutrons contains extra excitations over the Fermi surface which is not plotted in this case.}
                                                              \label{Fig_14}
\end{figure}

The last case in this series is the second in the oblate shape category. The calculation result for $^{118}${Rh} gives a configuration and ground state energy ($E=0$ solution) obtained via minimization, which shows that an ideal result happens not only in the prolate nuclei but also in oblate ones. The projections obtained here are $M=+3/2$ for protons and $M=-1/2$ for neutrons. The corresponding tilted Fermi surface diagrams are shown in Fig.~\ref{Fig_14} and thus we find as sought: $M = | 3/2 - 1/2 |=1 \leftrightarrow I = 1$. It should be mentioned that the oblate shape solution for $^{118}$Rh is similar to the prolate one for $^{106}$Rh in Fig.~\ref{Umb_106Rh} except the roles of the contributions are inverted:  $M=+3/2$ for neutrons and $M=-1/2$ for protons.

Needless to say, the apparent huge selection of various distributions of the bullets in the graphical illustrations -- and yet, resulting in the right total spin at the ground-state configurations for the nuclei selected following the experimental data stimulates our curiosity and further research  results are forthcoming.

Let us notice that the results of the configuration minimization with tilted Fermi surface algorithm depend on small differences between energies of the single-particle levels. Since no model is perfect, single-nucleon energy imperfections, even small, may induce final total energy resulting in energy at the correct experimental ground-state spin-parity configuration that is not the lowest. We would nevertheless like to examine under which conditions such a result can still be considered acceptable in the present mean-field context. Should such a result be obtained, we may still hope to deal just with imperfections (rather than totally incorrect schemes) -- but this requires that the obtained energy $E$ can be considered small  (close to obtained lowest energy). 


\subsection{Specific Issues Related to Hamiltonian Parametrization and Its Expected Universality}
\label{Sect-IV-D}

Let us begin by noticing that reproducing the structures underlying just the two ground-states with the $I^\pi=1^+$ spin-parity combinations was a matter of coherence/concordance among many single-point energy-positions in the umbrella-type $(e_\nu -{\rm vs.} - m_\nu)$ diagrams. Recall that the corresponding energies are eigenvalues of the mean-field Hamiltonian and even small misplacement of the points in the diagrams will likely impact such an agreement negatively. 

Moreover, each of the points is a function of the nuclear deformation entering the Hamiltonian and thus varying the nuclear shape leads to changing the relative positions of all the points simultaneously impacting in turn the agreement with the interpreted experimental results.

The commonly used concept of single-nucleon levels and wave-functions is a direct consequence of the independent particle  approximation, and a consequence of the nuclear mean-field theory. It follows that parameters of the employed Woods-Saxon mean-field potentials (both central and spin-orbit ones) must successfully control the evolution of relative positions of single nucleon energies both as functions of the proton and neutron numbers and the varying deformation -- and this, as can be deduced from reading the umbrella diagrams, all simultaneously. Needless to say, this introduces severe mathematical constrains on the structure of tested Hamiltonians and it will be instructive to add a few comments at this point, since as it seems, the Woods-Saxon potential parametrization  qualitatively differs from parametrization of many other interactions found in the literature. This is because the Woods-Saxon potential parameters: Nuclear radius, $R$, nuclear potential depth, $V$ and nuclear diffusivity, $a$, have unique features compared to parametrization of other Hamiltonians. 

Indeed, recall that according to rich experimental information, the diffusivity of the nuclear surface zone can be seen as practically independent of neither nucleon numbers nor deformations. Numerous verifications indicate (see e.g.~Ref.~\cite{WS01}) that the diffusivity parameters both for proton and neutron Woods-Saxon central potentials, $a \to a^c_\pi \leftrightarrow a^c_\nu$ and spin-orbit potentials, $ a^{so}_\pi \leftrightarrow a^{so}_\nu$ are found independent of nucleon numbers. This establishes encouraging parallels between the experimental indication that measured nuclear diffusivities are to a good approximation the same for all nuclei and  independent of $Z$ and $N$ -- and -- the fact that the potential parameters directly responsible for the discussed observable should be kept constant in the applications and in particular independent of nuclear shapes. 

We arrive at the observation that 4 diffusivity parameters
\begin{equation}
    a^c_\pi,  a^c_\nu, a^{so}_\pi \; {\rm and}\;  a^{so}_\nu
                                                                                                                                                              \label{eqn:03}
\end{equation}
among 12 originally considered Woods Saxon parameters, six for the central potential
\begin{equation}
    V^c_{\pi},\; R^c_\pi, a^c_\pi \; {\rm and}\;   V^c_{\nu},\; R^c_\nu, a^c_\nu,
                                                                                                                                                              \label{eqn:04}
\end{equation}
and six for the spin-orbit one
\begin{equation}
    V^{so}_{\pi},\; R^{so}_\pi, a^{so}_\pi \; {\rm and}\;   V^{so}_{\nu},\; R^{so}_\nu, a^{so}_\nu
                                                                                                                                                              \label{eqn:05}
\end{equation}
can be considered as universal constants, independent of neither shape nor particle numbers and applicable without modifications to all nuclei in the Atomic Mass Evaluation.

This property can be considered as a mathematical precondition contributing to a regular behavior of the single particle energies when proceeding with the calculations  from one nucleus to its neighbor.  In this context it will be natural to address the analogous question related to the remaining 8 parameters, which will be done next. Indeed, as it turns out, the nuclear effective radius and  potential depth parameters, are very regular functions of both $Z$ and $N$ and independent of nuclear shapes:
\begin{equation}
   R^c=r^c_0 \times (Z+N)^{1/3}\;\;{\rm and}\;\; R^{so}=r^{so}_0 \times (Z+N)^{1/3}
                                                                                                                                                              \label{eqn:06}
\end{equation} 
with constants $r^c_0$ and $r_0^{so}$ for the central and spin-orbit potentials, respectively,, typically of the order of 1.2 fm, whereas for the central
\begin{equation}
      V^c_{\pi,\nu} = V^c_0 \times \left[ 1 + \kappa^c \frac{N-Z}{N+Z} \right] ,
                                                                                                                                                              \label{eqn:07}
\end{equation}
and the spin orbit potential strength,
\begin{equation}
      V^{so}_{\pi,\nu} = V^{so}_0 \times \left[ 1 + \kappa^{so} \frac{N-Z}{N+Z} \right] ,
                                                                                                                                                              \label{eqn:08}
\end{equation}
parametrization involves very regular dependence in terms of $Z$ and $N$ with constant, deformation independent parameters $V^c_0$ and $V^{so}_0$ as well as  $\kappa^c$ and $\kappa^{so}$, and thus the Hamiltonian remain a very regular expression as function of its parameters. Recall that both nuclear radius and potential depth can be seen as directly controlled by their respective classes of experiments  with the ``one-to-one correspondence", what encourages expectation that resulting potentials represent a ``natural" realistic mathematical image of the experimental reality: nuclear diffuseness, effective radius and potential well vary very slowly from nucleus to nucleus, for the latter case cf.~specifically Ref.~\cite{Hornung2020} and references therein.

Let us emphasize that the discussed regularity in terms of parametric dependence  evokes expectations  in favor of regular evolution of single nucleon orbitals from one nucleus to its neighbors since such a regular behavior is believed to help in reconstructing independence of the $I^\pi=1^+$ on the ground-state single-nucleon structures. Indeed, similarities of this kind are present in experimental data; supporting examples can be found e.g.~in Ref.~\cite{Beck1984}, Fig.~7, where analogous spectral sequences in neighboring isotones are compared, or in~Ref.~\cite{deVoigt1983}, Fig.~33, illustrating similar regularity for a given excitation sequence repeated in various zones of the spectrum within a given nucleus.

To recapitulate, since the typical particle-hole spectral $E_I-{\rm vs.}-I$ sequences in neighboring axial nuclei observed experimentally manifest strong similarities/regularities, we expect that the underlying single nucleon energies generated by the mean-field Hamiltonian do not contain ``irregular jumps"  -- and consequently smooth dependence of the potentials on their parameters are welcome. However, these are only expectations, let us call them working hypotheses; they are not mathematical proofs. Consequently, to complete the argumentation, realistic numerical calculations have been performed and the corresponding results are presented in Appendices 1 and 2.

Let us mention in passing, that the arguments based on parametric regularity can be seen as a bit one-sided, since calculations performed with various model Hamiltonians show that single nucleon energies react sometimes in very different manners to  the parameter variations, usually only very few reacting strongly, with a majority of levels showing similar dependencies; interested reader may consult Fig.~8 in Ref.~\cite{Dudek2012} and surrounding discussion. This strengthens needs of effective and systematic realistic numerical calculations as just mentioned. 

Another aspect of interest in the context of discussing extensively one of the approaches, here, the universal Woods-Saxon, will be to formulate analogous remarks when modeling the same data with alternative Hamiltonians. In the mean-field theory comparisons the most natural alternative to the phenomenological realization of the mean filed would be to employ microscopic self-consistent approaches of the  Hartree-Fock or Hartree-Fock-Bogolyubov type. In the present context such an approach could be considered straightforward since in parallel with the construction of the universal Woods-Saxon numerical computer program (known to the users under the name WSODD), a series of microscopic Hartree Fock computer codes were constructed by the same collaborating teams using nearly the same numerical solutions and algorithms. Interested reader may consult the early articles presenting both the physical formulations  and numerical realizations of the program in question, called HFODD, in Refs.~\cite{Dobaczewski1997-1,Dobaczewski1997-2,Dobaczewski1997-3} as well as the series of publications presenting 8 versions in total, see the last in the series,  Ref.~\cite{Schunck2017} and references cited therein. While entering into details would bypass the framework of this article, it will be sufficient to mention that parametrization of the Hamiltonians treated by HFODD are numerous, significantly more complex compared with the one schematized in Eqs.~(\ref{eqn:03})-(\ref{eqn:08}), and  a priori not promising regularities discussed so far. Moreover, the microscopic approaches discussed are self-consistent, thus requiring the use of iterative algorithms, the latter both complicating the large scale calculations and introducing, even if small,  extra irregularities in the final single nucleon solutions.

For all these reasons the phenomenological realistic Woods-Saxon in its universal realization, Ref.~\cite{WS01,*WS02,*WS03,*WS04,*WS05}, has been selected for the present project. This choice was further supported by the favorable comparisons involving alternative modeling, see recent Ref.~\cite{Porter2022}, Fig.~4,  where the results of other authors with other standard often used methods, such as Finite Range Droplet Model and Hartree-Fock-Bogolyubov Skyrme are compared with the result favorable for the Woods-Saxon universal. Another encouraging argument can be found in the recent Ref.~\cite{Hornung2020}, where Fig.~6 illustrates favorably another comparison with experiment in terms of particle-hole excitations of particular interest for the present project.

Finally, anticipating possible case(s) where the perfect description is not achieved,  let us recall two general aspects of the nuclear modeling which are well known but not sufficiently often reminded of. Firstly, the complete final theory of the nuclear  interactions is not known, the effective Hamiltonian forms and parametrizations must be considered as “temporary” and the finally accepted and employed Hamiltonians depend on the parameters adjusted to experimental data, and, secondly, all experimental data are known within experimental uncertainties (error bars) what contributes even stronger the final uncertainties of the description of interactions used by our computer programs.

Combining these two facts brings us to the final observation that in any case, all forms of our modeling  must contain uncertainties — what implies that our Hamiltonian (as much as any other in our domain) is ``allowed to manifest limitations and imperfections". It is then a part of the everyday research to reveal the presence of such imperfections and formulate the next steps in searching for reasons and  possible  improvements. This is also how the imperfections in one case found in the present article can be interpreted; we are looking for further steps which will be left for a coming publication.



\subsection{Summary}

We have performed realistic phenomenological mean-field calculations using a ``universal'' Woods-Saxon Hamiltonian where only 9 independent parameters were fixed using experimental information about single particle levels in doubly magic spherical nuclei. This parametrization is applicable for all $\sim$3\,000 nuclei from the Atomic Mass Evaluation. Our calculations aim at interpreting the structure of the ground-state configurations of all the nuclei forming the longest known chain of odd-odd isotopes with the identical spin-parity assignment $I^{\pi}_{\rm g.s.}=1^+$ $\leftrightarrow$  $^{104-118}$Rh.

The Lagrange theorem plays a central role in the applied method of calculation by finding a minimum under constraint. To do so, we minimized the nuclear energy  over the nuclear deformation and the nucleonic configurations (occupation schemes of the proton and neutron levels, separately)  under the constraint of a fixed angular momentum. The method proposed originally by the Copenhagen School (in particular Bohr and Mottelson) leads to a very powerful and easily programmable algorithm often called ``tilted Fermi surface method''. 

The calculations explain the detailed occupation structures under the spin-parity independence condition $I^{\pi}_{\rm g.s.}=1^+$ for all the considered ground-states. They employ a parameter free minimization over deformation and configurations (Hamiltonian parameters are fixed once for all wherefrom the term ``universal"). The algorithm employed is more complex than presented since to determine the ground-state energy conditions for all isotopes concerned, we have performed the calculations of the yrast lines of all of them such that we can claim that the $I^\pi=1^+$ candidate state in question corresponds to the lowest energy along the yrast line (and represents indeed the ground-state).

In 3 among 8 considered cases the calculated energies of the $I^\pi=1^+$ states were not the lowest, in two cases, $^{106}$Rh and  $^{112}$Rh the deviations remaining of the order of a couple of hundreds of keV what can be qualified as imperfections, whereas in the case of   $^{116}$Rh discrepancy being higher than 1 MeV. This can be interpreted as a consequence of an approximate character of all Hamiltonians in nuclear physics, see Section \ref{Sect-IV-D}, very likely resulting either from the imperfections of the spin-orbit potential parametrization but also possibly related to the fact that we limited deformation space to two parameters only $\alpha_{20} \leftrightarrow \beta_2$ and $\alpha_{40} \leftrightarrow \beta_4$. 


\section{\label{subsec:Multiquasiparticle} Multi-quasiparticle blocking calculations}

In order to predict the structure of low-lying states in the odd-odd Rh nuclei, multi-quasiparticle blocking calculations were carried out for both prolate and oblate deformations by assuming axial symmetry. 
The energies of single-particle states were taken from the Woods-Saxon potential with the “universal” parametrization~\cite{WS05} and the adopted deformation parameters $\beta_{2}$, $\beta_{4}$ and $\beta_{6}$ are summarized in Table~\ref{tab:deformations}. 

\begin{table}[h!]
\caption{\label{tab:deformations} 
Deformation parameters $^{a)}$ used in the multi-quasiparticle blocking calculations.}
\begin{ruledtabular}
\begin{tabular}{ccccccc}
                        &  \multicolumn{3}{c}{prolate} & \multicolumn{3}{c}{oblate} \\
Nuclide            & $\beta_{2}$  & $\beta_{4}$   & $\beta_{6}$  &  $\beta_{2}$   & $\beta_{4}$   & $\beta_{6}$ \\
\hline
\\
$^{108}$Rh     &  +0.250        &  +0.011          &  -0.001         &   -0.248          &  -0.034          & +0.002         \\
$^{110}$Rh     &  +0.260        &  +0.005          &  -0.001         &   -0.248          &  -0.034          & +0.002         \\
$^{112}$Rh     &  +0.265        &  +0.005          &  -0.001         &   -0.248          &  -0.044          & +0.014         \\
$^{114}$Rh     &  +0.265        &  +0.005          &  -0.001         &   -0.258          &  -0.053          & +0.018         \\
$^{116}$Rh     &  +0.260        &  +0.005          &  -0.001         &   -0.258          &  -0.064          & +0.012         \\
\\
\end{tabular}
\end{ruledtabular}
$^{a)}$Values for $^{108}$Rh and $^{110,112,114,116}$Rh are from Ref.~\cite{2016Mo08}.
\end{table}

The pairing correlations were treated using the Lipkin-Nogami prescription~\cite{1985Na} with fixed strengths of G$_{\pi}$ = 24.0/A MeV (protons) and G$_{\nu}$ = 17.8/A MeV (neutrons) and they included the effect of blocking. Calculations did not include the effect of the residual proton-neutron interactions, and therefore, it was not possible to predict the excitation energies of the predicted states. Instead,  the so-called Gallagher-Moszkowski rule~\cite{GM58} was applied to establish the ordering of the $|\Omega_{p}-\Omega_{n}|$ and  $\Omega_{p}+\Omega_{n}$ states within a given $\Omega_{p}\otimes\Omega_{n}$ configuration, where  $\Omega$ is the projection of the angular momentum on the symmetry axis. 

The predicted most-likely configurations for the ground and isomeric states in the odd-odd Rh nuclei are given in Table~\ref{tab:multiquasiparticle}. Also in Table ~\ref{tab:multiquasiparticle} predictions based on the known experimental states in neighboring odd-Z, even-N Rh and even-Z, odd-N Ru nuclei~\cite{ENSDF,NUBASE} are given under the empirical column. The systematic of experimental values \cite{NUBASE} show that all odd-Z, even-N $^{107-115}$Rh nuclei have the I$^{\pi}$=7/2$^{+}$ ground state, associated with prolate shape and the $\pi$7/2[413] Nilsson orbital. At the same time, the experimental ground state of the odd-N, even-Z $^{107,109,111}$Ru (N=63-67) nuclei is assigned to be $5/2^+$, which correspond to a $\nu$5/2[413] Nilsson orbital. Thus, the lowest-energy configuration in the odd-odd $^{108,110,112}$Rh (N=63-67) nuclei is expected to be $\pi$7/2[413]$\otimes$$\nu$5/2[413] and in accordance with the Gallagher-Moszkowski rule~\cite{GM58},  I$^{\pi}$=1$^{+}$ for the ground states and I$^{\pi}$=6$^{+}$ for the isomers. Such assignments are in a relatively good agreement with the predictions from the multi-quasiparticle blocking calculations for prolate shapes. This is consistent with the calculation results presented in the previous section. 

The structure of the heavier $^{113,115}$Ru (N=69,71) is not well established and there are evidences for competing prolate and oblate configurations, also as evidenced by the calculation from the previous section. Recently, the $\nu$1/2[411] Nilsson orbital was proposed for the ground state of these nuclei, while the isomer was associated with the  $\nu$7/2[523] orbital~\cite{,07Ku23,10Ku25}. The expected configurations for the odd-odd $^{114,116}$Rh (N=69,71) nuclei are given in Table~\ref{tab:multiquasiparticle}. Once again, following the Gallagher-Moszkowski rule~\cite{GM58} and assuming a $\pi$7/2[413]$\otimes$$\nu$1/2[411] configuration for the ground state, a I$^{\pi}$=3$^{+}$ is expected for both $^{114,116}$Rh. This value differs from the observed I$^{\pi}$=1$^{+}$ \cite{NUBASE}. The isomeric state on the other hand appears to be in a $\pi$7/2[413]$\otimes$$\nu$7/2[523] configuration yielding a I$^{\pi}$=7$^{-}$. The isomeric state value of 7$^{-}$ do agree with the experimental value for $^{114}$Rh but not with the 6$^{-}$ value for $^{116}$Rh. Furthermore, these empirical configurations do not yield an inversion in the spin-parity of the ground and isomeric states. However, the multi-quasiparticle blocking calculations predictions for both the oblate and prolate shapes yield a higher spin for the ground state of $^{114,116}$Rh then their isomeric states. This contradicts the empirical configuration while agreeing with the observed yield ratio for $^{114}$Rh. Hence, there is a strong impetus for a direct spin determination using laser spectroscopy on neutron-rich, odd-odd Rh isotopes, in particular $^{114}$Rh.

\begin{table*}
\caption{\label{tab:multiquasiparticle} 
Predicted configuration for the ground states and the isomers in the odd-odd $^{108,110,112,114,116}$Rh nuclei.}
\begin{ruledtabular}
\begin{tabular}{ccccccc}
                        &  & {prolate} & &{oblate} & & {empirical} \\
Nuclide            & I$^{\pi}$                  & Configuration                                      & I$^{\pi}$     &  Configuration                                           & I$^{\pi}$      & Configuration \\
\hline
                         &                               &                                                             &                  &                                                                    &                    &                                                          \\
$^{108}$Rh      &  2$^{+}$                 & $\pi$1/2[301]$\otimes$$\nu$5/2[532] & 1$^{+}$     &  $\pi$3/2[411]$\otimes$$\nu$5/2[413]        & 1$^{+}$     & $\pi$7/2[413]$\otimes$$\nu$5/2[413] \\
                         &  6$^{-}$                  & $\pi$7/2[413]$\otimes$$\nu$5/2[532] &                  &                                                                    &                    &                                                          \\
$^{108m}$Rh   &  1$^{+}$                 & $\pi$7/2[413]$\otimes$$\nu$5/2[413] & 4$^{+}$     &  $\pi$3/2[411]$\otimes$$\nu$5/2[413]        & 6$^{+}$     & $\pi$7/2[413]$\otimes$$\nu$5/2[413]  \\
                         &  1$^{-}$                  & $\pi$7/2[413]$\otimes$$\nu$5/2[532] &                  &                                                                    & 6$^{-}$       & $\pi$7/2[413]$\otimes$$\nu$5/2[532]   \\
                         &  6$^{+}$                 & $\pi$7/2[413]$\otimes$$\nu$5/2[413] &                  &                                                                    &                    &                                                          \\
                        &                                &                                                            &                   &                                                                    &                  &                                                          \\
$^{110}$Rh      &  1$^{+}$                 & $\pi$7/2[413]$\otimes$$\nu$5/2[413] & 6$^{-}$       &  $\pi$3/2[411]$\otimes$$\nu$9/2[514]       & 1$^{+}$      & $\pi$7/2[413]$\otimes$$\nu$5/2[413] \\
                        &  6$^{-}$                  & $\pi$7/2[413]$\otimes$$\nu$5/2[532] &                    &                                                                   &                   &                                                          \\
$^{110m}$Rh   &  6$^{+}$                 & $\pi$7/2[413]$\otimes$$\nu$5/2[413] & 1$^{+}$      &  $\pi$3/2[411]$\otimes$$\nu$1/2[411]       & 6$^{+}$     & $\pi$7/2[413]$\otimes$$\nu$5/2[413]  \\
                        &  1$^{-}$                  & $\pi$7/2[413]$\otimes$$\nu$5/2[532] &  3$^{-}$      &  $\pi$3/2[411]$\otimes$$\nu$9/2[514]       & 6$^{-}$       & $\pi$7/2[413]$\otimes$$\nu$5/2[532]  \\
                        &                                &                                                             &                   &                                                                   &                    &                                                          \\ 
$^{112}$Rh      &  3$^{+}$                 & $\pi$7/2[413]$\otimes$$\nu$1/2[411]  & 1$^{+}$     &  $\pi$3/2[411]$\otimes$$\nu$1/2[411]        & 1$^{+}$      & $\pi$7/2[413]$\otimes$$\nu$5/2[413] \\
                        &                               &                                                             &                   &                                                                   &  3$^{+}$      &  $\pi$7/2[413]$\otimes$$\nu$1/2[411]         \\ 
$^{112m}$Rh   &  6$^{-}$                  & $\pi$7/2[413]$\otimes$$\nu$5/2[532]  & 6$^{-}$      & $\pi$3/2[411]$\otimes$$\nu$9/2[514]        &  6$^{+}$     & $\pi$7/2[413]$\otimes$$\nu$5/2[413] \\                         
                        &                               &                                                             &                   &                                                                   &                    &                                                          \\ 
$^{114}$Rh      &  6$^{+}$                 & $\pi$7/2[413]$\otimes$$\nu$5/2[402]  & 5$^{-}$       &  $\pi$3/2[411]$\otimes$$\nu$7/2[523]        &  3$^{+}$    & $\pi$7/2[413]$\otimes$$\nu$1/2[411] \\
                        &  7$^{-}$                  & $\pi$7/2[413]$\otimes$$\nu$7/2[523]  &                    &                                                                   &  & \\    
$^{114m}$Rh   &  1$^{+}$                 & $\pi$7/2[413]$\otimes$$\nu$5/2[402]  & 2$^{-}$       &  $\pi$3/2[411]$\otimes$$\nu$7/2[523]        &  7$^{-}$     & $\pi$7/2[413]$\otimes$$\nu$7/2[523] \\ 
                        &  0$^{-}$,1$^{-}$      & $\pi$7/2[413]$\otimes$$\nu$7/2[523]  &                   &                                                                   &  & \\   
                         &                                &                                                            &                   &                                                                   &                    &                                                          \\   
$^{116}$Rh      &  7$^{-}$                  & $\pi$7/2[413]$\otimes$$\nu$7/2[523]  & 4$^{-}$       &  $\pi$3/2[411]$\otimes$$\nu$5/2[512]      &  3$^{+}$    & $\pi$7/2[413]$\otimes$$\nu$1/2[411]  \\
                        &   6$^{+}$                & $\pi$7/2[413]$\otimes$$\nu$5/2[402]  &                    &                                                                   &                   & \\   
$^{116m}$Rh   &  0$^{-}$,1$^{-}$     & $\pi$7/2[413]$\otimes$$\nu$7/2[523]  & 1$^{-}$        &  $\pi$3/2[411]$\otimes$$\nu$5/2[512]      &  7$^{-}$     & $\pi$7/2[413]$\otimes$$\nu$7/2[523]   \\
                        &   1$^{+}$                & $\pi$7/2[413]$\otimes$$\nu$5/2[402]  &                    &                                                                   &                    & \\       
                        &                                &                                                            &                     &                                                                   &                    &                                                          \\                
\end{tabular}
\end{ruledtabular}
\end{table*}
\newpage
\section{\label{sec:summary}Conclusions}

The ground state and isomeric state(s) of $^{108, 110, 112, 114, 116}$Rh were identified and their masses were measured with high precision using the Canadian Penning Trap at the CARIBU facility. The obtained ground state mass excesses and isomeric state excitation energies are consistent with recent measurements from the JYFLTRAP Penning trap. Furthermore, a possible second isomeric state has also been found for $^{114}$Rh as reported in \cite{JYFLarxiv2}. 

The measured isotopes also happen to be part of the longest chain of odd-odd isotopes carrying identical spin-parity ($1^+$) across the whole nuclear chart amid known changes in shape. Detailed nuclear structure calculations were performed to investigate this peculiarity. This was accomplished employing for the first time the realistic phenomenological mean-field approach that uses a universal Woods-Saxon Hamiltonian. For all but three isotopes, the lowest calculated energy correspond to the $1^+$ state; in two cases the calculated energies lie a couple of hundreds of keV above the lowest ones what can be seen as imperfections rather than big discrepancies. The single case of bigger discrepancy, the $^{116}$Rh nucleus, can be a result of the limitations of the deformation space retained in the present calculations, composed of quadrupole and hexadecapole ones,   $\alpha_{20}=\beta_2$ and  $\alpha_{40}=\beta_4$ only. Given that the majority of the considered cases were reproduced or approximately reproduced with no parameter adjustments we may consider the $^{116}$Rh  case as a singularity to be examined elsewhere.


In addition, multi-quasiparticle blocking calculations were conducted to explore alternative spin-parity assignments different from the literature value, as a potential explanation for the anomalous yield of $^{114}$Rh. We found a different empirical ground state of $I^+=3^+$ that differs from the experimental evaluation \cite{NUBASE} while the empirical isomeric state yield a consistent value of $I^+=7^-$. However multi-quasiparticle blocking calculations assuming either prolate or oblate shape results in a spin inversion with the ground state having a higher spin than the isomeric state, which is consistent with what we have observed experimentally. 
Hence, there is a strong need for a direct measurement of the nuclear spin of $^{114}$Rh using laser spectroscopy to shed light on the situation. Such measurement on the other odd-odd isotopes in the $^{104-118}$Rh chain is also warranted to confirm the $I=1$ spin assignment of their ground states.
Finally, given only one half-life of the two isomeric states of $^{114}$Rh has been reported in literature, a new half-life measurement would be helpful to clarify the inversion of the fission yield and spin-parity assignment. \\

\section*{\label{sec:ac}Acknowledgments}

This work is supported in part by the National Science Foundation under Grant No. PHY-2310059; by the University of Notre Dame; and with resources of ANL’s ATLAS facility, an Office of Science User Facility; by the U.S. Department of Energy, Office of Nuclear Physics, under Contract No. DE-AC02-06CH11357; by NSERC (Canada), Application No. SAPPJ-2018-00028. Partial support from the French-Polish collaboration COPIN, No.~04-113 and No.~05-119, is acknowledged.


\section*{Appendix 1}
\label{Appendix-1}

Since we examine nuclear mean-field with axial symmetry, the corresponding Hamiltonian commutes with the operator of the third component of angular momentum 
\begin{equation}
    [\hat{H}, \hat{\jmath}_z] =0,
                                                               \label{Eqn_01}
\end{equation}
so that we can consider two simultaneous eigen-value equations
\begin{equation}
   \hat{H}\varphi_{\nu,m_\nu} = e_{\nu,m_\nu}\varphi_{\nu,m_\nu} ,
                                                               \label{Eqn_02}
\end{equation}
and
\begin{equation}
   \hat{\jmath}_z\varphi_{\nu,m_\nu} = m_{\nu}\varphi_{\nu,m_\nu} .
                                                               \label{Eqn_03}
\end{equation}
With these two equations solved we can construct the particle-hole excited configuration energies, $E^*_M$, corresponding to the angular momentum projection, $M$, as follows:
\begin{equation}
   E^*_M = \sum_{p} e_{p,m_p} - \sum_h  e_{h,m_h}\,,
                                                               \label{Eqn_04}
\end{equation}
where the index $p$ enumerates selected particle states and the index $h$, the related hole states; in analogy
\begin{equation}
   M = \sum_{p} m_{p} - \sum_h  m_{h}.
                                                               \label{Eqn_05}
\end{equation}
It has been argued in~\cite{deVoigt1983} that to an approximation, valid in particular at increasing angular momenta~$I$, one can state the
\begin{equation}
    I  \approx  | M | ,
                                                               \label{Eqn_06}
\end{equation} 
relation, sometimes referred to as ``maximum alignment hypothesis''. This relation facilitates significantly the calculations together with interpretation and comparison with experimental data.

There are certain advantages following from the above simple relations. Firstly, after solving numerically the Schr\"odinger equation with the Woods-Saxon Hamiltonian, we can proceed with calculating the particle-hole excitation energies and the associated projected angular momenta in a simpler manner. 
Secondly, and more importantly, according to the ``maximum alignment hypothesis'', the lowest energy states from Eq.~(\ref{Eqn_04}) correspond  possibly to the largest projections $M$, whereas the slightly higher neighboring energies correspond to slightly smaller projections, and all of these relations are under direct numerical control. It follows that we obtain directly the {\em yrast} and the next to {\em yrast} lines, and the associated energy vs.~spin fluctuations, in particular the yrast trap information. Even though our calculations were performed for the whole range of the spin values, here we limit our discussion to properties of the chain of the ground-states with $I^\pi_{\rm g.s.}=1^+$.

Let us notice that we are interested in finding the {\em configurations}, i.e.,~all the indices of the occupied proton and neutron levels, with which we minimize the energy $E^*_M$ at each requested $M$. For this reason we will rewrite the preceding relations using the terminology of the conditional minimization according to the Lagrange theorem presented below.

We would like to find the lowest energy excited configurations, say $\mathcal{C}$, the letter denoting the ensemble of indices of occupied states for protons and neutrons separately: 
\begin{equation}
   E^* = \sum_{\nu \in\{\mathcal{C}\}}  e_\nu ,
                                                               \label{Eqn_07}
\end{equation}
such that the number of particles $\mathcal{N}$ is $N$ or $Z$
\begin{equation}
   \sum_{\nu \in\{C\}} 1_\nu = \mathcal{N} ,
                                                               \label{Eqn_08}
\end{equation}
and the projected angular momentum $\mathcal{M}$ corresponds either to proton or to the neutron contributions, $M_Z$ or $M_N$, respectively
\begin{equation}
   \sum_{\nu \in\{C\}} m_\nu = \mathcal{M} \quad (M_Z\;{\rm or}\; M_N).
                                                               \label{Eqn_09}
\end{equation}
According to the Lagrange theorem, the minimization of the expression in Eq.~(\ref{Eqn_07}) under the conditions specified by Eqs.~(\ref{Eqn_08}) and (\ref{Eqn_09}) is equivalent to the minimization of an auxiliary expression defined, for each kind of particles separately, by
\begin{equation}
   \tilde{E}^*_M = \sum_{\nu \in\{C\}}  (e_\nu - \lambda \cdot 1_\nu - \omega \cdot m_\nu),
                                                               \label{Eqn_10}
\end{equation}
where the so called Lagrange multipliers $\lambda$ and $\omega$ are for the moment unknown.

\begin{figure}[h!]
     \centering
         \centering
         \includegraphics[width=0.45\textwidth,angle=-00]{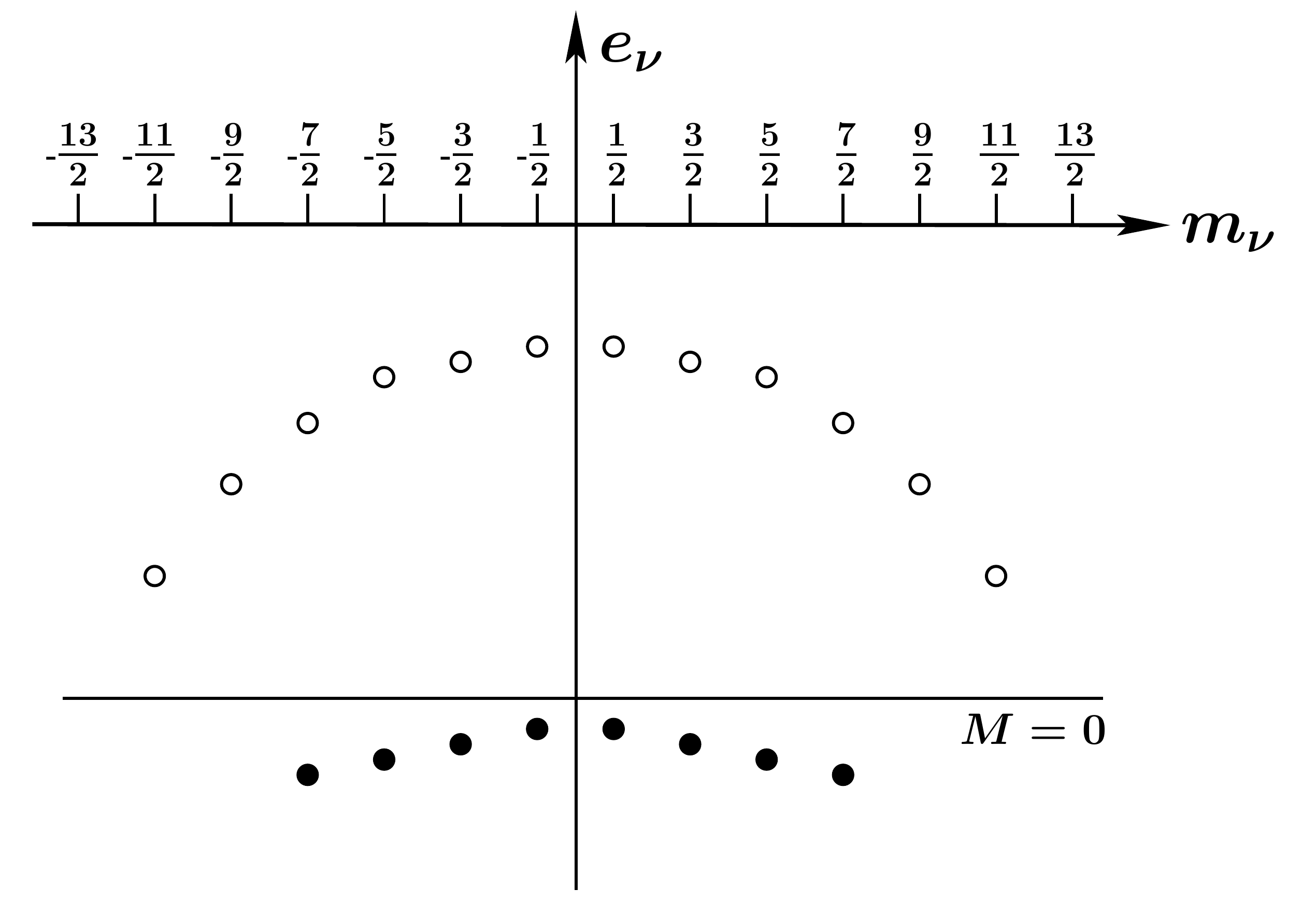}
         \caption{Schematic illustration of the applications of the Lagrange theorem 
                  about conditional minimization involving fictitious configuration
                  composed of two orbitals, lower one, assumed fully occupied, 
                  of $j=7/2$, and upper one, assumed empty, of $j=11/2$. Horizontal
                  line represents ``untilted'' Fermi surface, all levels below assumed
                  occupied (ground-state).
                 }
                                                                      \label{tilted1}
\end{figure}
One can easily show that finding the configuration $\mathcal{C}$ is geometrically equivalent to finding strictly all the points lying below the tilted straight line, $e(m)=\lambda +\omega \cdot m$, called ``tilted Fermi surface'' (with no empty one allowed). Several illustrations of the discussed approach can be found in Refs.~\cite{deVoigt1983, Hornung2020,Beck2021,Porter2022}, in which the Bohr and Mottelson maximum alignment hypothesis was shown to reproduce numerous experimental cases. 

\begin{figure}[h!]
         \centering
         \includegraphics[width=0.45\textwidth,angle=-00]{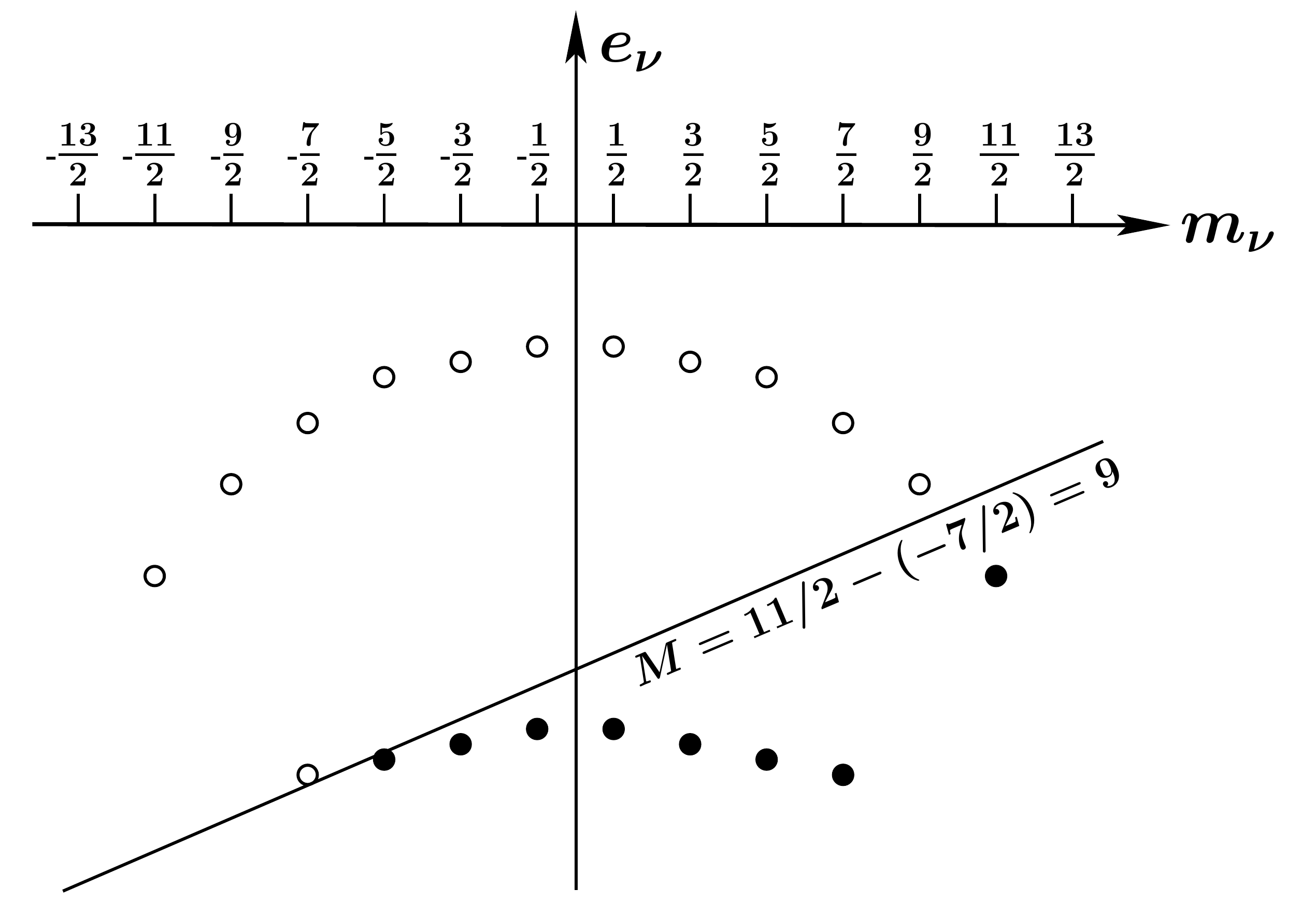}
         \caption{Schematic illustration similar to the preceding one, here all the particles
                  with the energies below the tilted 
                  Fermi surface are assumed occupied. We have one hole state with 
                  $(j,m)=(7/2,-7/2)$ and one particle state with 
                  $(j,m)=(11/2,+11/2)$ activated resulting with 
                  $M = 11/2 -(-7/2) =9$ $ (I \approx M)$.
                 }
                                                                      \label{tilted2}
\end{figure}
To facilitate the discussion, let us refer to figure \ref{tilted1}, which illustrates schematically a single nucleon spectrum composed of two orbitals only, $j=7/2$, assumed fully occupied and $j'=11/2$, assumed totally empty in the ground-state configuration. Please note the characteristic $e_\nu -{\rm vs.} -m_\nu$ dependence with the down-sloping branches. The diagrams like these are referred to as  ``umbrella plots''. Such a dependence is not accidental since it can be shown that for a single $j$-shell model, for the oblate quadrupole deformation, $e_\nu \propto - m^2_\nu$, cf.~Bohr and Mottelson, Ref.~\cite{BMVol2}.

Similar mathematical reasoning leads to the conclusion that for the single-$j$ orbitals corresponding to a prolate quadrupole deformation, the coefficient in front of $m_\nu^2$ changes sign 
\begin{equation} \!
   {\rm Oblate}\;\to\;\;e_\nu \propto - m^2_\nu\;
   \leftrightarrow 
   \;e_\nu \propto + m^2_\nu\; \leftarrow\; {\rm Prolate},
                                                               \label{Eqn_11}
\end{equation}
and the originally down-sloping branches become up-sloping (``inverted'' umbrellas); see  illustrations in Sec.~\ref{Sect-I-E} and the next appendix.

Figure~\ref{tilted2} illustrates the results of a slow tilting of the Fermi surface by increasing the Lagrange multiplier $\omega$ step by step so that at certain position the orbital with $m_h=-7/2$ becomes empty (hole state) whereas at the same time the orbital $m_p=11/2$ (particle state) becomes occupied, so that all the occupied levels lie below the (tilted) Fermi surface with the resulting projection $M=[11/2 -(-7/2)]=9$.

Let us emphasize at this point that the simple smooth-looking umbrella diagrams as in Figs.~\ref{tilted1} and \ref{tilted2} represent an ideal situation of a single orbital. Indeed, in the case of several $j$-orbitals in a deformed mean field potential, the  orbitals of same symmetry (here: parity) interact with each other, the resulting repulsion called ``Landau-Zener non-crossing rule'': the neighboring orbitals repel each other, with the resulting relatively complex patterns visible in Figs.~\ref{Umb_104Rh} and onward.   

\section*{Appendix 2}

This appendix discusses the ground state configurations of the Rh isotopes not discussed in the main text, namely $^{104}$Rh, $^{106}$Rh, $^{108}$Rh, $^{112}$Rh, $^{114}$Rh, and $^{116}$Rh.

\subsubsection{I$_{g.s.}^{\,\pi} = 1^+$ ground state in $^{104}${Rh}}
\label{Section-D1}

\begin{figure}[h!]
\begin{center}
         \includegraphics[width=0.48\textwidth,angle=-00]{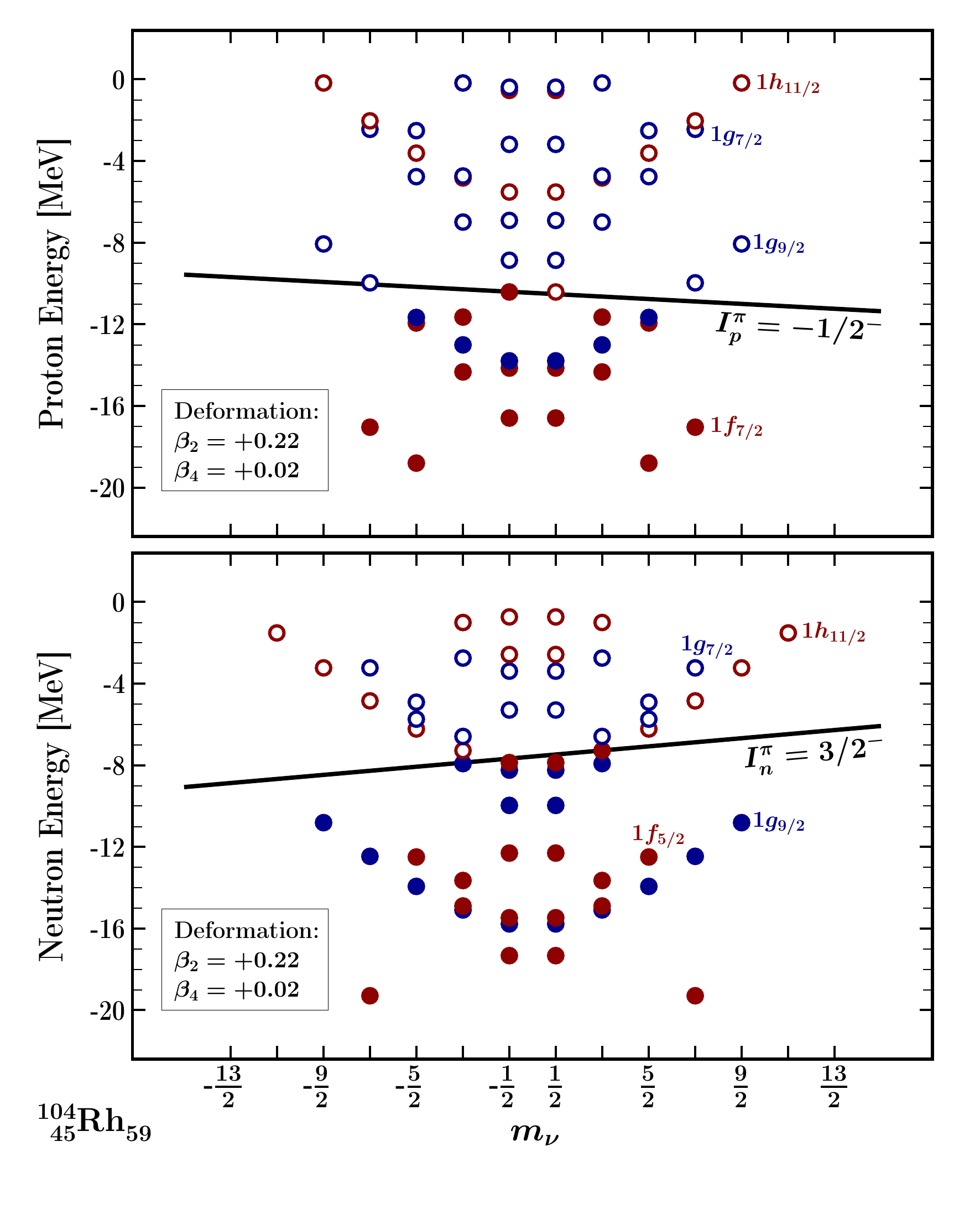}
         \end{center}
\caption{Single particle energies, $e_\nu$ vs.~angular momentum projection $m_\nu$ for protons, top, and neutrons, bottom. Notice that the resulting proton projection is $M=-1/2$ originating from the unpaired $p_{1/2,m=-1/2}$ orbital whereas the 59th neutron occupying $h_{11/2,m=3/2}$ orbital resulting in $M=3/2$. Both configurations being of negative parity result in the positive parity of the final product wave function $\varphi=\varphi_{prot.} \cdot \varphi_{neut.}$ with $M=|3/2-1/2|=1 \leftrightarrow I=1$.}
                                                              \label{Umb_104Rh}
\end{figure}

Figure~\ref{Umb_104Rh} shows the single-nucleon occupation schemes for $^{104}$Rh. The unpaired proton at $M=-1/2$ combined with the unpaired neutron at $M=3/2$ leads to an overall spin-parity $M \leftrightarrow I^\pi = 1^+$ for the ground state. This calculation corresponds to the ideal case where the ground state and parity were reproduced at the lowest energy.

Let us notice that in the series of the realistic (inverted) umbrella diagrams which follows beginning with Fig.~\ref{Umb_104Rh}, the positions of the bullets do not manifest the regularity of the schematic illustrations in the two preceding figures, with the deviations from the parabolic-like dependence caused by Landau-Zener mechanism mentioned earlier.


\subsubsection{I$_{g.s.}^{\,\pi} = 1^+$ ground state in $^{106}${Rh}}
\label{Section-D2}

\begin{figure}[h!]
\begin{center}
         \includegraphics[width=0.48\textwidth,angle=-00]{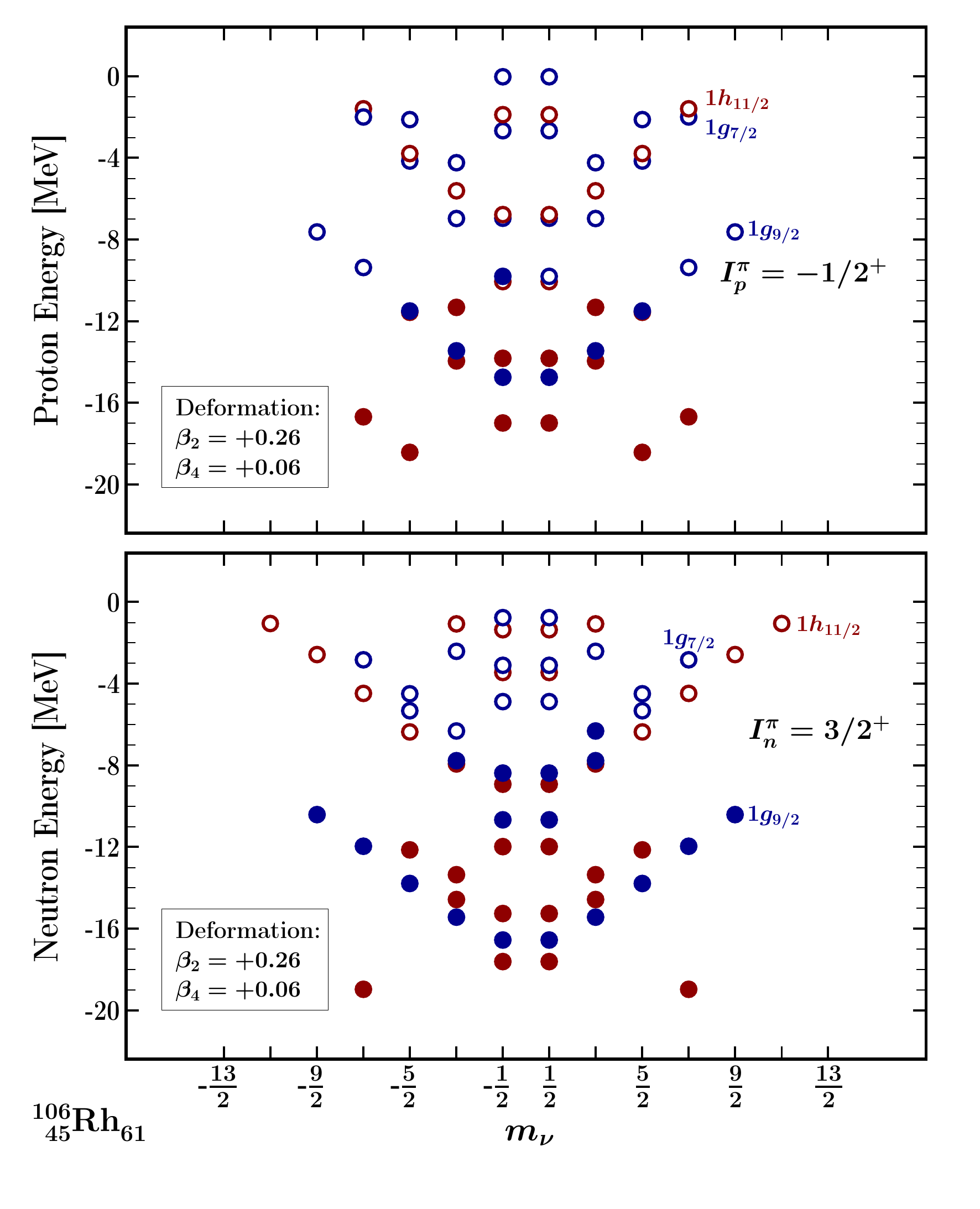}
         \end{center}
\caption{The single particle diagrams for $^{106}$Rh are similar to the ones for $^{104}$Rh but at slightly different deformations resulting from the minimizations, which are performed for each nucleus separately. In this case the titled Fermi lines are missing, since both configurations correspond to the 1-particle 1-hole excitations above the tilted Fermi lines. Again $M=|3/2-1/2|=1 \leftrightarrow I = 1$.}
                                                              \label{Umb_106Rh}
\end{figure}

The tilted Fermi surface diagrams in Fig.~\ref{Umb_106Rh} show the occupation pattern of the single particle orbitals giving again the spin estimate $M \leftrightarrow I = 1$ for $^{106}${Rh}.

Unlike the previous case, the contribution $M=-1/2$ is caused by the occupation of the level $g_{7/2,m=-1/2}$ by the 45th proton, whereas projection $M=+3/2$ is obtained  thanks to $g_{7/2,m=3/2}$ orbital occupied by the 61st neutron.

Unlike $^{104}$Rh, in this case we do not obtain $E=0$ but a slight discrepancy of a couple of hundreds of keV, which can be qualified as imperfection in terms of the single-nucleon level positions.

In this case the tilted Fermi surface lines are missing since the configurations retained correspond, in both cases, to 1-particle 1-hole excitations above the tilted lines.

\subsubsection{I$_{g.s.}^{\,\pi} = 1^+$ ground state in $^{108}${Rh}}
\label{Section-D3}

\begin{figure}[h!]
\begin{center}
         \includegraphics[width=0.50\textwidth,angle=-00]{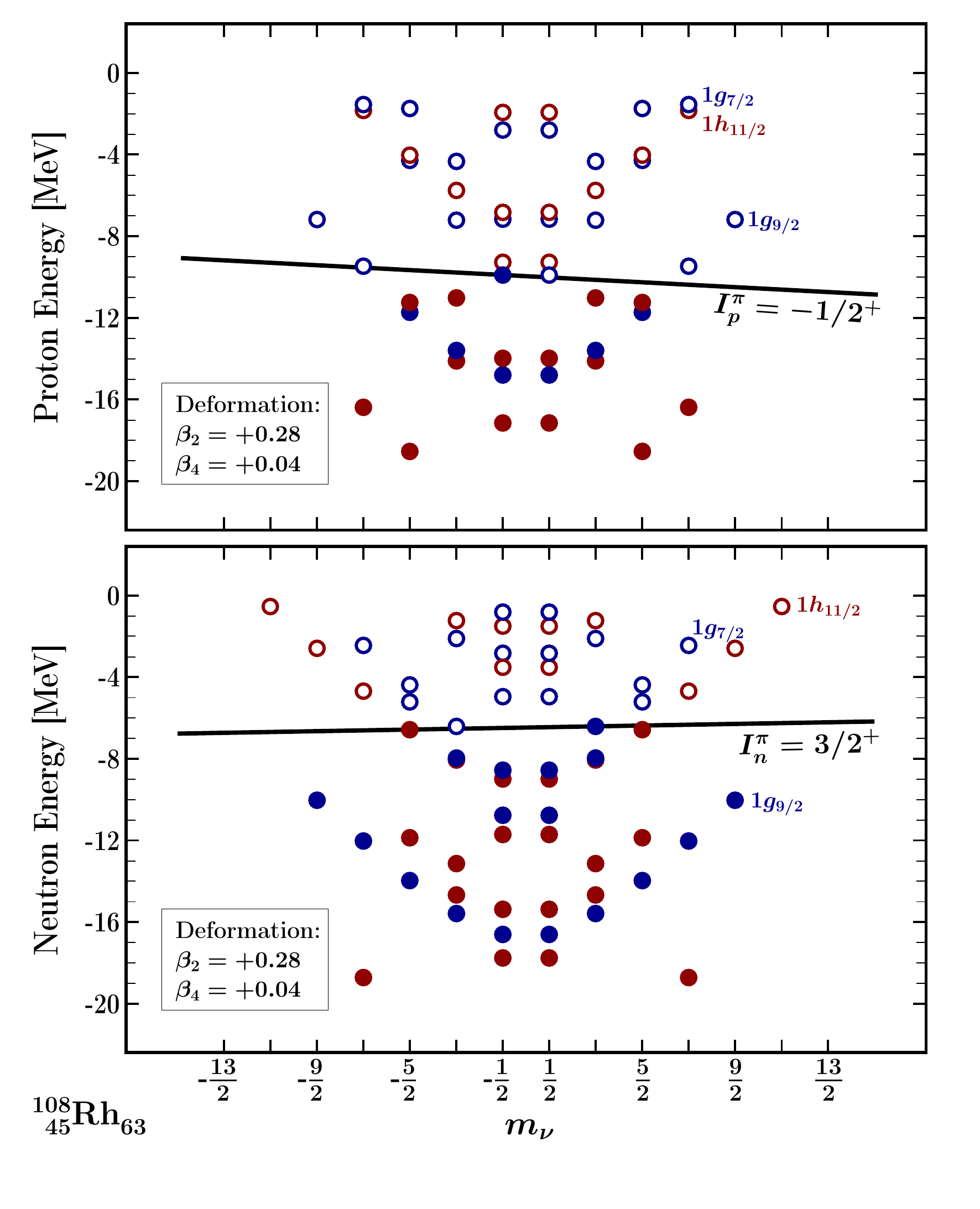}
         \end{center}
\caption{Single particle energies, $e_\nu$ vs.~angular momentum projection $m_\nu$ for protons, top, and neutrons, bottom. Notice that the resulting proton projection is $M=-1/2$ originating from the unpaired $g_{7/2,m=-1/2}$ orbital whereas the 63rd neutron occupying $g_{7/2,m=3/2}$ orbital resulting in $M=3/2$. Both configurations being of negative parity result in the positive parity of the final product wave function $\varphi=\varphi_{prot.} \cdot \varphi_{neut.}$. 
}
                                                              \label{Umb_108Rh}
\end{figure}

Results from the tilted Fermi surface calculations in Fig.~\ref{Umb_108Rh} for protons, top diagram and neutrons, bottom diagram, show that the single particle orbitals lead to the sought spin estimate $M = | 3/2 - 1/2 |=1  \leftrightarrow I = 1$. A projection $3/2^+$ is generated by the 63rd neutron occupying again the $g_{7/2,m=3/2}$ orbital. The couple formed by the  61st and the 62nd neutrons occupies a pair of opposite $m_\nu$-projection orbitals, $h_{{11/2},m=\pm5/2}$, and thus results in a null spin contribution to the final projection value. The 45th proton occupies the $g_{7/2,m=-1/2}$ orbital and is the least bound proton in this nucleus.

\subsubsection{I$_{g.s.}^{\,\pi} = 1^+$ ground state in $^{112}${Rh}}
\label{Section-D5}

\begin{figure}[h!]
\begin{center}
         \includegraphics[width=0.50\textwidth,angle=-00]{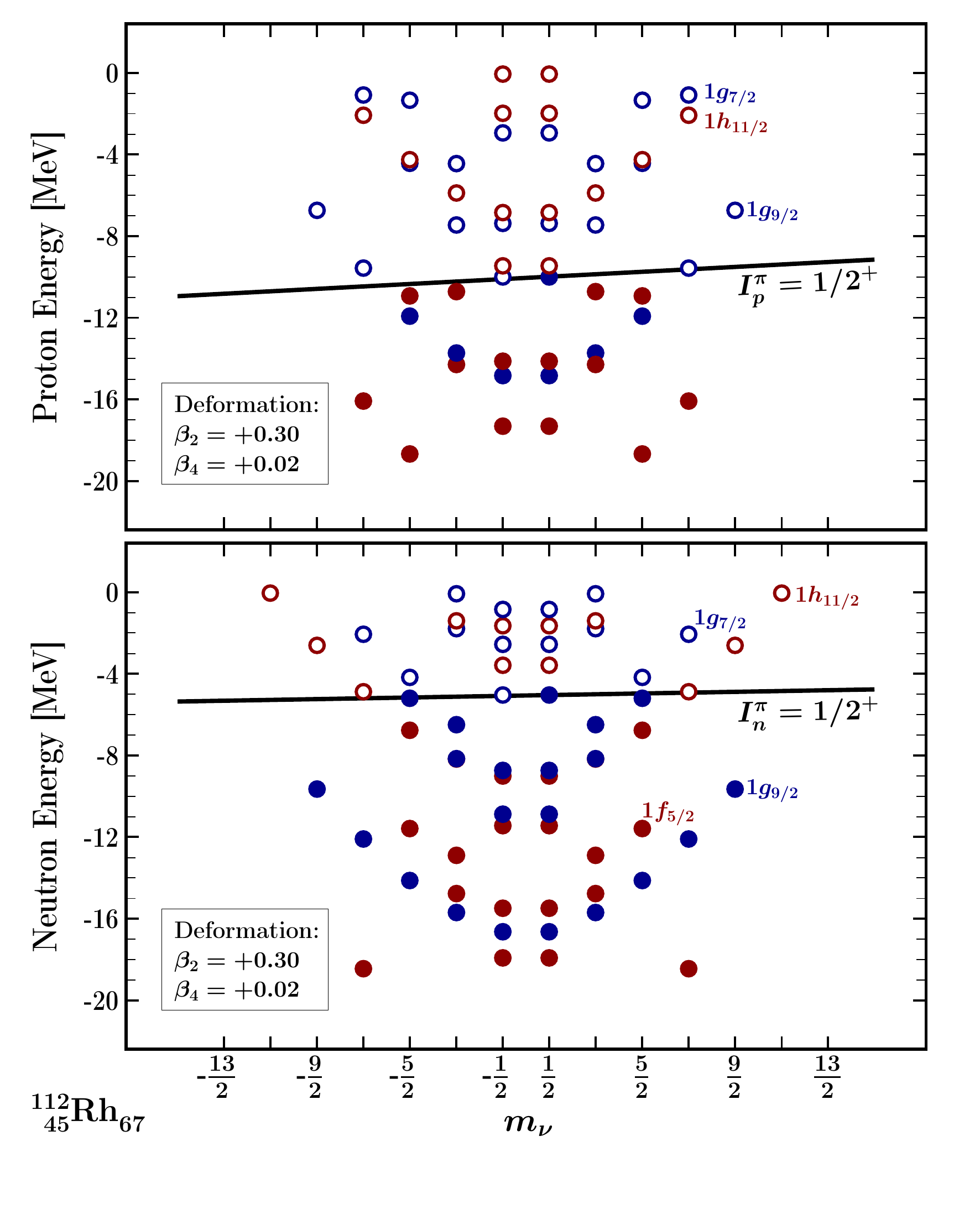}
         \end{center}
\caption{These single particle diagrams correspond to the largest quadrupole deformation in the series: $\alpha_{20} \leftrightarrow \beta_2=0.30$. Here we collect two smallest possible projections of $1/2^+$, once obtained by 67th neutron occupying the $s_{1/2,m=+1/2}$ orbital and once with the 45th proton occupying its $g_{7/2,m=+1/2}$ orbital, all in the zones of several orbitals interacting via well known Landau-Zener non-crossing rule. All the remaining valence neutrons occupy their orbitals with both $m_\nu$-projections pairwise occupied with the null contribution to the final $M$ value.}
                                                              \label{Umb_112Rh}
\end{figure}

The tilted Fermi surface diagrams, shown in Figs.~\ref{Umb_112Rh}, indicate that also $^{112}$Rh has the expected ground-state with the spin-parity estimate. 

The expected ground state spin $I_{\rm g.s.}^\pi =1$  is obtained in this case by summing the smallest positive contributions from protons and from neutrons: $M = ( 1/2 + 1/2 )=1  \leftrightarrow I = 1$, whereas all the remaining neutrons occupy their orbitals pairwise with opposite $m_\nu$-projections.

\subsubsection{I$_{g.s.}^{\,\pi} = 1^+$ ground state in $^{114}${Rh}}
\label{Section-D6}

Interestingly, the formal spin-projection contributions calculated for  $^{114}${Rh}
are the same as the ones for  $^{110}${Rh} but at a visibly smaller quadrupole deformation, here at $\beta_2=0.24$ compared to $\beta_2=0.28$. Whereas the proton occupation configurations are the same, despite visible differences in the distributions of the ``bullets'' -- in both cases we have the exact solutions of the tilted Fermi surface algorithm with all occupied levels lying totally below the corresponding straight lines in the single-particle diagrams -- in the neutron case, with two extra particles present, the tilted Fermi level algorithm is respected only in the $^{110}${Rh} case. This is because in the case of  $^{114}${Rh} we have $s_{1/2,m=\pm1/2}$ couple occupied, the latter disobeying the tilting pattern. 

\begin{figure}[h!]
\begin{center}
         \includegraphics[width=0.50\textwidth,angle=-00]{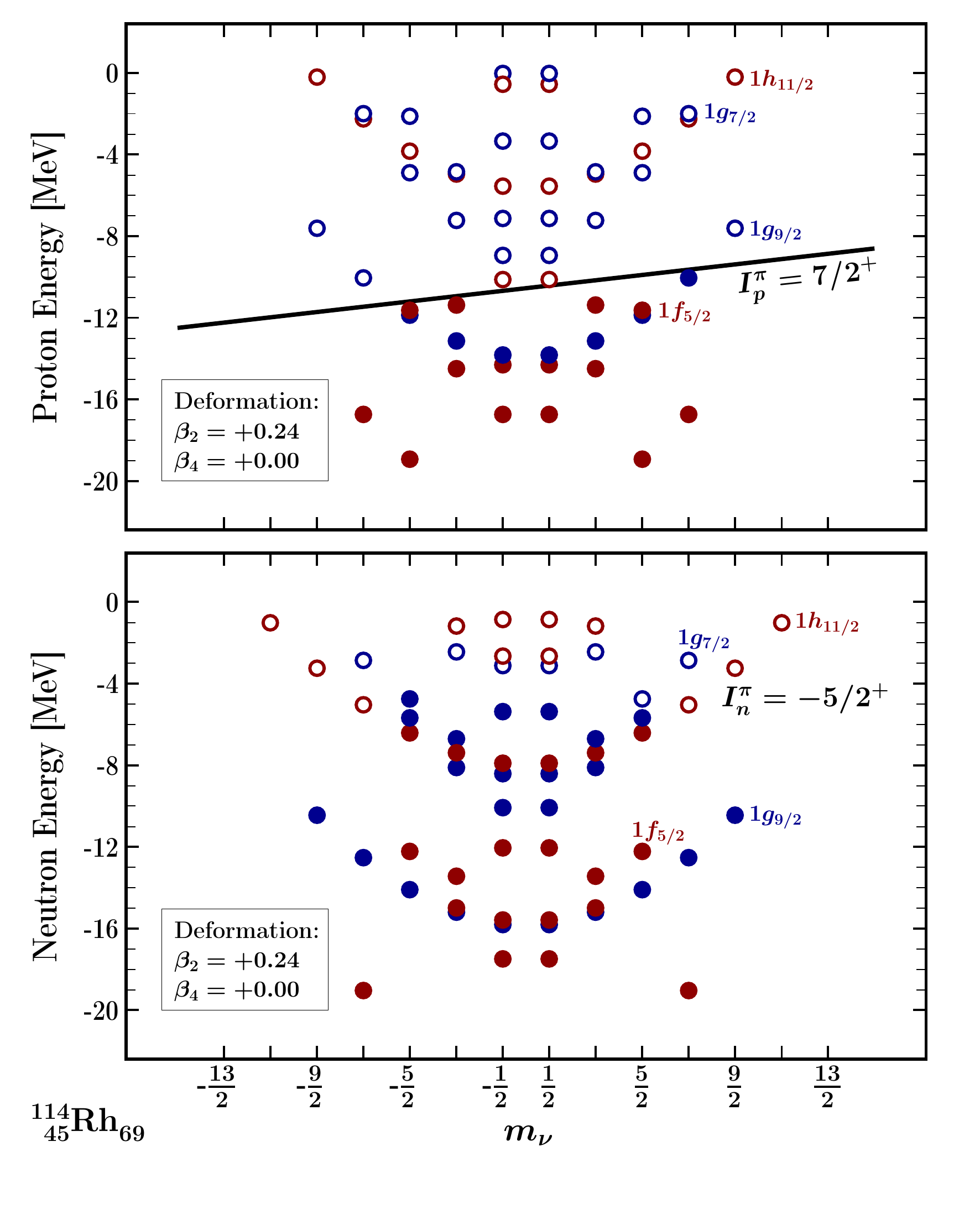}
         \end{center}
\caption{Here the negative projection $-5/2^+$ is obtained by the 69th neutron occupying the uncompensated $d_{5/2,m=-5/2}$ orbital in the zone of several orbitals interacting via well known Landau-Zener repulsion rules, whereas the remaining valence neutrons occupy their orbital pairwise with the null contribution to the final projection value. In contrast, proton occupation pattern follows clearly the tilted Fermi surface scenario with the unpaired occupation of $g_{9/2,m=7/2}$ level.}
                                                              \label{Umb_114Rh}
\end{figure}

At the end, the numerical values of the projections are equal in both cases, $M=1$; graphical illustration of the discussed patterns is to be seen from Fig.~\ref{Umb_114Rh}.  

Similarly as for $^{106}$Rh, we do not obtain $E=0$ but a slight discrepancy of a couple of hundreds of keV, that can be qualified as imperfection in terms of the single-nucleon level positions.

\subsubsection{I$_{g.s.}^{\,\pi} = 1^+$ ground state in $^{116}${Rh}}
\label{Section-D7}

\begin{figure}[h!]
\begin{center}
         \includegraphics[width=0.50\textwidth,angle=-00]{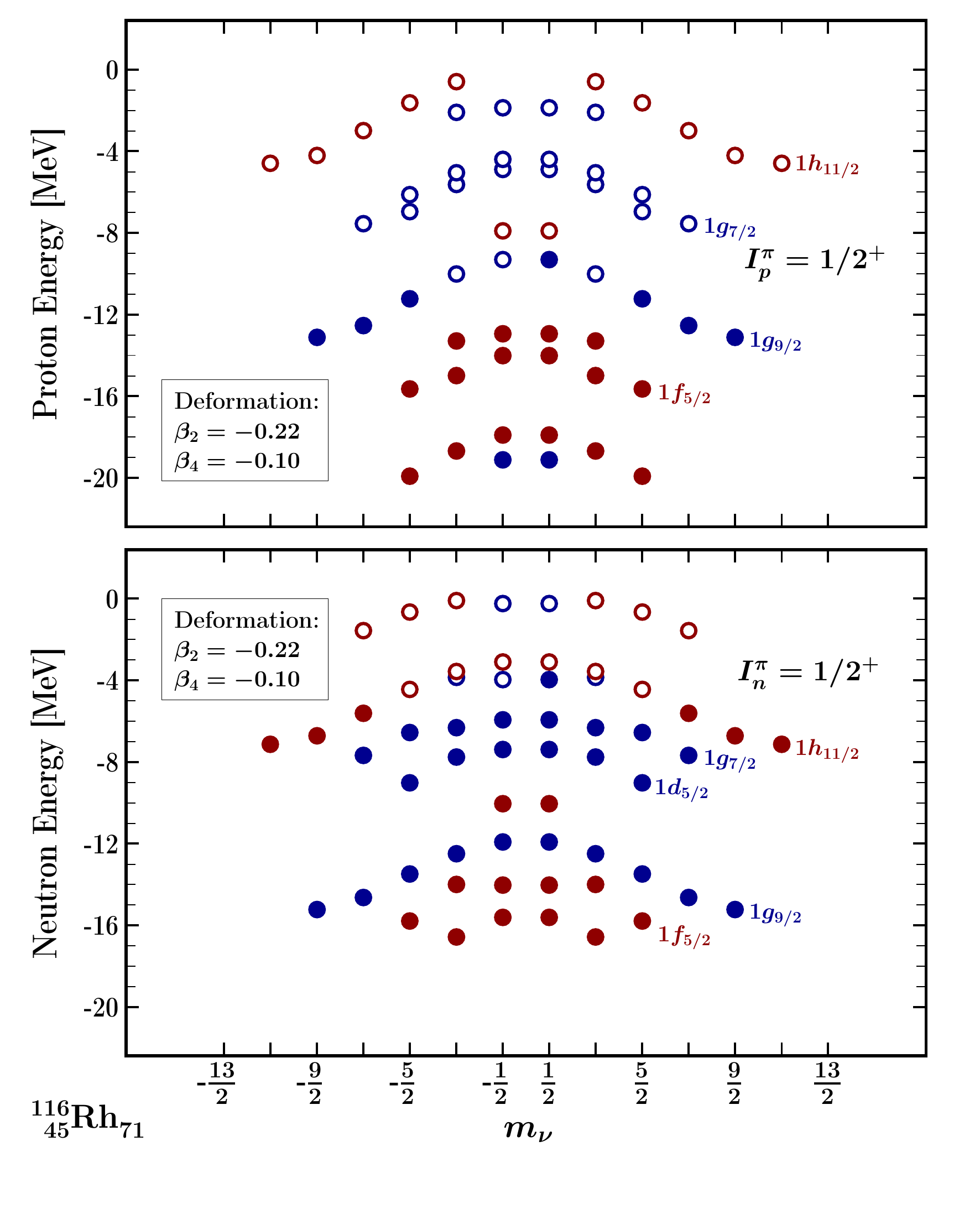}
         \end{center}
\caption{This $^{116}$Rh diagram correspond to the oblate shape of $\beta_2=-0.22$ as evidenced by the reversing of the umbrella shape.}
                                                              \label{Umb_116Rh}
\end{figure}

In the case of  $^{116}$Rh, we are entering particularly interesting results switching from the prolate shape to oblate shape configurations as indicated by the reversing in the umbrella plot shape in Fig.~\ref{Umb_116Rh}. In geometrical terms the occurring shape transition represents highly non-trivial transformation since we are passing from quadrupole deformations typically of the order of $\beta_2\approx +0.25$ to the strongly oblate ones of the order of $\beta_2\approx -0.25$. 

Our tilted Fermi surface solution is illustrated in Fig.~\ref{Umb_116Rh} with the diagrams which show that both protons and neutrons contribute with projection $m_\nu=+1/2$. Consequently, as the result of the contributions with the smallest angular momentum projections of single particle orbitals we find: $M = | 1/2 + 1/2 |=1 \leftrightarrow I = 1$.

$^{116}$Rh is the only case where the calculated energies are far above the ground state, slightly over 1 MeV. Here the energy expected to be the lowest in the yrast line, happens to be very high and cannot be qualified as the result of imperfections in the single-nucleon energies. 
It should be noted that for two more cases, $^{106,114}$Rh, the calculated energy of the ground state is not the lowest, but only by a small margin of a few 100s of keV.

\clearpage

\bibliography{apssamp}

\begin{thebibliography}{56}%
\makeatletter
\providecommand \@ifxundefined [1]{%
 \@ifx{#1\undefined}
}%
\providecommand \@ifnum [1]{%
 \ifnum #1\expandafter \@firstoftwo
 \else \expandafter \@secondoftwo
 \fi
}%
\providecommand \@ifx [1]{%
 \ifx #1\expandafter \@firstoftwo
 \else \expandafter \@secondoftwo
 \fi
}%
\providecommand \natexlab [1]{#1}%
\providecommand \enquote  [1]{``#1''}%
\providecommand \bibnamefont  [1]{#1}%
\providecommand \bibfnamefont [1]{#1}%
\providecommand \citenamefont [1]{#1}%
\providecommand \href@noop [0]{\@secondoftwo}%
\providecommand \href [0]{\begingroup \@sanitize@url \@href}%
\providecommand \@href[1]{\@@startlink{#1}\@@href}%
\providecommand \@@href[1]{\endgroup#1\@@endlink}%
\providecommand \@sanitize@url [0]{\catcode `\\12\catcode `\$12\catcode `\&12\catcode `\#12\catcode `\^12\catcode `\_12\catcode `\%12\relax}%
\providecommand \@@startlink[1]{}%
\providecommand \@@endlink[0]{}%
\providecommand \url  [0]{\begingroup\@sanitize@url \@url }%
\providecommand \@url [1]{\endgroup\@href {#1}{\urlprefix }}%
\providecommand \urlprefix  [0]{URL }%
\providecommand \Eprint [0]{\href }%
\providecommand \doibase [0]{http://dx.doi.org/}%
\providecommand \selectlanguage [0]{\@gobble}%
\providecommand \bibinfo  [0]{\@secondoftwo}%
\providecommand \bibfield  [0]{\@secondoftwo}%
\providecommand \translation [1]{[#1]}%
\providecommand \BibitemOpen [0]{}%
\providecommand \bibitemStop [0]{}%
\providecommand \bibitemNoStop [0]{.\EOS\space}%
\providecommand \EOS [0]{\spacefactor3000\relax}%
\providecommand \BibitemShut  [1]{\csname bibitem#1\endcsname}%
\let\auto@bib@innerbib\@empty
\bibitem [{\citenamefont {Garrett}\ \emph {et~al.}(2022)\citenamefont {Garrett}, \citenamefont {Zieli\'nska},\ and\ \citenamefont {Clement}}]{GARRETT2022}%
  \BibitemOpen
  \bibfield  {author} {\bibinfo {author} {\bibfnamefont {P.~E.}\ \bibnamefont {Garrett}}, \bibinfo {author} {\bibfnamefont {M.}~\bibnamefont {Zieli\'nska}}, \ and\ \bibinfo {author} {\bibfnamefont {E.}~\bibnamefont {Clement}},\ }\href {\doibase https://doi.org/10.1016/j.ppnp.2021.103931} {\bibfield  {journal} {\bibinfo  {journal} {Progress in Particle and Nuclear Physics}\ }\textbf {\bibinfo {volume} {124}},\ \bibinfo {pages} {103931} (\bibinfo {year} {2022})}\BibitemShut {NoStop}%
\bibitem [{\citenamefont {Orford}\ \emph {et~al.}(2020{\natexlab{a}})\citenamefont {Orford}, \citenamefont {Kondev}, \citenamefont {Savard}, \citenamefont {Clark}, \citenamefont {Porter}, \citenamefont {Ray}, \citenamefont {Buchinger}, \citenamefont {Burkey}, \citenamefont {Gorelov}, \citenamefont {Hartley}, \citenamefont {Klimes}, \citenamefont {Sharma}, \citenamefont {Valverde},\ and\ \citenamefont {Yan}}]{OrfordSpinTrap}%
  \BibitemOpen
  \bibfield  {author} {\bibinfo {author} {\bibfnamefont {R.}~\bibnamefont {Orford}}, \bibinfo {author} {\bibfnamefont {F.~G.}\ \bibnamefont {Kondev}}, \bibinfo {author} {\bibfnamefont {G.}~\bibnamefont {Savard}}, \bibinfo {author} {\bibfnamefont {J.~A.}\ \bibnamefont {Clark}}, \bibinfo {author} {\bibfnamefont {W.~S.}\ \bibnamefont {Porter}}, \bibinfo {author} {\bibfnamefont {D.}~\bibnamefont {Ray}}, \bibinfo {author} {\bibfnamefont {F.}~\bibnamefont {Buchinger}}, \bibinfo {author} {\bibfnamefont {M.~T.}\ \bibnamefont {Burkey}}, \bibinfo {author} {\bibfnamefont {D.~A.}\ \bibnamefont {Gorelov}}, \bibinfo {author} {\bibfnamefont {D.~J.}\ \bibnamefont {Hartley}}, \bibinfo {author} {\bibfnamefont {J.~W.}\ \bibnamefont {Klimes}}, \bibinfo {author} {\bibfnamefont {K.~S.}\ \bibnamefont {Sharma}}, \bibinfo {author} {\bibfnamefont {A.~A.}\ \bibnamefont {Valverde}}, \ and\ \bibinfo {author} {\bibfnamefont {X.~L.}\ \bibnamefont {Yan}},\ }\href {\doibase 10.1103/PhysRevC.102.011303} {\bibfield  {journal} {\bibinfo
  {journal} {Phys. Rev. C}\ }\textbf {\bibinfo {volume} {102}},\ \bibinfo {pages} {011303} (\bibinfo {year} {2020}{\natexlab{a}})}\BibitemShut {NoStop}%
\bibitem [{\citenamefont {Eliseev}\ \emph {et~al.}(2013)\citenamefont {Eliseev}, \citenamefont {Blaum}, \citenamefont {Block}, \citenamefont {Droese}, \citenamefont {Goncharov}, \citenamefont {Minaya~Ramirez}, \citenamefont {Nesterenko}, \citenamefont {Novikov},\ and\ \citenamefont {Schweikhard}}]{PhysRevLett.110.082501}%
  \BibitemOpen
  \bibfield  {author} {\bibinfo {author} {\bibfnamefont {S.}~\bibnamefont {Eliseev}}, \bibinfo {author} {\bibfnamefont {K.}~\bibnamefont {Blaum}}, \bibinfo {author} {\bibfnamefont {M.}~\bibnamefont {Block}}, \bibinfo {author} {\bibfnamefont {C.}~\bibnamefont {Droese}}, \bibinfo {author} {\bibfnamefont {M.}~\bibnamefont {Goncharov}}, \bibinfo {author} {\bibfnamefont {E.}~\bibnamefont {Minaya~Ramirez}}, \bibinfo {author} {\bibfnamefont {D.~A.}\ \bibnamefont {Nesterenko}}, \bibinfo {author} {\bibfnamefont {Y.~N.}\ \bibnamefont {Novikov}}, \ and\ \bibinfo {author} {\bibfnamefont {L.}~\bibnamefont {Schweikhard}},\ }\href {\doibase 10.1103/PhysRevLett.110.082501} {\bibfield  {journal} {\bibinfo  {journal} {Phys. Rev. Lett.}\ }\textbf {\bibinfo {volume} {110}},\ \bibinfo {pages} {082501} (\bibinfo {year} {2013})}\BibitemShut {NoStop}%
\bibitem [{\citenamefont {Kondev}\ \emph {et~al.}(2021)\citenamefont {Kondev}, \citenamefont {Wang}, \citenamefont {Huang}, \citenamefont {Naimi},\ and\ \citenamefont {Audi}}]{NUBASE}%
  \BibitemOpen
  \bibfield  {author} {\bibinfo {author} {\bibfnamefont {F.}~\bibnamefont {Kondev}}, \bibinfo {author} {\bibfnamefont {M.}~\bibnamefont {Wang}}, \bibinfo {author} {\bibfnamefont {W.}~\bibnamefont {Huang}}, \bibinfo {author} {\bibfnamefont {S.}~\bibnamefont {Naimi}}, \ and\ \bibinfo {author} {\bibfnamefont {G.}~\bibnamefont {Audi}},\ }\href {\doibase 10.1088/1674-1137/abddae} {\bibfield  {journal} {\bibinfo  {journal} {Chinese Physics C}\ }\textbf {\bibinfo {volume} {45}},\ \bibinfo {pages} {030001} (\bibinfo {year} {2021})}\BibitemShut {NoStop}%
\bibitem [{\citenamefont {Hukkanen}\ \emph {et~al.}(2023)\citenamefont {Hukkanen}, \citenamefont {Ryssens}, \citenamefont {Ascher}, \citenamefont {Bender}, \citenamefont {Eronen}, \citenamefont {Gr\'evy}, \citenamefont {Kankainen}, \citenamefont {Stryjczyk}, \citenamefont {Al~Ayoubi}, \citenamefont {Ayet}, \citenamefont {Beliuskina}, \citenamefont {Delafosse}, \citenamefont {Gins}, \citenamefont {Gerbaux}, \citenamefont {Husson}, \citenamefont {Jokinen}, \citenamefont {Nesterenko}, \citenamefont {Pohjalainen}, \citenamefont {Reponen}, \citenamefont {Rinta-Antila}, \citenamefont {de~Roubin},\ and\ \citenamefont {Weaver}}]{JYFLarxiv}%
  \BibitemOpen
  \bibfield  {author} {\bibinfo {author} {\bibfnamefont {M.}~\bibnamefont {Hukkanen}}, \bibinfo {author} {\bibfnamefont {W.}~\bibnamefont {Ryssens}}, \bibinfo {author} {\bibfnamefont {P.}~\bibnamefont {Ascher}}, \bibinfo {author} {\bibfnamefont {M.}~\bibnamefont {Bender}}, \bibinfo {author} {\bibfnamefont {T.}~\bibnamefont {Eronen}}, \bibinfo {author} {\bibfnamefont {S.}~\bibnamefont {Gr\'evy}}, \bibinfo {author} {\bibfnamefont {A.}~\bibnamefont {Kankainen}}, \bibinfo {author} {\bibfnamefont {M.}~\bibnamefont {Stryjczyk}}, \bibinfo {author} {\bibfnamefont {L.}~\bibnamefont {Al~Ayoubi}}, \bibinfo {author} {\bibfnamefont {S.}~\bibnamefont {Ayet}}, \bibinfo {author} {\bibfnamefont {O.}~\bibnamefont {Beliuskina}}, \bibinfo {author} {\bibfnamefont {C.}~\bibnamefont {Delafosse}}, \bibinfo {author} {\bibfnamefont {W.}~\bibnamefont {Gins}}, \bibinfo {author} {\bibfnamefont {M.}~\bibnamefont {Gerbaux}}, \bibinfo {author} {\bibfnamefont {A.}~\bibnamefont {Husson}}, \bibinfo {author} {\bibfnamefont {A.}~\bibnamefont
  {Jokinen}}, \bibinfo {author} {\bibfnamefont {D.~A.}\ \bibnamefont {Nesterenko}}, \bibinfo {author} {\bibfnamefont {I.}~\bibnamefont {Pohjalainen}}, \bibinfo {author} {\bibfnamefont {M.}~\bibnamefont {Reponen}}, \bibinfo {author} {\bibfnamefont {S.}~\bibnamefont {Rinta-Antila}}, \bibinfo {author} {\bibfnamefont {A.}~\bibnamefont {de~Roubin}}, \ and\ \bibinfo {author} {\bibfnamefont {A.~P.}\ \bibnamefont {Weaver}},\ }\href {\doibase 10.1103/PhysRevC.107.014306} {\bibfield  {journal} {\bibinfo  {journal} {Phys. Rev. C}\ }\textbf {\bibinfo {volume} {107}},\ \bibinfo {pages} {014306} (\bibinfo {year} {2023})}\BibitemShut {NoStop}%
\bibitem [{\citenamefont {Woods}\ and\ \citenamefont {Saxon}(1954)}]{WoodsSaxon}%
  \BibitemOpen
  \bibfield  {author} {\bibinfo {author} {\bibfnamefont {R.~D.}\ \bibnamefont {Woods}}\ and\ \bibinfo {author} {\bibfnamefont {D.~S.}\ \bibnamefont {Saxon}},\ }\href {\doibase 10.1103/PhysRev.95.577} {\bibfield  {journal} {\bibinfo  {journal} {Phys. Rev.}\ }\textbf {\bibinfo {volume} {95}},\ \bibinfo {pages} {577} (\bibinfo {year} {1954})}\BibitemShut {NoStop}%
\bibitem [{\citenamefont {Porter}\ \emph {et~al.}(2022)\citenamefont {Porter}, \citenamefont {Ashrafkhani}, \citenamefont {Bergmann}, \citenamefont {Brown}, \citenamefont {Brunner}, \citenamefont {Cardona}, \citenamefont {Curien}, \citenamefont {Dedes}, \citenamefont {Dickel}, \citenamefont {Dudek}, \citenamefont {Dunling}, \citenamefont {Gwinner}, \citenamefont {Hockenbery}, \citenamefont {Holt}, \citenamefont {Hornung}, \citenamefont {Izzo}, \citenamefont {Jacobs}, \citenamefont {Javaji}, \citenamefont {Kootte}, \citenamefont {Kripk\'o-Koncz}, \citenamefont {Lykiardopoulou}, \citenamefont {Miyagi}, \citenamefont {Mukul}, \citenamefont {Murb\"ock}, \citenamefont {Pla\ss{}}, \citenamefont {Reiter}, \citenamefont {Ringuette}, \citenamefont {Scheidenberger}, \citenamefont {Silwal}, \citenamefont {Walls}, \citenamefont {Wang}, \citenamefont {Wang}, \citenamefont {Yang}, \citenamefont {Dilling},\ and\ \citenamefont {Kwiatkowski}}]{Porter2022}%
  \BibitemOpen
  \bibfield  {author} {\bibinfo {author} {\bibfnamefont {W.~S.}\ \bibnamefont {Porter}}, \bibinfo {author} {\bibfnamefont {B.}~\bibnamefont {Ashrafkhani}}, \bibinfo {author} {\bibfnamefont {J.}~\bibnamefont {Bergmann}}, \bibinfo {author} {\bibfnamefont {C.}~\bibnamefont {Brown}}, \bibinfo {author} {\bibfnamefont {T.}~\bibnamefont {Brunner}}, \bibinfo {author} {\bibfnamefont {J.~D.}\ \bibnamefont {Cardona}}, \bibinfo {author} {\bibfnamefont {D.}~\bibnamefont {Curien}}, \bibinfo {author} {\bibfnamefont {I.}~\bibnamefont {Dedes}}, \bibinfo {author} {\bibfnamefont {T.}~\bibnamefont {Dickel}}, \bibinfo {author} {\bibfnamefont {J.}~\bibnamefont {Dudek}}, \bibinfo {author} {\bibfnamefont {E.}~\bibnamefont {Dunling}}, \bibinfo {author} {\bibfnamefont {G.}~\bibnamefont {Gwinner}}, \bibinfo {author} {\bibfnamefont {Z.}~\bibnamefont {Hockenbery}}, \bibinfo {author} {\bibfnamefont {J.~D.}\ \bibnamefont {Holt}}, \bibinfo {author} {\bibfnamefont {C.}~\bibnamefont {Hornung}}, \bibinfo {author} {\bibfnamefont
  {C.}~\bibnamefont {Izzo}}, \bibinfo {author} {\bibfnamefont {A.}~\bibnamefont {Jacobs}}, \bibinfo {author} {\bibfnamefont {A.}~\bibnamefont {Javaji}}, \bibinfo {author} {\bibfnamefont {B.}~\bibnamefont {Kootte}}, \bibinfo {author} {\bibfnamefont {G.}~\bibnamefont {Kripk\'o-Koncz}}, \bibinfo {author} {\bibfnamefont {E.~M.}\ \bibnamefont {Lykiardopoulou}}, \bibinfo {author} {\bibfnamefont {T.}~\bibnamefont {Miyagi}}, \bibinfo {author} {\bibfnamefont {I.}~\bibnamefont {Mukul}}, \bibinfo {author} {\bibfnamefont {T.}~\bibnamefont {Murb\"ock}}, \bibinfo {author} {\bibfnamefont {W.~R.}\ \bibnamefont {Pla\ss{}}}, \bibinfo {author} {\bibfnamefont {M.~P.}\ \bibnamefont {Reiter}}, \bibinfo {author} {\bibfnamefont {J.}~\bibnamefont {Ringuette}}, \bibinfo {author} {\bibfnamefont {C.}~\bibnamefont {Scheidenberger}}, \bibinfo {author} {\bibfnamefont {R.}~\bibnamefont {Silwal}}, \bibinfo {author} {\bibfnamefont {C.}~\bibnamefont {Walls}}, \bibinfo {author} {\bibfnamefont {H.~L.}\ \bibnamefont {Wang}}, \bibinfo {author}
  {\bibfnamefont {Y.}~\bibnamefont {Wang}}, \bibinfo {author} {\bibfnamefont {J.}~\bibnamefont {Yang}}, \bibinfo {author} {\bibfnamefont {J.}~\bibnamefont {Dilling}}, \ and\ \bibinfo {author} {\bibfnamefont {A.~A.}\ \bibnamefont {Kwiatkowski}},\ }\href {\doibase 10.1103/PhysRevC.105.L041301} {\bibfield  {journal} {\bibinfo  {journal} {Phys. Rev. C}\ }\textbf {\bibinfo {volume} {105}},\ \bibinfo {pages} {L041301} (\bibinfo {year} {2022})}\BibitemShut {NoStop}%
\bibitem [{\citenamefont {Savard}\ \emph {et~al.}(2008)\citenamefont {Savard}, \citenamefont {Baker}, \citenamefont {Davids}, \citenamefont {Levand}, \citenamefont {Moore}, \citenamefont {Pardo}, \citenamefont {Vondrasek}, \citenamefont {Zabransky},\ and\ \citenamefont {Zinkann}}]{CARIBU}%
  \BibitemOpen
  \bibfield  {author} {\bibinfo {author} {\bibfnamefont {G.}~\bibnamefont {Savard}}, \bibinfo {author} {\bibfnamefont {S.}~\bibnamefont {Baker}}, \bibinfo {author} {\bibfnamefont {C.}~\bibnamefont {Davids}}, \bibinfo {author} {\bibfnamefont {A.}~\bibnamefont {Levand}}, \bibinfo {author} {\bibfnamefont {E.}~\bibnamefont {Moore}}, \bibinfo {author} {\bibfnamefont {R.}~\bibnamefont {Pardo}}, \bibinfo {author} {\bibfnamefont {R.}~\bibnamefont {Vondrasek}}, \bibinfo {author} {\bibfnamefont {B.}~\bibnamefont {Zabransky}}, \ and\ \bibinfo {author} {\bibfnamefont {G.}~\bibnamefont {Zinkann}},\ }\href {\doibase https://doi.org/10.1016/j.nimb.2008.05.091} {\bibfield  {journal} {\bibinfo  {journal} {Nuclear Instruments and Methods in Physics Research Section B: Beam Interactions with Materials and Atoms}\ }\textbf {\bibinfo {volume} {266}},\ \bibinfo {pages} {4086} (\bibinfo {year} {2008})},\ \bibinfo {note} {proceedings of the XVth International Conference on Electromagnetic Isotope Separators and Techniques Related to
  their Applications}\BibitemShut {NoStop}%
\bibitem [{\citenamefont {Hartley}\ \emph {et~al.}(2018)\citenamefont {Hartley}, \citenamefont {Kondev}, \citenamefont {Orford}, \citenamefont {Clark}, \citenamefont {Savard}, \citenamefont {Ayangeakaa}, \citenamefont {Bottoni}, \citenamefont {Buchinger}, \citenamefont {Burkey}, \citenamefont {Carpenter}, \citenamefont {Copp}, \citenamefont {Gorelov}, \citenamefont {Hicks}, \citenamefont {Hoffman}, \citenamefont {Hu}, \citenamefont {Janssens}, \citenamefont {Klimes}, \citenamefont {Lauritsen}, \citenamefont {Sethi}, \citenamefont {Seweryniak}, \citenamefont {Sharma}, \citenamefont {Zhang}, \citenamefont {Zhu},\ and\ \citenamefont {Zhu}}]{Hartley2018}%
  \BibitemOpen
  \bibfield  {author} {\bibinfo {author} {\bibfnamefont {D.~J.}\ \bibnamefont {Hartley}}, \bibinfo {author} {\bibfnamefont {F.~G.}\ \bibnamefont {Kondev}}, \bibinfo {author} {\bibfnamefont {R.}~\bibnamefont {Orford}}, \bibinfo {author} {\bibfnamefont {J.~A.}\ \bibnamefont {Clark}}, \bibinfo {author} {\bibfnamefont {G.}~\bibnamefont {Savard}}, \bibinfo {author} {\bibfnamefont {A.~D.}\ \bibnamefont {Ayangeakaa}}, \bibinfo {author} {\bibfnamefont {S.}~\bibnamefont {Bottoni}}, \bibinfo {author} {\bibfnamefont {F.}~\bibnamefont {Buchinger}}, \bibinfo {author} {\bibfnamefont {M.~T.}\ \bibnamefont {Burkey}}, \bibinfo {author} {\bibfnamefont {M.~P.}\ \bibnamefont {Carpenter}}, \bibinfo {author} {\bibfnamefont {P.}~\bibnamefont {Copp}}, \bibinfo {author} {\bibfnamefont {D.~A.}\ \bibnamefont {Gorelov}}, \bibinfo {author} {\bibfnamefont {K.}~\bibnamefont {Hicks}}, \bibinfo {author} {\bibfnamefont {C.~R.}\ \bibnamefont {Hoffman}}, \bibinfo {author} {\bibfnamefont {C.}~\bibnamefont {Hu}}, \bibinfo {author} {\bibfnamefont
  {R.~V.~F.}\ \bibnamefont {Janssens}}, \bibinfo {author} {\bibfnamefont {J.~W.}\ \bibnamefont {Klimes}}, \bibinfo {author} {\bibfnamefont {T.}~\bibnamefont {Lauritsen}}, \bibinfo {author} {\bibfnamefont {J.}~\bibnamefont {Sethi}}, \bibinfo {author} {\bibfnamefont {D.}~\bibnamefont {Seweryniak}}, \bibinfo {author} {\bibfnamefont {K.~S.}\ \bibnamefont {Sharma}}, \bibinfo {author} {\bibfnamefont {H.}~\bibnamefont {Zhang}}, \bibinfo {author} {\bibfnamefont {S.}~\bibnamefont {Zhu}}, \ and\ \bibinfo {author} {\bibfnamefont {Y.}~\bibnamefont {Zhu}},\ }\href {\doibase 10.1103/PhysRevLett.120.182502} {\bibfield  {journal} {\bibinfo  {journal} {Phys. Rev. Lett.}\ }\textbf {\bibinfo {volume} {120}},\ \bibinfo {pages} {182502} (\bibinfo {year} {2018})}\BibitemShut {NoStop}%
\bibitem [{\citenamefont {Davids}\ and\ \citenamefont {Peterson}(2008)}]{Davids08}%
  \BibitemOpen
  \bibfield  {author} {\bibinfo {author} {\bibfnamefont {C.~N.}\ \bibnamefont {Davids}}\ and\ \bibinfo {author} {\bibfnamefont {D.}~\bibnamefont {Peterson}},\ }\href {\doibase https://doi.org/10.1016/j.nimb.2008.05.148} {\bibfield  {journal} {\bibinfo  {journal} {Nuclear Instruments and Methods in Physics Research Section B: Beam Interactions with Materials and Atoms}\ }\textbf {\bibinfo {volume} {266}},\ \bibinfo {pages} {4449} (\bibinfo {year} {2008})},\ \bibinfo {note} {proceedings of the XVth International Conference on Electromagnetic Isotope Separators and Techniques Related to their Applications}\BibitemShut {NoStop}%
\bibitem [{\citenamefont {Hirsh}\ \emph {et~al.}(2016)\citenamefont {Hirsh}, \citenamefont {Paul}, \citenamefont {Burkey}, \citenamefont {Aprahamian}, \citenamefont {Buchinger}, \citenamefont {Caldwell}, \citenamefont {Clark}, \citenamefont {Levand}, \citenamefont {Ying}, \citenamefont {Marley}, \citenamefont {Morgan}, \citenamefont {Nystrom}, \citenamefont {Orford}, \citenamefont {Galvan}, \citenamefont {Rohrer}, \citenamefont {Savard}, \citenamefont {Sharma},\ and\ \citenamefont {Siegl}}]{CARIBU-MRTOF}%
  \BibitemOpen
  \bibfield  {author} {\bibinfo {author} {\bibfnamefont {T.~Y.}\ \bibnamefont {Hirsh}}, \bibinfo {author} {\bibfnamefont {N.}~\bibnamefont {Paul}}, \bibinfo {author} {\bibfnamefont {M.}~\bibnamefont {Burkey}}, \bibinfo {author} {\bibfnamefont {A.}~\bibnamefont {Aprahamian}}, \bibinfo {author} {\bibfnamefont {F.}~\bibnamefont {Buchinger}}, \bibinfo {author} {\bibfnamefont {S.}~\bibnamefont {Caldwell}}, \bibinfo {author} {\bibfnamefont {J.~A.}\ \bibnamefont {Clark}}, \bibinfo {author} {\bibfnamefont {A.~F.}\ \bibnamefont {Levand}}, \bibinfo {author} {\bibfnamefont {L.~L.}\ \bibnamefont {Ying}}, \bibinfo {author} {\bibfnamefont {S.~T.}\ \bibnamefont {Marley}}, \bibinfo {author} {\bibfnamefont {G.~E.}\ \bibnamefont {Morgan}}, \bibinfo {author} {\bibfnamefont {A.}~\bibnamefont {Nystrom}}, \bibinfo {author} {\bibfnamefont {R.}~\bibnamefont {Orford}}, \bibinfo {author} {\bibfnamefont {A.~P.}\ \bibnamefont {Galvan}}, \bibinfo {author} {\bibfnamefont {J.}~\bibnamefont {Rohrer}}, \bibinfo {author} {\bibfnamefont
  {G.}~\bibnamefont {Savard}}, \bibinfo {author} {\bibfnamefont {K.~S.}\ \bibnamefont {Sharma}}, \ and\ \bibinfo {author} {\bibfnamefont {K.}~\bibnamefont {Siegl}},\ }\href {\doibase https://doi.org/10.1016/j.nimb.2015.12.037} {\bibfield  {journal} {\bibinfo  {journal} {Nuclear Instruments and Methods in Physics Research Section B: Beam Interactions with Materials and Atoms}\ }\textbf {\bibinfo {volume} {376}},\ \bibinfo {pages} {229} (\bibinfo {year} {2016})},\ \bibinfo {note} {proceedings of the XVIIth International Conference on Electromagnetic Isotope Separators and Related Topics (EMIS2015), Grand Rapids, MI, U.S.A., 11-15 May 2015}\BibitemShut {NoStop}%
\bibitem [{\citenamefont {König}\ \emph {et~al.}(1995)\citenamefont {König}, \citenamefont {Bollen}, \citenamefont {Kluge}, \citenamefont {Otto},\ and\ \citenamefont {Szerypo}}]{KONIG199595}%
  \BibitemOpen
  \bibfield  {author} {\bibinfo {author} {\bibfnamefont {M.}~\bibnamefont {König}}, \bibinfo {author} {\bibfnamefont {G.}~\bibnamefont {Bollen}}, \bibinfo {author} {\bibfnamefont {H.-J.}\ \bibnamefont {Kluge}}, \bibinfo {author} {\bibfnamefont {T.}~\bibnamefont {Otto}}, \ and\ \bibinfo {author} {\bibfnamefont {J.}~\bibnamefont {Szerypo}},\ }\href {\doibase https://doi.org/10.1016/0168-1176(95)04146-C} {\bibfield  {journal} {\bibinfo  {journal} {International Journal of Mass Spectrometry and Ion Processes}\ }\textbf {\bibinfo {volume} {142}},\ \bibinfo {pages} {95} (\bibinfo {year} {1995})}\BibitemShut {NoStop}%
\bibitem [{\citenamefont {Bollen}\ \emph {et~al.}(1990)\citenamefont {Bollen}, \citenamefont {Moore}, \citenamefont {Savard},\ and\ \citenamefont {Stolzenberg}}]{Bollen:204719}%
  \BibitemOpen
  \bibfield  {author} {\bibinfo {author} {\bibfnamefont {G.}~\bibnamefont {Bollen}}, \bibinfo {author} {\bibfnamefont {R.~B.}\ \bibnamefont {Moore}}, \bibinfo {author} {\bibfnamefont {G.}~\bibnamefont {Savard}}, \ and\ \bibinfo {author} {\bibfnamefont {H.}~\bibnamefont {Stolzenberg}},\ }\href {\doibase 10.1063/1.346185} {\bibfield  {journal} {\bibinfo  {journal} {J. Appl. Phys.}\ }\textbf {\bibinfo {volume} {68}},\ \bibinfo {pages} {4355} (\bibinfo {year} {1990})}\BibitemShut {NoStop}%
\bibitem [{\citenamefont {Orford}\ \emph {et~al.}(2020{\natexlab{b}})\citenamefont {Orford}, \citenamefont {Clark}, \citenamefont {Savard}, \citenamefont {Aprahamian}, \citenamefont {Buchinger}, \citenamefont {Burkey}, \citenamefont {Gorelov}, \citenamefont {Klimes}, \citenamefont {Morgan}, \citenamefont {Nystrom}, \citenamefont {Porter}, \citenamefont {Ray},\ and\ \citenamefont {Sharma}}]{ORFORD2020491}%
  \BibitemOpen
  \bibfield  {author} {\bibinfo {author} {\bibfnamefont {R.}~\bibnamefont {Orford}}, \bibinfo {author} {\bibfnamefont {J.}~\bibnamefont {Clark}}, \bibinfo {author} {\bibfnamefont {G.}~\bibnamefont {Savard}}, \bibinfo {author} {\bibfnamefont {A.}~\bibnamefont {Aprahamian}}, \bibinfo {author} {\bibfnamefont {F.}~\bibnamefont {Buchinger}}, \bibinfo {author} {\bibfnamefont {M.}~\bibnamefont {Burkey}}, \bibinfo {author} {\bibfnamefont {D.}~\bibnamefont {Gorelov}}, \bibinfo {author} {\bibfnamefont {J.}~\bibnamefont {Klimes}}, \bibinfo {author} {\bibfnamefont {G.}~\bibnamefont {Morgan}}, \bibinfo {author} {\bibfnamefont {A.}~\bibnamefont {Nystrom}}, \bibinfo {author} {\bibfnamefont {W.}~\bibnamefont {Porter}}, \bibinfo {author} {\bibfnamefont {D.}~\bibnamefont {Ray}}, \ and\ \bibinfo {author} {\bibfnamefont {K.}~\bibnamefont {Sharma}},\ }\href {\doibase https://doi.org/10.1016/j.nimb.2019.04.016} {\bibfield  {journal} {\bibinfo  {journal} {Nuclear Instruments and Methods in Physics Research Section B: Beam
  Interactions with Materials and Atoms}\ }\textbf {\bibinfo {volume} {463}},\ \bibinfo {pages} {491} (\bibinfo {year} {2020}{\natexlab{b}})}\BibitemShut {NoStop}%
\bibitem [{\citenamefont {Wang}\ \emph {et~al.}(2021)\citenamefont {Wang}, \citenamefont {Huang}, \citenamefont {Kondev}, \citenamefont {Audi},\ and\ \citenamefont {Naimi}}]{AME2020}%
  \BibitemOpen
  \bibfield  {author} {\bibinfo {author} {\bibfnamefont {M.}~\bibnamefont {Wang}}, \bibinfo {author} {\bibfnamefont {W.}~\bibnamefont {Huang}}, \bibinfo {author} {\bibfnamefont {F.}~\bibnamefont {Kondev}}, \bibinfo {author} {\bibfnamefont {G.}~\bibnamefont {Audi}}, \ and\ \bibinfo {author} {\bibfnamefont {S.}~\bibnamefont {Naimi}},\ }\href {\doibase 10.1088/1674-1137/abddaf} {\bibfield  {journal} {\bibinfo  {journal} {Chinese Physics C}\ }\textbf {\bibinfo {volume} {45}},\ \bibinfo {pages} {030003} (\bibinfo {year} {2021})}\BibitemShut {NoStop}%
\bibitem [{SCM()}]{SCM}%
  \BibitemOpen
  \href@noop {} {}\bibinfo {note} {The software SMC can be downloaded here \url{https://groups.frib.msu.edu/lebit/downloads/index.html}}\BibitemShut {NoStop}%
\bibitem [{\citenamefont {Hager}\ \emph {et~al.}(2007)\citenamefont {Hager}, \citenamefont {Elomaa}, \citenamefont {Eronen}, \citenamefont {Hakala}, \citenamefont {Jokinen}, \citenamefont {Kankainen}, \citenamefont {Rahaman}, \citenamefont {Rinta-Antila}, \citenamefont {Saastamoinen}, \citenamefont {Sonoda},\ and\ \citenamefont {\"Ayst\"o}}]{07Ha20}%
  \BibitemOpen
  \bibfield  {author} {\bibinfo {author} {\bibfnamefont {U.}~\bibnamefont {Hager}}, \bibinfo {author} {\bibfnamefont {V.-V.}\ \bibnamefont {Elomaa}}, \bibinfo {author} {\bibfnamefont {T.}~\bibnamefont {Eronen}}, \bibinfo {author} {\bibfnamefont {J.}~\bibnamefont {Hakala}}, \bibinfo {author} {\bibfnamefont {A.}~\bibnamefont {Jokinen}}, \bibinfo {author} {\bibfnamefont {A.}~\bibnamefont {Kankainen}}, \bibinfo {author} {\bibfnamefont {S.}~\bibnamefont {Rahaman}}, \bibinfo {author} {\bibfnamefont {S.}~\bibnamefont {Rinta-Antila}}, \bibinfo {author} {\bibfnamefont {A.}~\bibnamefont {Saastamoinen}}, \bibinfo {author} {\bibfnamefont {T.}~\bibnamefont {Sonoda}}, \ and\ \bibinfo {author} {\bibfnamefont {J.}~\bibnamefont {\"Ayst\"o}},\ }\href {\doibase 10.1103/PhysRevC.75.064302} {\bibfield  {journal} {\bibinfo  {journal} {Phys. Rev. C}\ }\textbf {\bibinfo {volume} {75}},\ \bibinfo {pages} {064302} (\bibinfo {year} {2007})}\BibitemShut {NoStop}%
\bibitem [{\citenamefont {Kolhinen}(2003)}]{03KoA}%
  \BibitemOpen
  \bibfield  {author} {\bibinfo {author} {\bibfnamefont {V.}~\bibnamefont {Kolhinen}},\ }\href {http://urn.fi/URN:ISBN:978-951-39-3141-4} {\bibinfo {type} {Phd thesis}},\ \bibinfo  {school} {University of Jyväskylä} (\bibinfo {year} {2003})\BibitemShut {NoStop}%
\bibitem [{\citenamefont {Jokinen}\ \emph {et~al.}(1991)\citenamefont {Jokinen}, \citenamefont {{\"A}yst{\"o}}, \citenamefont {Dendooven}, \citenamefont {Eskola}, \citenamefont {Janas}, \citenamefont {Jauho}, \citenamefont {Leino}, \citenamefont {Parmonen}, \citenamefont {Penttil{\"a}}, \citenamefont {Rykaczewski},\ and\ \citenamefont {Taskinen}}]{91Jo11}%
  \BibitemOpen
  \bibfield  {author} {\bibinfo {author} {\bibfnamefont {A.}~\bibnamefont {Jokinen}}, \bibinfo {author} {\bibfnamefont {J.}~\bibnamefont {{\"A}yst{\"o}}}, \bibinfo {author} {\bibfnamefont {P.}~\bibnamefont {Dendooven}}, \bibinfo {author} {\bibfnamefont {K.}~\bibnamefont {Eskola}}, \bibinfo {author} {\bibfnamefont {Z.}~\bibnamefont {Janas}}, \bibinfo {author} {\bibfnamefont {P.~P.}\ \bibnamefont {Jauho}}, \bibinfo {author} {\bibfnamefont {M.~E.}\ \bibnamefont {Leino}}, \bibinfo {author} {\bibfnamefont {J.~M.}\ \bibnamefont {Parmonen}}, \bibinfo {author} {\bibfnamefont {H.}~\bibnamefont {Penttil{\"a}}}, \bibinfo {author} {\bibfnamefont {K.}~\bibnamefont {Rykaczewski}}, \ and\ \bibinfo {author} {\bibfnamefont {P.}~\bibnamefont {Taskinen}},\ }\href {\doibase 10.1007/BF01284476} {\bibfield  {journal} {\bibinfo  {journal} {Zeitschrift f{\"u}r Physik A Hadrons and Nuclei}\ }\textbf {\bibinfo {volume} {340}},\ \bibinfo {pages} {21} (\bibinfo {year} {1991})}\BibitemShut {NoStop}%
\bibitem [{\citenamefont {Kratz}\ and\ \citenamefont {Pfeiffer}(2000)}]{00KrA}%
  \BibitemOpen
  \bibfield  {author} {\bibinfo {author} {\bibfnamefont {K.-L.}\ \bibnamefont {Kratz}}\ and\ \bibinfo {author} {\bibfnamefont {B.}~\bibnamefont {Pfeiffer}},\ }\href@noop {} {\enquote {\bibinfo {title} {private communication},}\ } (\bibinfo {year} {2000})\BibitemShut {NoStop}%
\bibitem [{\citenamefont {Knöbel}(2008)}]{08KnA}%
  \BibitemOpen
  \bibfield  {author} {\bibinfo {author} {\bibfnamefont {R.}~\bibnamefont {Knöbel}},\ }\href@noop {} {\bibinfo {type} {Phd thesis}},\ \bibinfo  {school} {JLU Giessen} (\bibinfo {year} {2008})\BibitemShut {NoStop}%
\bibitem [{\citenamefont {Stryjczyk}\ \emph {et~al.}(2024)\citenamefont {Stryjczyk}, \citenamefont {Jaries}, \citenamefont {Ryssens}, \citenamefont {Bender}, \citenamefont {Kankainen}, \citenamefont {Eronen}, \citenamefont {Ge}, \citenamefont {Moore}, \citenamefont {Mougeot}, \citenamefont {Raggio},\ and\ \citenamefont {Ruotsalainen}}]{JYFLarxiv2}%
  \BibitemOpen
  \bibfield  {author} {\bibinfo {author} {\bibfnamefont {M.}~\bibnamefont {Stryjczyk}}, \bibinfo {author} {\bibfnamefont {A.}~\bibnamefont {Jaries}}, \bibinfo {author} {\bibfnamefont {W.}~\bibnamefont {Ryssens}}, \bibinfo {author} {\bibfnamefont {M.}~\bibnamefont {Bender}}, \bibinfo {author} {\bibfnamefont {A.}~\bibnamefont {Kankainen}}, \bibinfo {author} {\bibfnamefont {T.}~\bibnamefont {Eronen}}, \bibinfo {author} {\bibfnamefont {Z.}~\bibnamefont {Ge}}, \bibinfo {author} {\bibfnamefont {I.~D.}\ \bibnamefont {Moore}}, \bibinfo {author} {\bibfnamefont {M.}~\bibnamefont {Mougeot}}, \bibinfo {author} {\bibfnamefont {A.}~\bibnamefont {Raggio}}, \ and\ \bibinfo {author} {\bibfnamefont {J.}~\bibnamefont {Ruotsalainen}},\ }\href {https://arxiv.org/abs/2409.15016} {\enquote {\bibinfo {title} {Discovery of a new long-lived isomer in $^{114}$rh via penning-trap mass spectrometry},}\ } (\bibinfo {year} {2024}),\ \Eprint {http://arxiv.org/abs/2409.15016} {arXiv:2409.15016 [nucl-ex]} \BibitemShut {NoStop}%
\bibitem [{\citenamefont {Äystö}\ \emph {et~al.}(1988)\citenamefont {Äystö}, \citenamefont {Davids}, \citenamefont {Hattula}, \citenamefont {Honkanen}, \citenamefont {Honkanen}, \citenamefont {Jauho}, \citenamefont {Julin}, \citenamefont {Juutinen}, \citenamefont {Kumpulainen}, \citenamefont {Lönnroth}, \citenamefont {Pakkanen}, \citenamefont {Passoja}, \citenamefont {Penttilä}, \citenamefont {Taskinen}, \citenamefont {Verho}, \citenamefont {Virtanen},\ and\ \citenamefont {Yoshii}}]{1988Ay02}%
  \BibitemOpen
  \bibfield  {author} {\bibinfo {author} {\bibfnamefont {J.}~\bibnamefont {Äystö}}, \bibinfo {author} {\bibfnamefont {C.}~\bibnamefont {Davids}}, \bibinfo {author} {\bibfnamefont {J.}~\bibnamefont {Hattula}}, \bibinfo {author} {\bibfnamefont {J.}~\bibnamefont {Honkanen}}, \bibinfo {author} {\bibfnamefont {K.}~\bibnamefont {Honkanen}}, \bibinfo {author} {\bibfnamefont {P.}~\bibnamefont {Jauho}}, \bibinfo {author} {\bibfnamefont {R.}~\bibnamefont {Julin}}, \bibinfo {author} {\bibfnamefont {S.}~\bibnamefont {Juutinen}}, \bibinfo {author} {\bibfnamefont {J.}~\bibnamefont {Kumpulainen}}, \bibinfo {author} {\bibfnamefont {T.}~\bibnamefont {Lönnroth}}, \bibinfo {author} {\bibfnamefont {A.}~\bibnamefont {Pakkanen}}, \bibinfo {author} {\bibfnamefont {A.}~\bibnamefont {Passoja}}, \bibinfo {author} {\bibfnamefont {H.}~\bibnamefont {Penttilä}}, \bibinfo {author} {\bibfnamefont {P.}~\bibnamefont {Taskinen}}, \bibinfo {author} {\bibfnamefont {E.}~\bibnamefont {Verho}}, \bibinfo {author} {\bibfnamefont {A.}~\bibnamefont
  {Virtanen}}, \ and\ \bibinfo {author} {\bibfnamefont {M.}~\bibnamefont {Yoshii}},\ }\href {\doibase https://doi.org/10.1016/0375-9474(88)90387-9} {\bibfield  {journal} {\bibinfo  {journal} {Nuclear Physics A}\ }\textbf {\bibinfo {volume} {480}},\ \bibinfo {pages} {104} (\bibinfo {year} {1988})}\BibitemShut {NoStop}%
\bibitem [{\citenamefont {Lhersonneau}\ \emph {et~al.}(2003)\citenamefont {Lhersonneau}, \citenamefont {Wang}, \citenamefont {Capote}, \citenamefont {Suhonen}, \citenamefont {Dendooven}, \citenamefont {Huikari}, \citenamefont {Per\"aj\"arvi},\ and\ \citenamefont {Wang}}]{2003Lh01}%
  \BibitemOpen
  \bibfield  {author} {\bibinfo {author} {\bibfnamefont {G.}~\bibnamefont {Lhersonneau}}, \bibinfo {author} {\bibfnamefont {Y.}~\bibnamefont {Wang}}, \bibinfo {author} {\bibfnamefont {R.}~\bibnamefont {Capote}}, \bibinfo {author} {\bibfnamefont {J.}~\bibnamefont {Suhonen}}, \bibinfo {author} {\bibfnamefont {P.}~\bibnamefont {Dendooven}}, \bibinfo {author} {\bibfnamefont {J.}~\bibnamefont {Huikari}}, \bibinfo {author} {\bibfnamefont {K.}~\bibnamefont {Per\"aj\"arvi}}, \ and\ \bibinfo {author} {\bibfnamefont {J.~C.}\ \bibnamefont {Wang}},\ }\href {\doibase 10.1103/PhysRevC.67.024303} {\bibfield  {journal} {\bibinfo  {journal} {Phys. Rev. C}\ }\textbf {\bibinfo {volume} {67}},\ \bibinfo {pages} {024303} (\bibinfo {year} {2003})}\BibitemShut {NoStop}%
\bibitem [{\citenamefont {{The {\em Universal Woods-Saxon Hamiltonian} and the associated `universal parametrization' has been developed in a series of articles and continue on being used without modifications to this day. J. Dudek, T. Werner}}(1978)}]{WS01}%
  \BibitemOpen
  \bibfield  {author} {\bibinfo {author} {\bibnamefont {{The {\em Universal Woods-Saxon Hamiltonian} and the associated `universal parametrization' has been developed in a series of articles and continue on being used without modifications to this day. J. Dudek, T. Werner}}},\ }\href {\doibase 10.1088/0305-4616/4/10/006} {\bibfield  {journal} {\bibinfo  {journal} {Journal of Physics G: Nuclear Physics}\ }\textbf {\bibinfo {volume} {4}},\ \bibinfo {pages} {1543} (\bibinfo {year} {1978})}\BibitemShut {NoStop}%
\bibitem [{\citenamefont {Dudek}\ \emph {et~al.}(1979)\citenamefont {Dudek}, \citenamefont {Majhofer}, \citenamefont {Skalski}, \citenamefont {Werner}, \citenamefont {Cwiok},\ and\ \citenamefont {Nazarewicz}}]{WS02}%
  \BibitemOpen
  \bibfield  {author} {\bibinfo {author} {\bibfnamefont {J.}~\bibnamefont {Dudek}}, \bibinfo {author} {\bibfnamefont {A.}~\bibnamefont {Majhofer}}, \bibinfo {author} {\bibfnamefont {J.}~\bibnamefont {Skalski}}, \bibinfo {author} {\bibfnamefont {T.}~\bibnamefont {Werner}}, \bibinfo {author} {\bibfnamefont {S.}~\bibnamefont {Cwiok}}, \ and\ \bibinfo {author} {\bibfnamefont {W.}~\bibnamefont {Nazarewicz}},\ }\href {\doibase 10.1088/0305-4616/5/10/014} {\bibfield  {journal} {\bibinfo  {journal} {Journal of Physics G: Nuclear Physics}\ }\textbf {\bibinfo {volume} {5}},\ \bibinfo {pages} {1359} (\bibinfo {year} {1979})}\BibitemShut {NoStop}%
\bibitem [{\citenamefont {Dudek}\ \emph {et~al.}(1980)\citenamefont {Dudek}, \citenamefont {Nazarewicz},\ and\ \citenamefont {Werner}}]{WS03}%
  \BibitemOpen
  \bibfield  {author} {\bibinfo {author} {\bibfnamefont {J.}~\bibnamefont {Dudek}}, \bibinfo {author} {\bibfnamefont {W.}~\bibnamefont {Nazarewicz}}, \ and\ \bibinfo {author} {\bibfnamefont {T.}~\bibnamefont {Werner}},\ }\href {\doibase https://doi.org/10.1016/0375-9474(80)90312-7} {\bibfield  {journal} {\bibinfo  {journal} {Nuclear Physics A}\ }\textbf {\bibinfo {volume} {341}},\ \bibinfo {pages} {253} (\bibinfo {year} {1980})}\BibitemShut {NoStop}%
\bibitem [{\citenamefont {Dudek}\ \emph {et~al.}(1981)\citenamefont {Dudek}, \citenamefont {Szyma\ifmmode~\acute{n}\else \'{n}\fi{}ski},\ and\ \citenamefont {Werner}}]{WS04}%
  \BibitemOpen
  \bibfield  {author} {\bibinfo {author} {\bibfnamefont {J.}~\bibnamefont {Dudek}}, \bibinfo {author} {\bibfnamefont {Z.}~\bibnamefont {Szyma\ifmmode~\acute{n}\else \'{n}\fi{}ski}}, \ and\ \bibinfo {author} {\bibfnamefont {T.}~\bibnamefont {Werner}},\ }\href {\doibase 10.1103/PhysRevC.23.920} {\bibfield  {journal} {\bibinfo  {journal} {Phys. Rev. C}\ }\textbf {\bibinfo {volume} {23}},\ \bibinfo {pages} {920} (\bibinfo {year} {1981})}\BibitemShut {NoStop}%
\bibitem [{\citenamefont {\'Cwiok}\ \emph {et~al.}(1987)\citenamefont {\'Cwiok}, \citenamefont {Dudek}, \citenamefont {Nazarewicz}, \citenamefont {Skalski},\ and\ \citenamefont {Werner}}]{WS05}%
  \BibitemOpen
  \bibfield  {author} {\bibinfo {author} {\bibfnamefont {S.}~\bibnamefont {\'Cwiok}}, \bibinfo {author} {\bibfnamefont {J.}~\bibnamefont {Dudek}}, \bibinfo {author} {\bibfnamefont {W.}~\bibnamefont {Nazarewicz}}, \bibinfo {author} {\bibfnamefont {J.}~\bibnamefont {Skalski}}, \ and\ \bibinfo {author} {\bibfnamefont {T.}~\bibnamefont {Werner}},\ }\href {\doibase https://doi.org/10.1016/0010-4655(87)90093-2} {\bibfield  {journal} {\bibinfo  {journal} {Computer Physics Communications}\ }\textbf {\bibinfo {volume} {46}},\ \bibinfo {pages} {379} (\bibinfo {year} {1987})}\BibitemShut {NoStop}%
\bibitem [{\citenamefont {{To illustrate the frequent use of the `universal' Woods-Saxon phenomenological mean-field Hamiltonian we quote below the articles which appeared only in one year (2013) and only in one journal: D. S. Delion, and J. Suhonen}}(2013)}]{Y2013-01}%
  \BibitemOpen
  \bibfield  {author} {\bibinfo {author} {\bibnamefont {{To illustrate the frequent use of the `universal' Woods-Saxon phenomenological mean-field Hamiltonian we quote below the articles which appeared only in one year (2013) and only in one journal: D. S. Delion, and J. Suhonen}}},\ }\href {\doibase 10.1103/PhysRevC.87.024309} {\bibfield  {journal} {\bibinfo  {journal} {Phys. Rev. C}\ }\textbf {\bibinfo {volume} {87}},\ \bibinfo {pages} {024309} (\bibinfo {year} {2013})}\BibitemShut {NoStop}%
\bibitem [{\citenamefont {Rissanen}\ \emph {et~al.}(2013)\citenamefont {Rissanen}, \citenamefont {Clark}, \citenamefont {Gregorich}, \citenamefont {Gates}, \citenamefont {Campbell}, \citenamefont {Crawford}, \citenamefont {Cromaz}, \citenamefont {Esker}, \citenamefont {Fallon}, \citenamefont {Forsberg}, \citenamefont {Gothe}, \citenamefont {Lee}, \citenamefont {Liu}, \citenamefont {Machiavelli}, \citenamefont {Mudder}, \citenamefont {Nitsche}, \citenamefont {Pang}, \citenamefont {Rice}, \citenamefont {Rudolph}, \citenamefont {Stoyer}, \citenamefont {Wiens},\ and\ \citenamefont {Xu}}]{Y2013-02}%
  \BibitemOpen
  \bibfield  {author} {\bibinfo {author} {\bibfnamefont {J.}~\bibnamefont {Rissanen}}, \bibinfo {author} {\bibfnamefont {R.~M.}\ \bibnamefont {Clark}}, \bibinfo {author} {\bibfnamefont {K.~E.}\ \bibnamefont {Gregorich}}, \bibinfo {author} {\bibfnamefont {J.~M.}\ \bibnamefont {Gates}}, \bibinfo {author} {\bibfnamefont {C.~M.}\ \bibnamefont {Campbell}}, \bibinfo {author} {\bibfnamefont {H.~L.}\ \bibnamefont {Crawford}}, \bibinfo {author} {\bibfnamefont {M.}~\bibnamefont {Cromaz}}, \bibinfo {author} {\bibfnamefont {N.~E.}\ \bibnamefont {Esker}}, \bibinfo {author} {\bibfnamefont {P.}~\bibnamefont {Fallon}}, \bibinfo {author} {\bibfnamefont {U.}~\bibnamefont {Forsberg}}, \bibinfo {author} {\bibfnamefont {O.}~\bibnamefont {Gothe}}, \bibinfo {author} {\bibfnamefont {I.-Y.}\ \bibnamefont {Lee}}, \bibinfo {author} {\bibfnamefont {H.~L.}\ \bibnamefont {Liu}}, \bibinfo {author} {\bibfnamefont {A.~O.}\ \bibnamefont {Machiavelli}}, \bibinfo {author} {\bibfnamefont {P.}~\bibnamefont {Mudder}}, \bibinfo {author}
  {\bibfnamefont {H.}~\bibnamefont {Nitsche}}, \bibinfo {author} {\bibfnamefont {G.}~\bibnamefont {Pang}}, \bibinfo {author} {\bibfnamefont {A.}~\bibnamefont {Rice}}, \bibinfo {author} {\bibfnamefont {D.}~\bibnamefont {Rudolph}}, \bibinfo {author} {\bibfnamefont {M.~A.}\ \bibnamefont {Stoyer}}, \bibinfo {author} {\bibfnamefont {A.}~\bibnamefont {Wiens}}, \ and\ \bibinfo {author} {\bibfnamefont {F.~R.}\ \bibnamefont {Xu}},\ }\href {\doibase 10.1103/PhysRevC.88.044313} {\bibfield  {journal} {\bibinfo  {journal} {Phys. Rev. C}\ }\textbf {\bibinfo {volume} {88}},\ \bibinfo {pages} {044313} (\bibinfo {year} {2013})}\BibitemShut {NoStop}%
\bibitem [{\citenamefont {Brodzi\ifmmode~\acute{n}\else \'{n}\fi{}ski}\ and\ \citenamefont {Skalski}(2013)}]{Y2013-03}%
  \BibitemOpen
  \bibfield  {author} {\bibinfo {author} {\bibfnamefont {W.}~\bibnamefont {Brodzi\ifmmode~\acute{n}\else \'{n}\fi{}ski}}\ and\ \bibinfo {author} {\bibfnamefont {J.}~\bibnamefont {Skalski}},\ }\href {\doibase 10.1103/PhysRevC.88.044307} {\bibfield  {journal} {\bibinfo  {journal} {Phys. Rev. C}\ }\textbf {\bibinfo {volume} {88}},\ \bibinfo {pages} {044307} (\bibinfo {year} {2013})}\BibitemShut {NoStop}%
\bibitem [{\citenamefont {Lalkovski}\ \emph {et~al.}(2013)\citenamefont {Lalkovski}, \citenamefont {Bruce}, \citenamefont {Denis~Bacelar}, \citenamefont {G\'orska}, \citenamefont {Pietri}, \citenamefont {Podoly\'ak}, \citenamefont {Bednarczyk}, \citenamefont {Caceres}, \citenamefont {Casarejos}, \citenamefont {Cullen}, \citenamefont {Doornenbal}, \citenamefont {Farrelly}, \citenamefont {Garnsworthy}, \citenamefont {Geissel}, \citenamefont {Gelletly}, \citenamefont {Gerl}, \citenamefont {Grebosz}, \citenamefont {Hinke}, \citenamefont {Ilie}, \citenamefont {Ivanova}, \citenamefont {Jaworski}, \citenamefont {Kisyov}, \citenamefont {Kojouharov}, \citenamefont {Kurz}, \citenamefont {Minkov}, \citenamefont {Myalski}, \citenamefont {Palacz}, \citenamefont {Petkov}, \citenamefont {Prokopowicz}, \citenamefont {Regan}, \citenamefont {Schaffner}, \citenamefont {Steer}, \citenamefont {Tashenov}, \citenamefont {Walker},\ and\ \citenamefont {Wollersheim}}]{Y2013-04}%
  \BibitemOpen
  \bibfield  {author} {\bibinfo {author} {\bibfnamefont {S.}~\bibnamefont {Lalkovski}}, \bibinfo {author} {\bibfnamefont {A.~M.}\ \bibnamefont {Bruce}}, \bibinfo {author} {\bibfnamefont {A.~M.}\ \bibnamefont {Denis~Bacelar}}, \bibinfo {author} {\bibfnamefont {M.}~\bibnamefont {G\'orska}}, \bibinfo {author} {\bibfnamefont {S.}~\bibnamefont {Pietri}}, \bibinfo {author} {\bibfnamefont {Z.}~\bibnamefont {Podoly\'ak}}, \bibinfo {author} {\bibfnamefont {P.}~\bibnamefont {Bednarczyk}}, \bibinfo {author} {\bibfnamefont {L.}~\bibnamefont {Caceres}}, \bibinfo {author} {\bibfnamefont {E.}~\bibnamefont {Casarejos}}, \bibinfo {author} {\bibfnamefont {I.~J.}\ \bibnamefont {Cullen}}, \bibinfo {author} {\bibfnamefont {P.}~\bibnamefont {Doornenbal}}, \bibinfo {author} {\bibfnamefont {G.~F.}\ \bibnamefont {Farrelly}}, \bibinfo {author} {\bibfnamefont {A.~B.}\ \bibnamefont {Garnsworthy}}, \bibinfo {author} {\bibfnamefont {H.}~\bibnamefont {Geissel}}, \bibinfo {author} {\bibfnamefont {W.}~\bibnamefont {Gelletly}}, \bibinfo
  {author} {\bibfnamefont {J.}~\bibnamefont {Gerl}}, \bibinfo {author} {\bibfnamefont {J.}~\bibnamefont {Grebosz}}, \bibinfo {author} {\bibfnamefont {C.}~\bibnamefont {Hinke}}, \bibinfo {author} {\bibfnamefont {G.}~\bibnamefont {Ilie}}, \bibinfo {author} {\bibfnamefont {D.}~\bibnamefont {Ivanova}}, \bibinfo {author} {\bibfnamefont {G.}~\bibnamefont {Jaworski}}, \bibinfo {author} {\bibfnamefont {S.}~\bibnamefont {Kisyov}}, \bibinfo {author} {\bibfnamefont {I.}~\bibnamefont {Kojouharov}}, \bibinfo {author} {\bibfnamefont {N.}~\bibnamefont {Kurz}}, \bibinfo {author} {\bibfnamefont {N.}~\bibnamefont {Minkov}}, \bibinfo {author} {\bibfnamefont {S.}~\bibnamefont {Myalski}}, \bibinfo {author} {\bibfnamefont {M.}~\bibnamefont {Palacz}}, \bibinfo {author} {\bibfnamefont {P.}~\bibnamefont {Petkov}}, \bibinfo {author} {\bibfnamefont {W.}~\bibnamefont {Prokopowicz}}, \bibinfo {author} {\bibfnamefont {P.~H.}\ \bibnamefont {Regan}}, \bibinfo {author} {\bibfnamefont {H.}~\bibnamefont {Schaffner}}, \bibinfo {author}
  {\bibfnamefont {S.}~\bibnamefont {Steer}}, \bibinfo {author} {\bibfnamefont {S.}~\bibnamefont {Tashenov}}, \bibinfo {author} {\bibfnamefont {P.~M.}\ \bibnamefont {Walker}}, \ and\ \bibinfo {author} {\bibfnamefont {H.~J.}\ \bibnamefont {Wollersheim}},\ }\href {\doibase 10.1103/PhysRevC.88.024302} {\bibfield  {journal} {\bibinfo  {journal} {Phys. Rev. C}\ }\textbf {\bibinfo {volume} {88}},\ \bibinfo {pages} {024302} (\bibinfo {year} {2013})}\BibitemShut {NoStop}%
\bibitem [{\citenamefont {Liu}\ and\ \citenamefont {Xu}(2013)}]{Y2013-05}%
  \BibitemOpen
  \bibfield  {author} {\bibinfo {author} {\bibfnamefont {H.~L.}\ \bibnamefont {Liu}}\ and\ \bibinfo {author} {\bibfnamefont {F.~R.}\ \bibnamefont {Xu}},\ }\href {\doibase 10.1103/PhysRevC.87.067304} {\bibfield  {journal} {\bibinfo  {journal} {Phys. Rev. C}\ }\textbf {\bibinfo {volume} {87}},\ \bibinfo {pages} {067304} (\bibinfo {year} {2013})}\BibitemShut {NoStop}%
\bibitem [{\citenamefont {Delion}\ and\ \citenamefont {Liotta}(2013)}]{Y2013-06}%
  \BibitemOpen
  \bibfield  {author} {\bibinfo {author} {\bibfnamefont {D.~S.}\ \bibnamefont {Delion}}\ and\ \bibinfo {author} {\bibfnamefont {R.~J.}\ \bibnamefont {Liotta}},\ }\href {\doibase 10.1103/PhysRevC.87.041302} {\bibfield  {journal} {\bibinfo  {journal} {Phys. Rev. C}\ }\textbf {\bibinfo {volume} {87}},\ \bibinfo {pages} {041302} (\bibinfo {year} {2013})}\BibitemShut {NoStop}%
\bibitem [{\citenamefont {Delion}\ \emph {et~al.}(2013)\citenamefont {Delion}, \citenamefont {Liotta},\ and\ \citenamefont {Wyss}}]{Y2013-07}%
  \BibitemOpen
  \bibfield  {author} {\bibinfo {author} {\bibfnamefont {D.~S.}\ \bibnamefont {Delion}}, \bibinfo {author} {\bibfnamefont {R.~J.}\ \bibnamefont {Liotta}}, \ and\ \bibinfo {author} {\bibfnamefont {R.}~\bibnamefont {Wyss}},\ }\href {\doibase 10.1103/PhysRevC.87.034328} {\bibfield  {journal} {\bibinfo  {journal} {Phys. Rev. C}\ }\textbf {\bibinfo {volume} {87}},\ \bibinfo {pages} {034328} (\bibinfo {year} {2013})}\BibitemShut {NoStop}%
\bibitem [{\citenamefont {Deleanu}\ \emph {et~al.}(2013)\citenamefont {Deleanu}, \citenamefont {Balabanski}, \citenamefont {Venkova}, \citenamefont {Bucurescu}, \citenamefont {Marginean}, \citenamefont {Ganioglu}, \citenamefont {Cata-Danil}, \citenamefont {Atanasova}, \citenamefont {Cata-Danil}, \citenamefont {Detistov}, \citenamefont {Filipescu}, \citenamefont {Ghita}, \citenamefont {Glodariu}, \citenamefont {Ivascu}, \citenamefont {Marginean}, \citenamefont {Mihai}, \citenamefont {Negret}, \citenamefont {Pascu}, \citenamefont {Sava}, \citenamefont {Stroe}, \citenamefont {Suliman},\ and\ \citenamefont {Zamfir}}]{Y2013-08}%
  \BibitemOpen
  \bibfield  {author} {\bibinfo {author} {\bibfnamefont {D.}~\bibnamefont {Deleanu}}, \bibinfo {author} {\bibfnamefont {D.~L.}\ \bibnamefont {Balabanski}}, \bibinfo {author} {\bibfnamefont {T.}~\bibnamefont {Venkova}}, \bibinfo {author} {\bibfnamefont {D.}~\bibnamefont {Bucurescu}}, \bibinfo {author} {\bibfnamefont {N.}~\bibnamefont {Marginean}}, \bibinfo {author} {\bibfnamefont {E.}~\bibnamefont {Ganioglu}}, \bibinfo {author} {\bibfnamefont {G.}~\bibnamefont {Cata-Danil}}, \bibinfo {author} {\bibfnamefont {L.}~\bibnamefont {Atanasova}}, \bibinfo {author} {\bibfnamefont {I.}~\bibnamefont {Cata-Danil}}, \bibinfo {author} {\bibfnamefont {P.}~\bibnamefont {Detistov}}, \bibinfo {author} {\bibfnamefont {D.}~\bibnamefont {Filipescu}}, \bibinfo {author} {\bibfnamefont {D.}~\bibnamefont {Ghita}}, \bibinfo {author} {\bibfnamefont {T.}~\bibnamefont {Glodariu}}, \bibinfo {author} {\bibfnamefont {M.}~\bibnamefont {Ivascu}}, \bibinfo {author} {\bibfnamefont {R.}~\bibnamefont {Marginean}}, \bibinfo {author} {\bibfnamefont
  {C.}~\bibnamefont {Mihai}}, \bibinfo {author} {\bibfnamefont {A.}~\bibnamefont {Negret}}, \bibinfo {author} {\bibfnamefont {S.}~\bibnamefont {Pascu}}, \bibinfo {author} {\bibfnamefont {T.}~\bibnamefont {Sava}}, \bibinfo {author} {\bibfnamefont {L.}~\bibnamefont {Stroe}}, \bibinfo {author} {\bibfnamefont {G.}~\bibnamefont {Suliman}}, \ and\ \bibinfo {author} {\bibfnamefont {N.~V.}\ \bibnamefont {Zamfir}},\ }\href {\doibase 10.1103/PhysRevC.87.014329} {\bibfield  {journal} {\bibinfo  {journal} {Phys. Rev. C}\ }\textbf {\bibinfo {volume} {87}},\ \bibinfo {pages} {014329} (\bibinfo {year} {2013})}\BibitemShut {NoStop}%
\bibitem [{\citenamefont {Dedes}\ and\ \citenamefont {Dudek}(2019)}]{Ded19}%
  \BibitemOpen
  \bibfield  {author} {\bibinfo {author} {\bibfnamefont {I.}~\bibnamefont {Dedes}}\ and\ \bibinfo {author} {\bibfnamefont {J.}~\bibnamefont {Dudek}},\ }\href {\doibase 10.1103/PhysRevC.99.054310} {\bibfield  {journal} {\bibinfo  {journal} {Phys. Rev. C}\ }\textbf {\bibinfo {volume} {99}},\ \bibinfo {pages} {054310} (\bibinfo {year} {2019})}\BibitemShut {NoStop}%
\bibitem [{\citenamefont {Gaamouci}\ \emph {et~al.}(2021)\citenamefont {Gaamouci}, \citenamefont {Dedes}, \citenamefont {Dudek}, \citenamefont {Baran}, \citenamefont {Benhamouda}, \citenamefont {Curien}, \citenamefont {Wang},\ and\ \citenamefont {Yang}}]{Gaa2021}%
  \BibitemOpen
  \bibfield  {author} {\bibinfo {author} {\bibfnamefont {A.}~\bibnamefont {Gaamouci}}, \bibinfo {author} {\bibfnamefont {I.}~\bibnamefont {Dedes}}, \bibinfo {author} {\bibfnamefont {J.}~\bibnamefont {Dudek}}, \bibinfo {author} {\bibfnamefont {A.}~\bibnamefont {Baran}}, \bibinfo {author} {\bibfnamefont {N.}~\bibnamefont {Benhamouda}}, \bibinfo {author} {\bibfnamefont {D.}~\bibnamefont {Curien}}, \bibinfo {author} {\bibfnamefont {H.~L.}\ \bibnamefont {Wang}}, \ and\ \bibinfo {author} {\bibfnamefont {J.}~\bibnamefont {Yang}},\ }\href {\doibase 10.1103/PhysRevC.103.054311} {\bibfield  {journal} {\bibinfo  {journal} {Phys. Rev. C}\ }\textbf {\bibinfo {volume} {103}},\ \bibinfo {pages} {054311} (\bibinfo {year} {2021})}\BibitemShut {NoStop}%
\bibitem [{\citenamefont {Dedes~Nonell}(2017)}]{PhDedes17}%
  \BibitemOpen
  \bibfield  {author} {\bibinfo {author} {\bibfnamefont {I.}~\bibnamefont {Dedes~Nonell}},\ }\emph {\bibinfo {title} {{Stochastic approach to the problem of predictive power in the theoretical modeling of the mean-field}}},\ \href {https://theses.hal.science/tel-01724641} {\bibinfo {type} {Phd thesis}},\ \bibinfo  {school} {{Universit{\'e} de Strasbourg}} (\bibinfo {year} {2017})\BibitemShut {NoStop}%
\bibitem [{\citenamefont {Hornung}\ \emph {et~al.}(2020)\citenamefont {Hornung}, \citenamefont {Amanbayev}, \citenamefont {Dedes}, \citenamefont {Kripko-Koncz}, \citenamefont {Miskun}, \citenamefont {Shimizu}, \citenamefont {{Ayet San Andres}}, \citenamefont {Bergmann}, \citenamefont {Dickel}, \citenamefont {Dudek}, \citenamefont {Ebert}, \citenamefont {Geissel}, \citenamefont {Gorska}, \citenamefont {Grawe}, \citenamefont {Greiner}, \citenamefont {Haettner}, \citenamefont {Otsuka}, \citenamefont {Plass}, \citenamefont {Purushothaman}, \citenamefont {Rink}, \citenamefont {Scheidenberger}, \citenamefont {Weick}, \citenamefont {Bagchi}, \citenamefont {Blazhev}, \citenamefont {Charviakova}, \citenamefont {Curien}, \citenamefont {Finlay}, \citenamefont {Kaur}, \citenamefont {Lippert}, \citenamefont {Otto}, \citenamefont {Patyk}, \citenamefont {Pietri}, \citenamefont {Tanaka}, \citenamefont {Tsunoda},\ and\ \citenamefont {Winfield}}]{Hornung2020}%
  \BibitemOpen
  \bibfield  {author} {\bibinfo {author} {\bibfnamefont {C.}~\bibnamefont {Hornung}}, \bibinfo {author} {\bibfnamefont {D.}~\bibnamefont {Amanbayev}}, \bibinfo {author} {\bibfnamefont {I.}~\bibnamefont {Dedes}}, \bibinfo {author} {\bibfnamefont {G.}~\bibnamefont {Kripko-Koncz}}, \bibinfo {author} {\bibfnamefont {I.}~\bibnamefont {Miskun}}, \bibinfo {author} {\bibfnamefont {N.}~\bibnamefont {Shimizu}}, \bibinfo {author} {\bibfnamefont {S.}~\bibnamefont {{Ayet San Andres}}}, \bibinfo {author} {\bibfnamefont {J.}~\bibnamefont {Bergmann}}, \bibinfo {author} {\bibfnamefont {T.}~\bibnamefont {Dickel}}, \bibinfo {author} {\bibfnamefont {J.}~\bibnamefont {Dudek}}, \bibinfo {author} {\bibfnamefont {J.}~\bibnamefont {Ebert}}, \bibinfo {author} {\bibfnamefont {H.}~\bibnamefont {Geissel}}, \bibinfo {author} {\bibfnamefont {M.}~\bibnamefont {Gorska}}, \bibinfo {author} {\bibfnamefont {H.}~\bibnamefont {Grawe}}, \bibinfo {author} {\bibfnamefont {F.}~\bibnamefont {Greiner}}, \bibinfo {author} {\bibfnamefont
  {E.}~\bibnamefont {Haettner}}, \bibinfo {author} {\bibfnamefont {T.}~\bibnamefont {Otsuka}}, \bibinfo {author} {\bibfnamefont {W.~R.}\ \bibnamefont {Plass}}, \bibinfo {author} {\bibfnamefont {S.}~\bibnamefont {Purushothaman}}, \bibinfo {author} {\bibfnamefont {A.-K.}\ \bibnamefont {Rink}}, \bibinfo {author} {\bibfnamefont {C.}~\bibnamefont {Scheidenberger}}, \bibinfo {author} {\bibfnamefont {H.}~\bibnamefont {Weick}}, \bibinfo {author} {\bibfnamefont {S.}~\bibnamefont {Bagchi}}, \bibinfo {author} {\bibfnamefont {A.}~\bibnamefont {Blazhev}}, \bibinfo {author} {\bibfnamefont {O.}~\bibnamefont {Charviakova}}, \bibinfo {author} {\bibfnamefont {D.}~\bibnamefont {Curien}}, \bibinfo {author} {\bibfnamefont {A.}~\bibnamefont {Finlay}}, \bibinfo {author} {\bibfnamefont {S.}~\bibnamefont {Kaur}}, \bibinfo {author} {\bibfnamefont {W.}~\bibnamefont {Lippert}}, \bibinfo {author} {\bibfnamefont {J.-H.}\ \bibnamefont {Otto}}, \bibinfo {author} {\bibfnamefont {Z.}~\bibnamefont {Patyk}}, \bibinfo {author} {\bibfnamefont
  {S.}~\bibnamefont {Pietri}}, \bibinfo {author} {\bibfnamefont {Y.~K.}\ \bibnamefont {Tanaka}}, \bibinfo {author} {\bibfnamefont {Y.}~\bibnamefont {Tsunoda}}, \ and\ \bibinfo {author} {\bibfnamefont {J.~S.}\ \bibnamefont {Winfield}},\ }\href {\doibase https://doi.org/10.1016/j.physletb.2020.135200} {\bibfield  {journal} {\bibinfo  {journal} {Physics Letters B}\ }\textbf {\bibinfo {volume} {802}},\ \bibinfo {pages} {135200} (\bibinfo {year} {2020})}\BibitemShut {NoStop}%
\bibitem [{\citenamefont {Beck}\ \emph {et~al.}(1984)\citenamefont {Beck}, \citenamefont {Dudek}, \citenamefont {Haas}, \citenamefont {Merdinger}, \citenamefont {Nourreddine}, \citenamefont {Schutz}, \citenamefont {Vivien}, \citenamefont {Hubert}, \citenamefont {Dassi{\'e}}, \citenamefont {Bastin}, \citenamefont {Nguyen}, \citenamefont {Thibaud},\ and\ \citenamefont {Nazarewicz}}]{Beck1984}%
  \BibitemOpen
  \bibfield  {author} {\bibinfo {author} {\bibfnamefont {F.~A.}\ \bibnamefont {Beck}}, \bibinfo {author} {\bibfnamefont {J.}~\bibnamefont {Dudek}}, \bibinfo {author} {\bibfnamefont {B.}~\bibnamefont {Haas}}, \bibinfo {author} {\bibfnamefont {J.~C.}\ \bibnamefont {Merdinger}}, \bibinfo {author} {\bibfnamefont {A.}~\bibnamefont {Nourreddine}}, \bibinfo {author} {\bibfnamefont {Y.}~\bibnamefont {Schutz}}, \bibinfo {author} {\bibfnamefont {J.~P.}\ \bibnamefont {Vivien}}, \bibinfo {author} {\bibfnamefont {P.}~\bibnamefont {Hubert}}, \bibinfo {author} {\bibfnamefont {D.}~\bibnamefont {Dassi{\'e}}}, \bibinfo {author} {\bibfnamefont {G.}~\bibnamefont {Bastin}}, \bibinfo {author} {\bibfnamefont {L.}~\bibnamefont {Nguyen}}, \bibinfo {author} {\bibfnamefont {J.~P.}\ \bibnamefont {Thibaud}}, \ and\ \bibinfo {author} {\bibfnamefont {W.}~\bibnamefont {Nazarewicz}},\ }\href {\doibase 10.1007/BF01415624} {\bibfield  {journal} {\bibinfo  {journal} {Zeitschrift f{\"u}r Physik A Atoms and Nuclei}\ }\textbf {\bibinfo {volume}
  {319}},\ \bibinfo {pages} {119} (\bibinfo {year} {1984})}\BibitemShut {NoStop}%
\bibitem [{\citenamefont {de~Voigt}\ \emph {et~al.}(1983)\citenamefont {de~Voigt}, \citenamefont {Dudek},\ and\ \citenamefont {Szyma\ifmmode~\acute{n}\else \'{n}\fi{}ski}}]{deVoigt1983}%
  \BibitemOpen
  \bibfield  {author} {\bibinfo {author} {\bibfnamefont {M.~J.~A.}\ \bibnamefont {de~Voigt}}, \bibinfo {author} {\bibfnamefont {J.}~\bibnamefont {Dudek}}, \ and\ \bibinfo {author} {\bibfnamefont {Z.}~\bibnamefont {Szyma\ifmmode~\acute{n}\else \'{n}\fi{}ski}},\ }\href {\doibase 10.1103/RevModPhys.55.949} {\bibfield  {journal} {\bibinfo  {journal} {Rev. Mod. Phys.}\ }\textbf {\bibinfo {volume} {55}},\ \bibinfo {pages} {949} (\bibinfo {year} {1983})}\BibitemShut {NoStop}%
\bibitem [{\citenamefont {Dudek}\ \emph {et~al.}(2012)\citenamefont {Dudek}, \citenamefont {Szpak}, \citenamefont {Dromard}, \citenamefont {Porquet}, \citenamefont {Fornal},\ and\ \citenamefont {Gozdz}}]{Dudek2012}%
  \BibitemOpen
  \bibfield  {author} {\bibinfo {author} {\bibfnamefont {J.}~\bibnamefont {Dudek}}, \bibinfo {author} {\bibfnamefont {B.}~\bibnamefont {Szpak}}, \bibinfo {author} {\bibfnamefont {A.}~\bibnamefont {Dromard}}, \bibinfo {author} {\bibfnamefont {M.-G.}\ \bibnamefont {Porquet}}, \bibinfo {author} {\bibfnamefont {B.}~\bibnamefont {Fornal}}, \ and\ \bibinfo {author} {\bibfnamefont {A.}~\bibnamefont {Gozdz}},\ }\href {\doibase 10.1142/S021830131250053X} {\bibfield  {journal} {\bibinfo  {journal} {International Journal of Modern Physics E}\ }\textbf {\bibinfo {volume} {21}},\ \bibinfo {pages} {1250053} (\bibinfo {year} {2012})},\ \Eprint {http://arxiv.org/abs/https://doi.org/10.1142/S021830131250053X} {https://doi.org/10.1142/S021830131250053X} \BibitemShut {NoStop}%
\bibitem [{\citenamefont {Dobaczewski}\ and\ \citenamefont {Dudek}(1997{\natexlab{a}})}]{Dobaczewski1997-1}%
  \BibitemOpen
  \bibfield  {author} {\bibinfo {author} {\bibfnamefont {J.}~\bibnamefont {Dobaczewski}}\ and\ \bibinfo {author} {\bibfnamefont {J.}~\bibnamefont {Dudek}},\ }\href {\doibase https://doi.org/10.1016/S0010-4655(97)00004-0} {\bibfield  {journal} {\bibinfo  {journal} {Computer Physics Communications}\ }\textbf {\bibinfo {volume} {102}},\ \bibinfo {pages} {166} (\bibinfo {year} {1997}{\natexlab{a}})}\BibitemShut {NoStop}%
\bibitem [{\citenamefont {Dobaczewski}\ and\ \citenamefont {Dudek}(1997{\natexlab{b}})}]{Dobaczewski1997-2}%
  \BibitemOpen
  \bibfield  {author} {\bibinfo {author} {\bibfnamefont {J.}~\bibnamefont {Dobaczewski}}\ and\ \bibinfo {author} {\bibfnamefont {J.}~\bibnamefont {Dudek}},\ }\href {\doibase https://doi.org/10.1016/S0010-4655(97)00005-2} {\bibfield  {journal} {\bibinfo  {journal} {Computer Physics Communications}\ }\textbf {\bibinfo {volume} {102}},\ \bibinfo {pages} {183} (\bibinfo {year} {1997}{\natexlab{b}})}\BibitemShut {NoStop}%
\bibitem [{\citenamefont {Dobaczewski}\ and\ \citenamefont {Dudek}(2000)}]{Dobaczewski1997-3}%
  \BibitemOpen
  \bibfield  {author} {\bibinfo {author} {\bibfnamefont {J.}~\bibnamefont {Dobaczewski}}\ and\ \bibinfo {author} {\bibfnamefont {J.}~\bibnamefont {Dudek}},\ }\href {\doibase https://doi.org/10.1016/S0010-4655(00)00121-1} {\bibfield  {journal} {\bibinfo  {journal} {Computer Physics Communications}\ }\textbf {\bibinfo {volume} {131}},\ \bibinfo {pages} {164} (\bibinfo {year} {2000})}\BibitemShut {NoStop}%
\bibitem [{\citenamefont {Schunck}\ \emph {et~al.}(2017)\citenamefont {Schunck}, \citenamefont {Dobaczewski}, \citenamefont {Satuła}, \citenamefont {Baczyk}, \citenamefont {Dudek}, \citenamefont {Gao}, \citenamefont {Konieczka}, \citenamefont {Sato}, \citenamefont {Shi}, \citenamefont {Wang},\ and\ \citenamefont {Werner}}]{Schunck2017}%
  \BibitemOpen
  \bibfield  {author} {\bibinfo {author} {\bibfnamefont {N.}~\bibnamefont {Schunck}}, \bibinfo {author} {\bibfnamefont {J.}~\bibnamefont {Dobaczewski}}, \bibinfo {author} {\bibfnamefont {W.}~\bibnamefont {Satuła}}, \bibinfo {author} {\bibfnamefont {P.}~\bibnamefont {Baczyk}}, \bibinfo {author} {\bibfnamefont {J.}~\bibnamefont {Dudek}}, \bibinfo {author} {\bibfnamefont {Y.}~\bibnamefont {Gao}}, \bibinfo {author} {\bibfnamefont {M.}~\bibnamefont {Konieczka}}, \bibinfo {author} {\bibfnamefont {K.}~\bibnamefont {Sato}}, \bibinfo {author} {\bibfnamefont {Y.}~\bibnamefont {Shi}}, \bibinfo {author} {\bibfnamefont {X.}~\bibnamefont {Wang}}, \ and\ \bibinfo {author} {\bibfnamefont {T.}~\bibnamefont {Werner}},\ }\href {\doibase https://doi.org/10.1016/j.cpc.2017.03.007} {\bibfield  {journal} {\bibinfo  {journal} {Computer Physics Communications}\ }\textbf {\bibinfo {volume} {216}},\ \bibinfo {pages} {145} (\bibinfo {year} {2017})}\BibitemShut {NoStop}%
\bibitem [{\citenamefont {M\"oller}\ \emph {et~al.}(2016)\citenamefont {M\"oller}, \citenamefont {Sierk}, \citenamefont {Ichikawa},\ and\ \citenamefont {Sagawa}}]{2016Mo08}%
  \BibitemOpen
  \bibfield  {author} {\bibinfo {author} {\bibfnamefont {P.}~\bibnamefont {M\"oller}}, \bibinfo {author} {\bibfnamefont {A.}~\bibnamefont {Sierk}}, \bibinfo {author} {\bibfnamefont {T.}~\bibnamefont {Ichikawa}}, \ and\ \bibinfo {author} {\bibfnamefont {H.}~\bibnamefont {Sagawa}},\ }\href {\doibase https://doi.org/10.1016/j.adt.2015.10.002} {\bibfield  {journal} {\bibinfo  {journal} {Atomic Data and Nuclear Data Tables}\ }\textbf {\bibinfo {volume} {109-110}},\ \bibinfo {pages} {1} (\bibinfo {year} {2016})}\BibitemShut {NoStop}%
\bibitem [{\citenamefont {Nazarewicz}\ \emph {et~al.}(1985)\citenamefont {Nazarewicz}, \citenamefont {Dudek}, \citenamefont {Bengtsson}, \citenamefont {Bengtsson},\ and\ \citenamefont {Ragnarsson}}]{1985Na}%
  \BibitemOpen
  \bibfield  {author} {\bibinfo {author} {\bibfnamefont {W.}~\bibnamefont {Nazarewicz}}, \bibinfo {author} {\bibfnamefont {J.}~\bibnamefont {Dudek}}, \bibinfo {author} {\bibfnamefont {R.}~\bibnamefont {Bengtsson}}, \bibinfo {author} {\bibfnamefont {T.}~\bibnamefont {Bengtsson}}, \ and\ \bibinfo {author} {\bibfnamefont {I.}~\bibnamefont {Ragnarsson}},\ }\href {\doibase https://doi.org/10.1016/0375-9474(85)90471-3} {\bibfield  {journal} {\bibinfo  {journal} {Nuclear Physics A}\ }\textbf {\bibinfo {volume} {435}},\ \bibinfo {pages} {397} (\bibinfo {year} {1985})}\BibitemShut {NoStop}%
\bibitem [{\citenamefont {Gallagher}\ and\ \citenamefont {Moszkowski}(1958)}]{GM58}%
  \BibitemOpen
  \bibfield  {author} {\bibinfo {author} {\bibfnamefont {C.~J.}\ \bibnamefont {Gallagher}}\ and\ \bibinfo {author} {\bibfnamefont {S.~A.}\ \bibnamefont {Moszkowski}},\ }\href {\doibase 10.1103/PhysRev.111.1282} {\bibfield  {journal} {\bibinfo  {journal} {Phys. Rev.}\ }\textbf {\bibinfo {volume} {111}},\ \bibinfo {pages} {1282} (\bibinfo {year} {1958})}\BibitemShut {NoStop}%
\bibitem [{ENS()}]{ENSDF}%
  \BibitemOpen
  \href@noop {} {}\bibinfo {note} {From ENSDF database as of April 9 2024, Version available at \url{http://www.nndc.bnl.gov}}\BibitemShut {NoStop}%
\bibitem [{\citenamefont {Kurpeta}\ \emph {et~al.}(2007)\citenamefont {Kurpeta}, \citenamefont {Urban}, \citenamefont {Droste}, \citenamefont {P{\l}ochocki}, \citenamefont {Rohozi{\'{n}}ski}, \citenamefont {Rz\c{a}ca-Urban}, \citenamefont {Morek}, \citenamefont {Pr{\'o}chniak}, \citenamefont {Starosta}, \citenamefont {{\"A}yst{\"o}}, \citenamefont {Penttil{\"a}}, \citenamefont {Durell}, \citenamefont {Smith}, \citenamefont {Lhersonneau},\ and\ \citenamefont {Ahmad}}]{07Ku23}%
  \BibitemOpen
  \bibfield  {author} {\bibinfo {author} {\bibfnamefont {J.}~\bibnamefont {Kurpeta}}, \bibinfo {author} {\bibfnamefont {W.}~\bibnamefont {Urban}}, \bibinfo {author} {\bibfnamefont {C.}~\bibnamefont {Droste}}, \bibinfo {author} {\bibfnamefont {A.}~\bibnamefont {P{\l}ochocki}}, \bibinfo {author} {\bibfnamefont {S.~G.}\ \bibnamefont {Rohozi{\'{n}}ski}}, \bibinfo {author} {\bibfnamefont {T.}~\bibnamefont {Rz\c{a}ca-Urban}}, \bibinfo {author} {\bibfnamefont {T.}~\bibnamefont {Morek}}, \bibinfo {author} {\bibfnamefont {L.}~\bibnamefont {Pr{\'o}chniak}}, \bibinfo {author} {\bibfnamefont {K.}~\bibnamefont {Starosta}}, \bibinfo {author} {\bibfnamefont {J.}~\bibnamefont {{\"A}yst{\"o}}}, \bibinfo {author} {\bibfnamefont {H.}~\bibnamefont {Penttil{\"a}}}, \bibinfo {author} {\bibfnamefont {J.~L.}\ \bibnamefont {Durell}}, \bibinfo {author} {\bibfnamefont {A.~G.}\ \bibnamefont {Smith}}, \bibinfo {author} {\bibfnamefont {G.}~\bibnamefont {Lhersonneau}}, \ and\ \bibinfo {author} {\bibfnamefont {I.}~\bibnamefont {Ahmad}},\
  }\href {\doibase 10.1140/epja/i2006-10464-2} {\bibfield  {journal} {\bibinfo  {journal} {The European Physical Journal A}\ }\textbf {\bibinfo {volume} {33}},\ \bibinfo {pages} {307} (\bibinfo {year} {2007})}\BibitemShut {NoStop}%
\bibitem [{\citenamefont {Kurpeta}\ \emph {et~al.}(2010)\citenamefont {Kurpeta}, \citenamefont {Rissanen}, \citenamefont {P\l{}ochocki}, \citenamefont {Urban}, \citenamefont {Elomaa}, \citenamefont {Eronen}, \citenamefont {Hakala}, \citenamefont {Jokinen}, \citenamefont {Kankainen}, \citenamefont {Karvonen}, \citenamefont {Ma\l{}kiewicz}, \citenamefont {Moore}, \citenamefont {Penttil\"a}, \citenamefont {Saastamoinen}, \citenamefont {Simpson}, \citenamefont {Weber},\ and\ \citenamefont {\"Ayst\"o}}]{10Ku25}%
  \BibitemOpen
  \bibfield  {author} {\bibinfo {author} {\bibfnamefont {J.}~\bibnamefont {Kurpeta}}, \bibinfo {author} {\bibfnamefont {J.}~\bibnamefont {Rissanen}}, \bibinfo {author} {\bibfnamefont {A.}~\bibnamefont {P\l{}ochocki}}, \bibinfo {author} {\bibfnamefont {W.}~\bibnamefont {Urban}}, \bibinfo {author} {\bibfnamefont {V.-V.}\ \bibnamefont {Elomaa}}, \bibinfo {author} {\bibfnamefont {T.}~\bibnamefont {Eronen}}, \bibinfo {author} {\bibfnamefont {J.}~\bibnamefont {Hakala}}, \bibinfo {author} {\bibfnamefont {A.}~\bibnamefont {Jokinen}}, \bibinfo {author} {\bibfnamefont {A.}~\bibnamefont {Kankainen}}, \bibinfo {author} {\bibfnamefont {P.}~\bibnamefont {Karvonen}}, \bibinfo {author} {\bibfnamefont {T.}~\bibnamefont {Ma\l{}kiewicz}}, \bibinfo {author} {\bibfnamefont {I.~D.}\ \bibnamefont {Moore}}, \bibinfo {author} {\bibfnamefont {H.}~\bibnamefont {Penttil\"a}}, \bibinfo {author} {\bibfnamefont {A.}~\bibnamefont {Saastamoinen}}, \bibinfo {author} {\bibfnamefont {G.~S.}\ \bibnamefont {Simpson}}, \bibinfo {author}
  {\bibfnamefont {C.}~\bibnamefont {Weber}}, \ and\ \bibinfo {author} {\bibfnamefont {J.}~\bibnamefont {\"Ayst\"o}},\ }\href {\doibase 10.1103/PhysRevC.82.064318} {\bibfield  {journal} {\bibinfo  {journal} {Phys. Rev. C}\ }\textbf {\bibinfo {volume} {82}},\ \bibinfo {pages} {064318} (\bibinfo {year} {2010})}\BibitemShut {NoStop}%
\bibitem [{\citenamefont {Beck}\ \emph {et~al.}(2021)\citenamefont {Beck}, \citenamefont {Kootte}, \citenamefont {Dedes}, \citenamefont {Dickel}, \citenamefont {Kwiatkowski}, \citenamefont {Lykiardopoulou}, \citenamefont {Pla\ss{}}, \citenamefont {Reiter}, \citenamefont {Andreoiu}, \citenamefont {Bergmann}, \citenamefont {Brunner}, \citenamefont {Curien}, \citenamefont {Dilling}, \citenamefont {Dudek}, \citenamefont {Dunling}, \citenamefont {Flowerdew}, \citenamefont {Gaamouci}, \citenamefont {Graham}, \citenamefont {Gwinner}, \citenamefont {Jacobs}, \citenamefont {Klawitter}, \citenamefont {Lan}, \citenamefont {Leistenschneider}, \citenamefont {Minkov}, \citenamefont {Monier}, \citenamefont {Mukul}, \citenamefont {Paul}, \citenamefont {Scheidenberger}, \citenamefont {Thompson}, \citenamefont {Tracy}, \citenamefont {Vansteenkiste}, \citenamefont {Wang}, \citenamefont {Wieser}, \citenamefont {Will},\ and\ \citenamefont {Yang}}]{Beck2021}%
  \BibitemOpen
  \bibfield  {author} {\bibinfo {author} {\bibfnamefont {S.}~\bibnamefont {Beck}}, \bibinfo {author} {\bibfnamefont {B.}~\bibnamefont {Kootte}}, \bibinfo {author} {\bibfnamefont {I.}~\bibnamefont {Dedes}}, \bibinfo {author} {\bibfnamefont {T.}~\bibnamefont {Dickel}}, \bibinfo {author} {\bibfnamefont {A.~A.}\ \bibnamefont {Kwiatkowski}}, \bibinfo {author} {\bibfnamefont {E.~M.}\ \bibnamefont {Lykiardopoulou}}, \bibinfo {author} {\bibfnamefont {W.~R.}\ \bibnamefont {Pla\ss{}}}, \bibinfo {author} {\bibfnamefont {M.~P.}\ \bibnamefont {Reiter}}, \bibinfo {author} {\bibfnamefont {C.}~\bibnamefont {Andreoiu}}, \bibinfo {author} {\bibfnamefont {J.}~\bibnamefont {Bergmann}}, \bibinfo {author} {\bibfnamefont {T.}~\bibnamefont {Brunner}}, \bibinfo {author} {\bibfnamefont {D.}~\bibnamefont {Curien}}, \bibinfo {author} {\bibfnamefont {J.}~\bibnamefont {Dilling}}, \bibinfo {author} {\bibfnamefont {J.}~\bibnamefont {Dudek}}, \bibinfo {author} {\bibfnamefont {E.}~\bibnamefont {Dunling}}, \bibinfo {author} {\bibfnamefont
  {J.}~\bibnamefont {Flowerdew}}, \bibinfo {author} {\bibfnamefont {A.}~\bibnamefont {Gaamouci}}, \bibinfo {author} {\bibfnamefont {L.}~\bibnamefont {Graham}}, \bibinfo {author} {\bibfnamefont {G.}~\bibnamefont {Gwinner}}, \bibinfo {author} {\bibfnamefont {A.}~\bibnamefont {Jacobs}}, \bibinfo {author} {\bibfnamefont {R.}~\bibnamefont {Klawitter}}, \bibinfo {author} {\bibfnamefont {Y.}~\bibnamefont {Lan}}, \bibinfo {author} {\bibfnamefont {E.}~\bibnamefont {Leistenschneider}}, \bibinfo {author} {\bibfnamefont {N.}~\bibnamefont {Minkov}}, \bibinfo {author} {\bibfnamefont {V.}~\bibnamefont {Monier}}, \bibinfo {author} {\bibfnamefont {I.}~\bibnamefont {Mukul}}, \bibinfo {author} {\bibfnamefont {S.~F.}\ \bibnamefont {Paul}}, \bibinfo {author} {\bibfnamefont {C.}~\bibnamefont {Scheidenberger}}, \bibinfo {author} {\bibfnamefont {R.~I.}\ \bibnamefont {Thompson}}, \bibinfo {author} {\bibfnamefont {J.~L.}\ \bibnamefont {Tracy}}, \bibinfo {author} {\bibfnamefont {M.}~\bibnamefont {Vansteenkiste}}, \bibinfo {author}
  {\bibfnamefont {H.-L.}\ \bibnamefont {Wang}}, \bibinfo {author} {\bibfnamefont {M.~E.}\ \bibnamefont {Wieser}}, \bibinfo {author} {\bibfnamefont {C.}~\bibnamefont {Will}}, \ and\ \bibinfo {author} {\bibfnamefont {J.}~\bibnamefont {Yang}},\ }\href {\doibase 10.1103/PhysRevLett.127.112501} {\bibfield  {journal} {\bibinfo  {journal} {Phys. Rev. Lett.}\ }\textbf {\bibinfo {volume} {127}},\ \bibinfo {pages} {112501} (\bibinfo {year} {2021})}\BibitemShut {NoStop}%
\bibitem [{\citenamefont {Bohr}\ and\ \citenamefont {Mottelson}(1969)}]{BMVol2}%
  \BibitemOpen
  \bibfield  {author} {\bibinfo {author} {\bibfnamefont {A.}~\bibnamefont {Bohr}}\ and\ \bibinfo {author} {\bibfnamefont {B.}~\bibnamefont {Mottelson}},\ }\href {https://books.google.pl/books?id=PpofAQAAMAAJ} {\emph {\bibinfo {title} {Nuclear Structure: Volume II (Nuclear Deformations)}}},\ \bibinfo {series} {Nuclear Structure}, Vol.~\bibinfo {volume} {2}\ (\bibinfo  {publisher} {Basic Books},\ \bibinfo {year} {1969})\BibitemShut {NoStop}%
\end{thebibliography}%

\end{document}